\documentclass[aps,prb,reprint,superscriptaddress,longbibliography]{revtex4-1}

\usepackage{graphicx}
\usepackage{latexsym}
\usepackage{amsmath}
\usepackage{amssymb}
\usepackage{amsfonts}
\usepackage{color}
\usepackage{bm}
\usepackage{verbatim}
\usepackage[]{color}
\definecolor{red}{rgb}{1,0,0}
\usepackage{framed}
\definecolor{shadecolor}{RGB}{222,222,221}

\definecolor{MS-color}{RGB}{128,0,128}

\begin{document}

\title{Magnetoelectric effects in Josephson junctions}

 \date{\today}
 
\author{I. V. Bobkova}
\affiliation{Institute of Solid State Physics, Chernogolovka, Moscow
  reg., 142432 Russia}
\affiliation{Moscow Institute of Physics and Technology, Dolgoprudny, 141700 Russia}
\affiliation{National Research University Higher School of Economics, Moscow, 101000 Russia}

\author{A. M. Bobkov}
\affiliation{Institute of Solid State Physics, Chernogolovka, Moscow reg., 142432 Russia}

\author{M.A.~Silaev}
\affiliation{Independent researcher }

 \begin{abstract}

The review is devoted to the fundamental aspects and characteristic features of the magnetoelectric effects, reported in the literature on Josephson junctions (JJs). The main focus of the review is on the manifestations of the direct and inverse magnetoelectric effects in various types of Josephson systems. They provide a coupling of the magnetization in superconductor/ferromagnet/superconductor JJs to the Josephson current. The direct magnetoelectric effect is a driving force of spin torques acting on the ferromagnet inside the JJ. Therefore it is of key importance for the electrical control of the magnetization. The inverse magnetoelectric effect accounts  for  the  back  action  of  the  magnetization  dynamics  on  the  Josephson  subsystem, in particular, making  the  JJ  to be  in  the  resistive  state  in  the  presence  of  the  magnetization dynamics of any origin. The perspectives of the coupling of the magnetization in JJs with ferromagnetic interlayers to the Josephson current via the magnetoelectric effects are discussed. 
 
 \end{abstract}

 \maketitle
 
\tableofcontents{}

\section{Introduction}

Since the discovery of the Josephson effect in 1962 \cite{Josephson1962,Josephson1964,Josephson1965}, there has been a growing interest in the fundamental physics \cite{Golubov2004} and applications of this effect. The achievements in Josephson-junction technology enabled the development of sensors for detecting ultralow magnetic fields and weak electromagnetic radiation, ultrafast digital rapid single flux quantum
(RSFQ) circuits, the design of large-scale integrated circuits for signal processing and general-purpose
computing as well as  adiabatic superconducting cells operating as an artificial neuron and synapse\cite{Likharev1991,Likharev2000,Soloviev2018}.

Theoretical investigations of hybrid structures involving superconductors and ferromagnets and subsequent experimental realization of superconductor/ferromagnet/superconductor Josephson junctions \cite{Golubov2004,Buzdin2005,Bergeret2005} have led to the discovery of spin-triplet Cooper pairs, thus giving rise to a synergy between superconductivity
and spintronics. The emergent new field was called superconducting spintronics and is being actively developed now \cite{Linder2015,Eschrig2015}. One of the key  effects in spintronics is the so-called magnetoelectric effects. In the most general sense the field embrace all the effects related to the coupling and interconversion of the charge and spin degrees of freedom. The field  already went beyond the framework of the fundamental physics only and, in particular, a scalable energy-efficient magnetoelectric spin–orbit logic (MESO) has been proposed \cite{Manipatruni2019} thus potentially opening new technology paradigm for improving energy efficiency
in beyond-CMOS computing devices. 

Here our goal is to review of current understanding of fundamental aspects of magnetoelectric effects in Josephson junctions, which potentially open new perspectives in superconducting spintronics. As an introduction, we discuss  the fundamental aspects of the related magnetoelectric effects  in nonsuperconducting systems briefly and then their analogues in superconducting materials and structures. The main part of the review is devoted to the magnetoelectric effects in a particular type of superconducting hybrids - Josephson junctions (JJs). In Sec.~\ref{main_properties} we discuss the manifestations of the direct and inverse magnetoelectric effects in different types of JJs, Sec.~\ref{dynamics} is devoted to the role of the magnetoelectric effects in the magnetization dynamics and electrical control of the ferromagnet magnetization in the JJs via ferromagnets. A specific for superconductivity  magnetoelectric effect - generation of triplet superconductivity by a moving condensate is discussed in Sec.~\ref{sec:triplets}. Sec.~\ref{conclusions} provides a short summary of the current situation in the field. 
Studies of magnetoelectric effects have a long history\cite{ODell1970,LL8,Fiebig2005}. The most common view is that the magnetoelectric media are characterized by unconventional equilibrium responses
to an electric field $\bm E$ and a magnetic field $\bm B$. While $\bm E$ induces only an electric polarization
in ordinary materials, it also creates a magnetization in magnetoelectric materials. Similarly, the magnetic field $\bm B$ in magnetoelectric materials generates an electric polarization in addition to a magnetization. Further, the advent of
multiferroic materials \cite{Fiebig2016,Spaldin2017} with their large magnetoelectric couplings has
greatly boosted current interest in  magnetoelectricity. But we do not touch on this physics in the review. Here we focus on the related phenomena of current-induced spin polarization
and the inverse effects. In this field besides the well-established spin Hall effect and inverse spin Hall effect \cite{Dyakonov1971,Dyakonov1971_2, Chazalviel1975,Hirsch1999,Mishchenko2004,Kato2004,Kato2004_2,Wunderlich2005,Raimondi2006,Raimondi2012,Valenzuela2006,Morota2011,Isasa2015} the direct and inverse magnetoelectric effects are also known. The essence of the direct magnetoelectric effect is creating a stationary spin
density $S_a$ along the $a$ direction in spin space in response to an electric field
$E_k$ applied in the $k$ direction in the real space:
\begin{eqnarray}
S^a = \sigma_k^a E_k .
\label{II_1}
\end{eqnarray}
The effect is known for a wide class of systems. It is theoretically investigated and measured for spin-orbit coupled materials \cite{Aronov1989,Edelstein1990,Kato2004,Silov2004}, where it is also called the Edelstein effect. In this case the Edelstein conductivity $\sigma_k^a$ is proportional to the SOC constant of the material. The mutual orientation of the applied electric field and the induced electron spin polarization is determined by the particular form of the  SOC. Let's consider the examples of Rashba and Dresselhaus SOC. These types of SOC were originally discussed for noncentrosymmetric zinc-blende or wurtzite semiconductors by Dresselhaus \cite{Dresselhaus1955}, and Rashba \cite{Rashba1960}. The Rashba-type SOC also arises due to the structural inversion asymmetry (SIA). SIA typically occurs at the surfaces or interfaces. An important realization of a system with Rashba-type spin-orbit coupling is a
2D electron gas in doped semiconductor heterostuctures \cite{Vasko1979,Bychkov1984}, that support an electron gas at the interface between
two materials. Another possibility to study the Rashba-effect in 2DEG are surfaces that support a surface
state, e.g. in Au(111)\cite{Lashell1996}: the electrons of the surface state move in a potential gradient that is provided by the surface itself. The hamiltonian term accounting for the Rashba SOC takes the form $\hat H_R = \alpha \hat z (\bm \sigma \times \bm p)$, where $\alpha$ is the Rashba constant and $\hat z$ is the unit vector along the $z$-axis,  chosen along the polar vector of the material, which determines the direction of the broken inversion symmetry. $\bm \sigma = (\sigma_x, \sigma_y, \sigma_z)^T$ is the vector of Pauli matrices in spin space and $\bm p$ is the electron momentum. For Rashba SOC $\sigma_x^y = -\sigma_y^x$, while the other components of $\sigma_k^a$ are zero. Therefore, for this case the induced spin polarization lies in the plane perpendicular to the polar vector and is perpendicular to the applied electric field. The simplest form of the Dresselhaus SOC hamiltonian, realized in the presence of strain along the (001) direction is $\hat H_D = \beta_D (p_x \sigma_x - p_y \sigma_y)$, where $\beta$ is the Dresselhaus SOC constant. If the current direction coincides with $x$ or $y$ axes, the induced spin polarization is directed along the current.

\begin{figure}[!tbh]
         \centerline{\includegraphics[clip=true,width=2.5in]{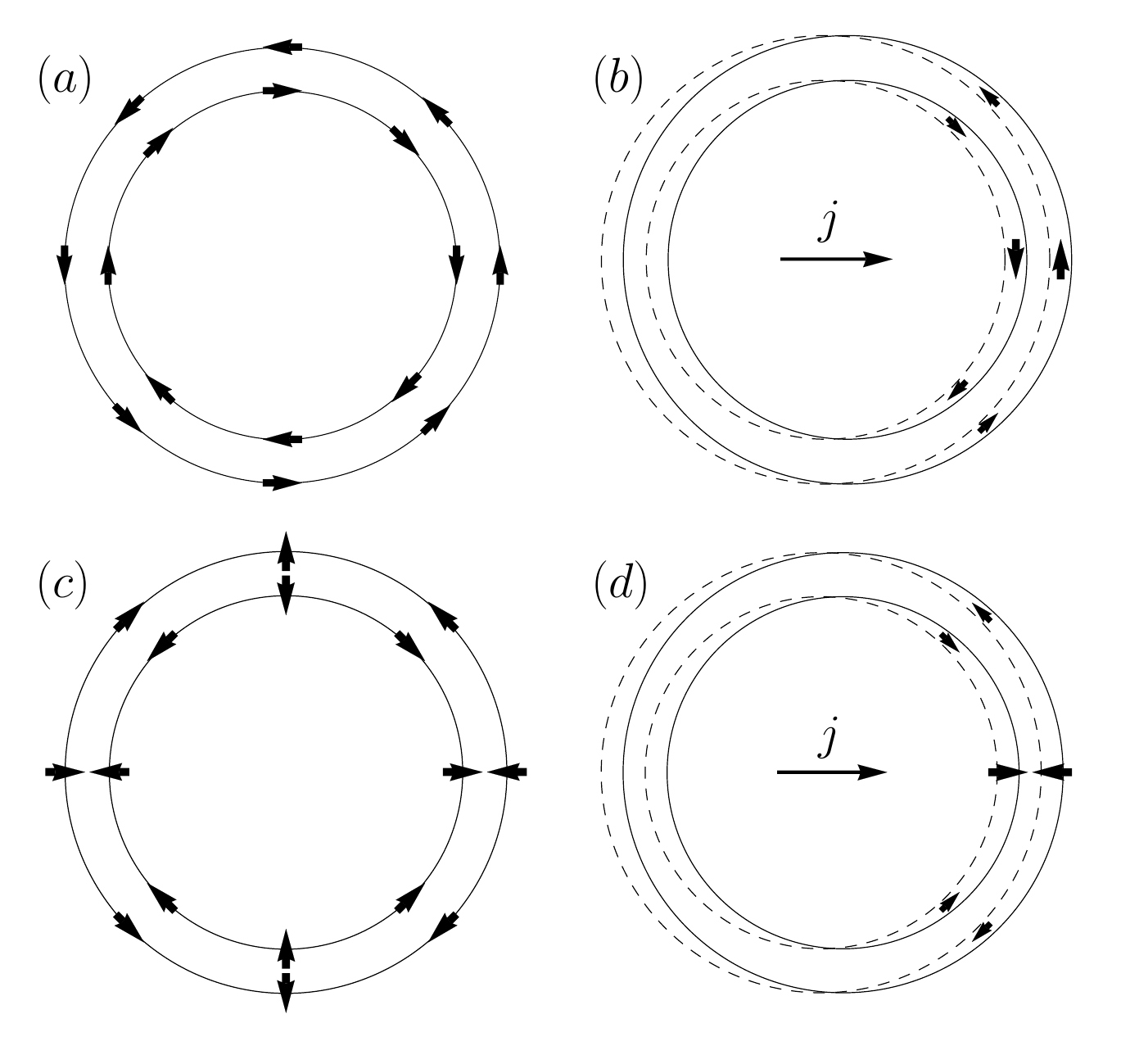}}
        \caption{Spin-split helical Fermi surfaces for (a) Rashba  and (c) Dresselhaus  spin-orbit coupled materials. (b) and (d) Current-induced shift of the Fermi surface. The original Fermi surface in the absence of the applied current is shown by the dashed lines. The direction of the accumulated spin in each of the subbands is shown by arrows. }
\label{Direct_SO_1}
\end{figure}

The reason for the electrically induced spin polarization and qualitative understanding of its direction with respect to the current in normal quasi-2D systems is clearly seen from Fig.~\ref{Direct_SO_1}, where the helical Fermi surfaces of the Rashba and Dresselhaus materials are demonstrated. The electric current results in the total shift of all the Fermi surfaces by $q$ along the current direction. This leads to the nonzero average spin polarization of the corresponding electronic states in each of the helical bands. Due to the different spin structure of the helical Fermi surfaces the directions of the resulting electron polarizations in the Rashba and Dresselhaus cases are different. The average polarization is perpendicular to the applied current for the Rashba case and can have different mutual orientations with the current for the Dresselhaus SOC depending on the orientation of the current with respect to the crystal axes. The split helical Fermi surfaces contribute to the polarization in opposite directions, as it is seen from Fig.~\ref{Direct_SO_1}. This leads to a great reduction of the total polarization, the resulting effect is nonzero only due to the difference between the Fermi momenta for the helical subbands. Therefore, the current induced spin polarization is always proportional to the ratio $\Delta_{so}/\varepsilon_F$, where $\Delta_{so}$ is the energy splitting of the helical subbands, $\Delta_{so} \sim \alpha (\beta_D) p_F$ for the Rashba (Dresselhaus) case.

 The direct magnetoelectric effect has also been predicted and measured in topological insulators\cite{Burkov2010,Culcer2010,Yazyev2010,Li2014,Li2016}, where the mutual orientation of the spin polarization and the current is the same as for the case of Rashba materials. In addition, the direct magnetoelectric effect  also exists in spin-textured ferromagnets, where the induced spin polarization takes the form:
\begin{eqnarray}
\bm S_\perp = -\frac{b_J j}{ J_{sd}M^2} \bm M \times \partial_x \bm M + \frac{c_J j}{ J_{sd}M} \partial_x \bm M ,
\label{II_2}
\end{eqnarray}
which results in the well-known spin transfer torque \cite{Tsymbal2012} acting on the magnetization according to 
\begin{eqnarray}
\bm T =  J_{sd} \bm M \times \bm S_\perp = b_J j \partial_x \bm M - \frac{c_J j}{M} \bm M \times \partial_x \bm M ,
\label{II_2a}
\end{eqnarray}
where $M$ is the saturation magnetization in the ferromagnet, $J_{sd}$ is the coupling constant of the exchange interaction between the $s$-band conduction electrons and $d$-band localized electrons responsible for the magnetism. $\bm S_\perp$ means the component of the current-induced spin polarization, perpendicular to the magnetization direction, because it is this component that leads to a torque on the magnetization. 

There is also an inverse magnetoelectric effect (it is also called by the spin-galvanic effect), which
consists of generating a charge current $j_k$ by a steady spin
imbalance, which can be induced, for example, by a time-dependent magnetic field via the
paramagnetic effect \cite{Shen2014}:
\begin{equation}
j_k = \sigma_k^a (g \mu_B \dot B^a),
\label{II_3}    
\end{equation}
where $g$ is the Lande factor, $\mu_B$ is the Bohr magneton, and $\dot B^a$ is
the time derivative of the magnetic field component along the $a$ axis.
The inverse magnetoelectric effect effect has been observed in experiments
with spin-orbit coupled materials \cite{Ganichev2002,Sanchez2013} and topological insulators\cite{Sanchez2016,Zhang2016}. 

The inverse magnetoelectric effect also takes place in spin-textured metallic ferromagnets. In this case, it manifests itself as the so-called electromotive force (emf) induced by the magnetization dynamics  \cite{Stern1992,Stone1996,Volovik1987,Berger1986,Barnes2007,Duine2008,Saslow2007,Tserkovnyak2008,Zhang2009,Yang2009,Yang2010}
\begin{eqnarray}
F_i = \frac{\hbar}{2}\Bigl[ \bm m (\partial_t \bm m \times \nabla_i \bm m) + \beta (\partial_t \bm m  \nabla_i \bm m)\Bigr],
\label{II_3a}
\end{eqnarray}
where $\bm m(\bm r, t)$ is the unit vector in the direction of the magnetization and $\beta$ is a phenomenological parameter. Due to the existence of the electromotive force the magnetization dynamics leads to appearance of an additional voltage drop. This voltage can vary in the range from $nV$ to $\mu V$ \cite{Yang2009} and in special situations can be used for electrical detection of the presence of magnetization dynamics \cite{Barnes2006}. 

Conceptually the same magnetoelectric effects also take place in superconducting systems.  However, here the physical situation is somewhat different because of the presence of the
superconducting condensate. In contrast to the normal case, in a superconductor
an equilibrium electric supercurrent can flow in the absence
of an external electric field and is directly related to the gauge invariant superconducting  condensate phase $\bm j \propto \bm v_s \propto \nabla \varphi - (2e/c)\bm A$, where $v_s$ is the condensate velocity, $\varphi$  is the phase of the superconducting order parameter and $\bm A$ is the vector potential. That leads to two consequences: (i) a supercurrent can generate an
equilibrium spin polarization in the presence of intrinsic SOC \cite{Edelstein1995,Edelstein2005,Sanz-Fernandes2019,Ilic2020}, extrinsic impurity-induced SOC \cite{Bergeret2016,Virtanen2021}, in topological insulator-based superconducting heterostructures \cite{Bobkova2016,Bobkova2017_4} and in superconductor/ferromagnet hybrids with spin-textured ferromagnets \cite{Rabinovich2019, Meng2019} and (ii) in contrast to the normal case, in superconductors a static Zeeman field B can induce a supercurrent $j_k$:
\begin{equation}
j_k = \chi_k^a h^a .
\label{II_4}    
\end{equation}
where $h^a = (1/2)g \mu_B B^a$. This effect
has been obtained for a case of a 2D superconductor with Rashba SOC  \cite{Yip2002}. It was also discussed  for heterostructures consisting of the superconducting and ferromagnetic layers with SOC \cite{Bobkova2004,Pershoguba2015,Malshukov2016,Mironov2017,Malshukov2020_2} or for a ferromagnet/superconducting TI hybrid structures \cite{Malshukov2020_3}, when the exchange field is not induced by the externally applied magnetic field, but is generated by the proximity to the ferromagnet. In the case of heterostructures with thick enough superconducting layer (the thickness should be much larger than the superconducting coherence length) \cite{Bobkova2004,Mironov2017} such a state with a spontaneous supercurrent flowing along the S/F interface and decaying into the depth of the superconductor can be a true ground state of the system. The same is valid  if the exchange field is spatially inhomogeneous \cite{Pershoguba2015,Malshukov2016} and even a topologically nontrivial vortex states can appear under the appropriate conditions \cite{Malshukov2020_2,Malshukov2020_3}. 

But for the case of the superconductors with an intrinsic SOC in the homogeneous Zeeman field the state carrying homogeneous nonzero supercurrent is not the true ground state. In the true ground state the superconducting phase gradient is developed in order to compensate the supercurrent. The resulting state is characterized by the zero supercurrent and {\it nonzero} superconducting phase gradient. It is called by the helical state and is a specific for superconducting systems manifestation of the inverse magnetoelectric effect. There is an important difference between the helical state and the well-known inhomogeneous FFLO state \cite{Fulde1964,Larkin1965,Mironov2012,Mironov2018}. While in the phase-modulated FFLO state the direction of the superconducting phase gradient does not depend on the direction of the exchange field, in the helical phase they are directly related. This fact results in strong coupling between the magnetization and the condensate phase. Therefore, it leads to the possibilities of the electrical control of the magnetization dynamics, which look perspective from the point of view of spintronics applications.
The helical state has been predicted for superconductors with intrinsic Rashba SOC under the applied Zeeman field \cite{Edelstein1989,Barzykin2002,Samokhin2004,Kaur2005,Dimitrova2007,Houzet2015} and superconducting hybrids with spin-textured ferromagnets \cite{Rabinovich2019,Meng2019}, where the magnetic inhomogeneity plays a role of the effective SOC.

The helical state is a kind of inverse magnetoelectric effect, which is realized in the simply connected superconducting systems. Similar effects occur
also in an S-X-S Josephson junction, between two 
superconductors and a normal or ferromagnetic interlayer
X with an intrinsic SOC or if the interlayer is a spin-textured ferromagnet. In a Josephson junction the
supercurrent depends on the phase difference $\varphi$ between
the superconducting electrodes. Similar to the simply connected superconductors a Zeeman 
field may induce a supercurrent through the junction at zero phase difference between the superconductors according to Eq.~(\ref{II_4}). In the ground state of the junction this "anomalous supercurrent", generated by the inverse magnetoelectric effect, is compensated by the phase shift $\varphi_0 \neq 0,\pi$. It is called by the anomalous ground state phase shift and the Josephson junctions manifesting this effect are called $\varphi_0$-junctions. 

The $\varphi_0$-junctions have been predicted in a wide class of systems including S/F/S junctions with intrinsic SOC, S/N/S junctions with intrinsic SOC under applied Zeeman field\cite{Krive2004,Nesterov2016,Reynoso2008,Buzdin2008,Zazunov2009,Brunetti2013,Yokoyama2014,Bergeret2015,Campagnano2015,Konschelle2015,Kuzmanovski2016,Malshukov2010}, S/topological insulator/S junctions under the applied Zeeman field or if the Zeeman field in the topological insulator surface states is induced by the proximity to a ferromagnet (S/TI-F/S junctions) \cite{Tanaka2009,Linder2010,Dolcini2015,Zyuzin2016,Lu2015} and also in S/F/S junctions with spin-textured interlayers\cite{Braude2007,Asano2007,Liu2010,Alidoust2013,Mironov2015,Kulagina2014,Eschrig2008,Grein2009,Moor2015,Moor2015_2,Silaev2017,Bobkova2017_3,Rabinovich2018,Meng2019}. Below we will discuss all the mentioned classes of systems in more details. The anomalous phase shift has been observed experimentally for Al/InAs/Al Josephson junctions (JJ) \cite{Mayer2020}, in JJs via nanowire quantum dots \cite{Szombati2016}, in $Bi_2Se_3$ JJs \cite{Assouline2019} and in JJs via bismuth nanowires \cite{Murani2017} under the applied magnetic field. 
The magnetoelectric nature of the anomalous phase shift has been unveiled in Ref.~\onlinecite{Konschelle2015}. It has  also been reported very recently that the anomalous phase shift is a key ingredient of the mechanism providing an extremely long-range interaction of magnetic moments in a coupled system of JJs with magnetic interlayers\cite{Bobkov2022}.  There is also a recent review \cite{Shukrinov2021}, specially devoted to the physics of the anomalous phase shift.

Naturally, the equilibrium direct magnetoelectric effect discussed above can also occur in Josephson junctions. Indeed, it has been predicted for JJs via normal interlayers with Rashba SOC in diffusive \cite{Malshukov2008} and ballistic \cite{Bobkova2017_2} systems and via TI interlayers \cite{Bobkova2016}. The effect plays a key role in the electrical control of magnetization in S/F/S Josephson junctions, which is also discussed in detail below.

For completeness, we briefly mention other reported effects in superconducting hybrids, which can also be viewed as magnetoelectric ones but are not discussed in detail in this review. One group of effects is related to  different types of quasiparticle nonequilibrium in the system. Among them are  spin-charge conversion effects in superconducting hybrids. They involve a nonequilibrium spin polarization and spin current pumped into a superconducting system by some external source. It has been shown that for the Rashba SOC and SOC caused by spin-orbit impurities such a nonequilibrium spin distribution
can generate the electric current and electric potential in superconductors \cite{Malshukov2017,Espedal2017,Takahashi2008}. This nonequilibrium situation resembles much the analogous inverse magnetoelectric and spin Hall effects in normal systems.

\section{Direct and inverse magnetoelectric effects in Josephson junctions: main properties and physical systems}

\label{main_properties}

\subsection{Anomalous phase shift - a realization of inverse magnetoelectric effect in Josephson junctions with spin-orbit coupling and Zeeman field}

\label{af_FSO}

The minimal form of the  current-phase relation (CPR)
characterizing the dc Josephson effect 
is given by $ j (\varphi)= j_{c}\sin\varphi$. Here
$j$ is the total superconducting current flowing across the junction,  $|j_c|$ is the critical current and 
$\varphi$ is the phase difference between superconducting electrodes \cite{Josephson1965,Golubov2004}.
 The ordinary Josephson junctions have $j_c>0$ 
  yielding the zero phase difference  ground state $\varphi =0$.  
 In certain cases $j_c<0$ 
  leading to the ground state  
 $\varphi =\pi$. Such $\pi$-junctions are realized   
 in S/F/S JJs \cite{buzdin1982critical,ryazanov2001coupling,oboznov2006thickness,Buzdin2005}, 
 non-equilibrium S/N/S JJs \cite{baselmans1999reversing}, non-equilibrium S/F/S systems \cite{golikova2020controllable},
 $d$-wave superconductors\cite{van1995phase,hilgenkamp2003ordering},  semiconductor nanowires \cite{van2006supercurrent}, 
 gated carbon nanotubes\cite{cleuziou2006carbon} or multi-terminal Josephson systems \cite{vischi2017coherent}. 
 The $\pi$-junctions can be used in scalable superconducting logic and quantum computers \cite{terzioglu1998complementary, ustinov2003rapid,
 Feofanov2010, Gingrich2016}.
    
 Even more exotic situation occurs  in systems with 
 magnetoelectric effects where the 
  $\varphi_0$-junctions are realized. 
  They are described by the CPR \cite{Buzdin2008}
 \begin{equation} \label{Eq:CPRgen}
 j (\varphi)= j_{c}\sin(\varphi+\varphi_0)\; .  
 \end{equation}
 with {\it anomalous (spontaneous) phase shift } $\varphi_0\neq 0, \pi$. 
 In this case 
 there is a finite supercurrent
  at zero phase difference  $j_{an} = j_{c}\sin\varphi_0$ 
   called  the {\it anomalous (spontaneous) current}.  

 The Josephson energy 
 $E_J = j_c [1-\cos(\varphi+\varphi_0)]$ yields the ground state with non-trivial phase difference  
 $\varphi=-\varphi_0$ and zero current $j(\varphi_0)=0$. 
 Such a phase-shifted ground state is analogous to the helical state in the 
 homogeneous superconductor discussed above. 
 The general symmetry requirements for obtaining the 
 $\varphi_0$-Josephson junctions are the same as 
 for having magnetoelectric
 coupling in the homogeneous superconductor. 
 That is, we need to combine the two symmetry 
 breaking mechanism.
First, the time-reversal symmetry is to be broken by the  Zeeman field $\bm h$ inside the Josephson junction. 
 Second, the orbital and spin degrees of freedom should be coupled so that it is impossible to invert the magnetic moment by spin rotation independently from  the orbital coordinates. By analogy with spontaneous current (\ref{II_4}) one can construct the phenomenological expression for the anomalous phase shift $\varphi_0 \propto \chi_x^a h^a$, where $\bm x$ is the axis across Josephson junction.  
 This type of symmetry breaking is enabled e.g. by Rashba-type SOC  
 $\chi_c^a \propto \varepsilon_{c b a} n_b $  with anisotropy vector $\bm n$. 
 In this case the spontaneous current is 
 $\bm j \propto \bm n \times \bm h  $ 
 and the anomalous phase shift is $\varphi_0 \propto \bm x \cdot (\bm n \times \bm h )$. 
 
 Theoretical description of the
 $\varphi_0$-junctions  
is  significantly more challenging than that of the $0$ and $\pi$-junctions.      
 To obtain the anomalous phase shift one has to include 
the magnetoelectric coupling which is beyond the standard quasiclassical approximation \cite{Bergeret2015,
tokatly2017usadel,huang2018extrinsic,Konschelle2015,Bergeret2016,Bobkova2016, 
Silaev2017,
Bobkova2017, 
Bobkova2017_2, 
Bobkova2018,
Rabinovich2018,
Nashaat2019,
Rabinovich2019_2,
Rabinovich2020,
Bobkova2020,
Sanz-Fernandes2019,
Meng2019,
meng2019josephson}. 
For Rashba-type SOC described by the hamiltonian $H_{R} = \alpha [\bm p \times \bm n] \bm \sigma$ in the ballistic regime and for large Rashba constant $\alpha$, the anomalous phase
shift is given by \cite{Buzdin2008} 
\begin{eqnarray}
\varphi_{0b} = \frac{4 h \alpha d}{(\hbar v_F)^2} ,
\label{phi0_ballistic}
\end{eqnarray}
where $d$ is the length of the Josephson junction interlayer and $v_F$ is the Fermi velocity of the electrons in the interlayer. In the diffusive regime for weak $\alpha$, highly transparent interfaces and neglecting spin-relaxation, the predicted result for the anomalous anomalous phase shift is
\begin{eqnarray}
\varphi_{0d} = \frac{\tau m^{*2} h (\alpha d)^3}{3 \hbar^6 D} ,
\label{phi0_diffusive}
\end{eqnarray}
where $\tau$ is the elastic scattering time, $m^*$ is the effective electron mass and $D$ is the diffusion constant \cite{Bergeret2015}.

\begin{figure}[!tbh]
         \centerline{\includegraphics[clip=true,width=3.6in]{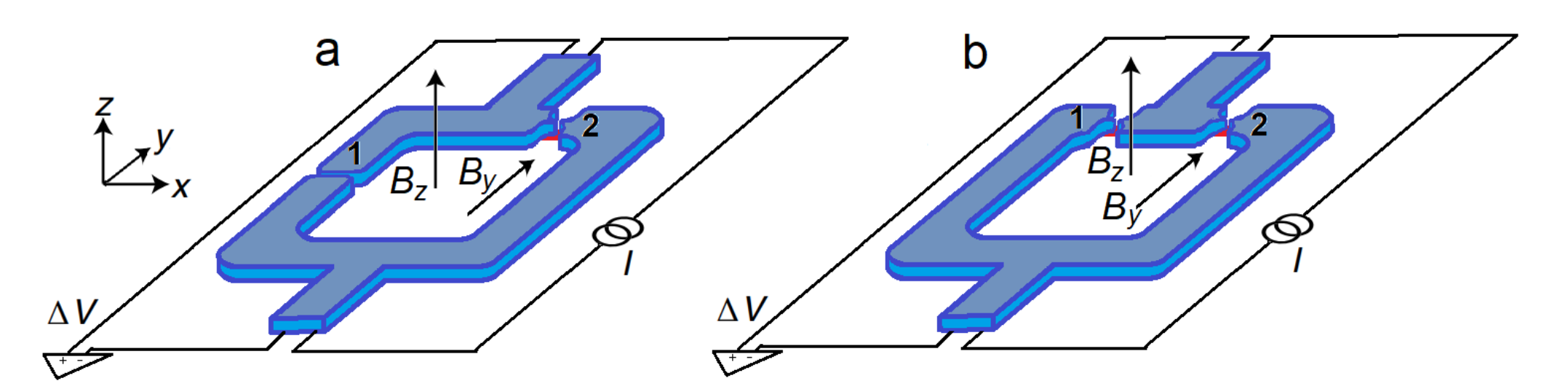}}
        \caption{Sketches of experimentally used SQUID-based setups for the anomalous phase shift measurements. (a) Asymmetric SQUID. The critical current of the reference JJ 1, without an anomalous phase shift, is much higher than the critical current of the $\varphi_0$-JJ 2. (b) Symmetric SQUID. Both JJs 1 and 2 have the same critical currents. The anomalous phase shifts for these junctions are opposite due to the current flowing in the opposite directions.}
\label{squid_sketch}
\end{figure}

The anomalous Josephson effect based on the systems with SOC has been found in  recent experiments. Anomalous phase junctions were demonstrated in $Bi$-nanowires \cite{Murani2017}, $InSb$
nanowires in a quantum dot geometry \cite{Szombati2016}, in JJ
using $Bi_2Se_3$ \cite{Assouline2019}, in  heterostructures formed by $InAs$ and
epitaxial superconducting $Al$ \cite{Mayer2020} and also in JJs via $InAs$ nanowires \cite{strambini2020josephson}. In the quantum dot realization \cite{Szombati2016} and in Ref.~\onlinecite{Mayer2020} the gate-tunable phase has been achieved. The JJs of Ref.~\onlinecite{Szombati2016} support a
few modes and consequently exhibit small critical currents, and the structures investigated in Ref.~\onlinecite{Mayer2020} have high interface transparency and large critical currents. In $Bi_2Se_3$, which is a
topological insulator, large planar $\varphi_0$-junction are possible \cite{Assouline2019}, however, they are not gate-tunable. In all the experimental works, the exchange field inside the interlayer of the JJ, which is required for the generation of the anomalous phase shift, has been created by the externally applied in-plane magnetic field via the Zeeman effect $h = (1/2)g \mu_B B_{y}$, where $g$-is the electron $g$-factor for the corresponding material and $B_{y}$ is the magnetic field component, perpendicular to the anisotropy vector $\bm n$ (along $z$, see Fig.~\ref{squid_sketch}) and to the Josephson current direction.

The anomalous phase shift has been observed directly through measurements of the current-phase relationship in a Josephson interferometer. A  typical current-biased measurement of a single JJ shows no measurable signature. Under the applied current, the phase difference across the JJ changes to maximise the critical current. This means that any phase shift applied to such a system will be invisible. Therefore, the experimental works  use SQUID geometry, whose primary property is phase sensitivity. 

\begin{figure}[!tbh]
         \centerline{\includegraphics[clip=true,width=3.6in]{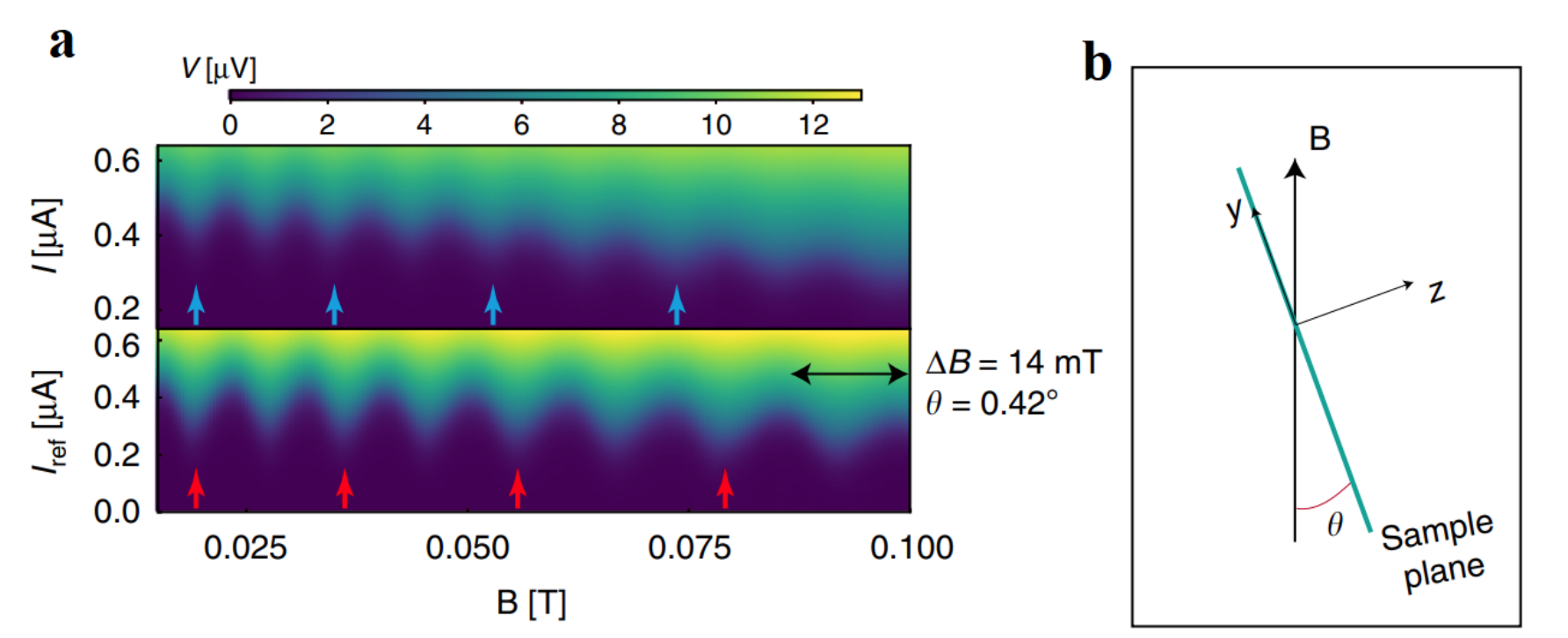}}
        \caption{(a) Voltage map showing the critical
current oscillations of the anomalous device (upper picture) and of the reference device (bottom picture) as a function of magnetic field $B$. The critical current of both devices oscillates due to the perpendicular component of the magnetic field $B_z = B \sin \theta$, as it is sketched in panel (b). Due to the anomalous phase shift, the frequency of the anomalous device is larger than the reference one. The oscillation frequency can be changed by mechanically tilting the sample, i.e by changing the angle
$\Theta$ between the plane containing the superconducting loop and the magnetic field $B$.  The colored arrows guide  the eyes to help visualise the increased phase shift in the anomalous device. Adopted from Ref.~\onlinecite{Assouline2019}.}
\label{oscillations}
\end{figure}

One scheme of measurements, which was realized in Refs.~\onlinecite{Murani2017,Assouline2019,Szombati2016}, is based on the asymmetric SQUID configuration. The SQUID consists of two junctions in parallel with very different critical currents $I_{c1} \gg I_{c2}$, where $I_{c1}$ is the critical current of the reference JJ and $I_{c2}$ is the critical current of the investigated JJ. The sketch of the asymmetric SQUID is presented in Fig.~\ref{squid_sketch}(a). The phase differences $\varphi_1$ and $\varphi_2$ for the two junctions are linked by the relation $\varphi_1 - \varphi_2 = 2\pi \Phi/\Phi_0$, where $\Phi = B_z S$ is the magnetic flux enclosed in
the SQUID of surface $S$, $B_z$ is a magnetic field component perpendicular to the sample, i.e. along $\bm e_z$, and $\Phi_0$ is the flux quantum. As the critical
current $I_{c1}$ is much higher than $I_{c2}$, 
then $\varphi_1 = \pi/2$ and $I_c = I_{c1}+I_{c2}\cos [2\pi \Phi/\Phi_0 - \varphi_0]$. Thus, a measurement of the critical current $I_c$ as function of $B_z$ provides a measure of the current $I_{2}$ as function of $\varphi_2$, i.e. the CPR. 
In the experiments the Zeeman field $\bm h$ has been induced by the externally applied in-plane magnetic field $h = (1/2)g\mu_B B_y$, where $B_y = B\cos \theta$ is the in-plane component of the applied magnetic field, see Fig.~\ref{oscillations}(b). Then the anomalous phase results in the increased oscillation frequency of the critical current as a function of the applied magnetic field \cite{Assouline2019} $2\pi \Phi/\Phi_0 - \varphi_0 = \omega B$, where $\omega = 2 \pi S \sin \theta/\Phi_0 - \varphi_0/B$. The corresponding experimental results adopted from Ref.~\onlinecite{Assouline2019} are presented in Fig.~\ref{oscillations}(a).

\begin{figure}[!tbh]
         \centerline{\includegraphics[clip=true,width=3.6in]
         {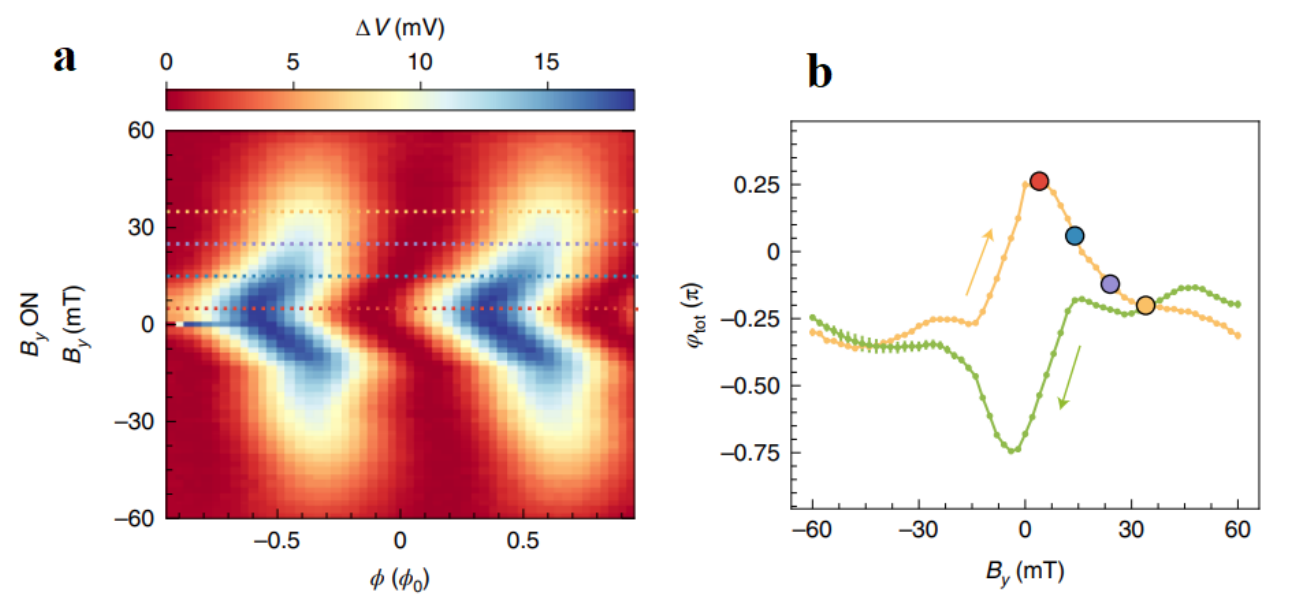}}
        \caption{(a) Voltage drop $\Delta V(\Phi)$ at constant current bias $I = 1 \mu A$ versus in-plane magnetic field $B_y$ applied
orthogonal to the nanowire axis. 
(b) Phase shift $\varphi_{tot}$ extracted from the data  (a) with back (blue) and forth (red) sweeps in $B_y$.  
The colored points in panel (b) correspond to the field values marked by the dashed lines of the same color in panel (a). 
Figure is adopted from  Ref.~\onlinecite{strambini2020josephson}. }
\label{phi0_symmetric}
\end{figure}

The other measurement scheme is to consider the symmetric SQUID, where the JJs has the same critical currents $I_{c1} = I_{c2} = I_c$ \cite{strambini2020josephson}. In this case the total supercurrent through the interferometer is
\begin{eqnarray}
I_s = 2I_c \sin \delta_0 \cos \bigl[ \frac{1}{2}\bigl( 2\pi \frac{\Phi}{\Phi_0}+ \varphi_{tot} \bigr) \bigr],
\label{js_symmetric}
\end{eqnarray}
where $\delta_0 = (\varphi_0^{(1)}+\varphi_0^{(2)})/2 - (\varphi_1 + \varphi_2)/2$ and $\varphi_{tot} = \varphi_0^{(1)}-\varphi_0^{(2)}$ is the total anomalous phase built in the interferometer. With the geometry realized in Ref.~\onlinecite{strambini2020josephson} and shown in Fig.~\ref{squid_sketch}(b), the two junctions experience the same in-plane magnetic field orientation but the supercurrents flow in opposite directions resulting in $\varphi_0^{(1)} = - \varphi_0^{(2)}$ and $\varphi_{tot} = 2 \varphi_0$. The stable state configuration of the SQUID is achieved by minimizing the total Josephson free energy obtained at $\delta_0 = \pi/2$. Therefore, the maximum supercurrent, which can be sustained  by the system, is 
\begin{eqnarray}
I_S(\Phi) = 2I_c \Bigl | \cos \bigl[ \pi \frac{\Phi}{\Phi_0}+ \frac{1}{2}\varphi_{tot}  \bigr] \Bigr|.
\label{jc_symmetric}
\end{eqnarray}
Therefore, the dependence $I_S(\Phi)$ contains the anomalous phase shift. Notably, there is a replica of the $I_S(\Phi)$ oscillations in the 
voltage drop $\Delta V (\Phi)$ when $I>I_S$, and the SQUID operates in the dissipative regime, as conventionally realized with strongly overdamped JJs. This replica is used in Ref.~\onlinecite{strambini2020josephson} for measurements of the anomalous phase shift. The results, adopted from Ref.~\onlinecite{strambini2020josephson}, are represented in Fig.~\ref{phi0_symmetric}. Fig.~\ref{phi0_symmetric}(a) demonstrates evolution of $\Delta V(\Phi)$ at constant current bias as a function of  the applied in-plane magnetic field. Fig.~\ref{phi0_symmetric}(b) shows the resulting anomalous phase shift extracted from the data in Fig.~\ref{phi0_symmetric}(a) according to the resistively shunted junction (RSJ) model relation $\Delta V = (R/2)\sqrt{I^2 - 4 I_{c}^2 \cos(\pi \Phi/\Phi_0 + \varphi_{tot}/2)^2}$.

Experimental data provide rather large values of the anomalous phase shift of the order of $\pi$. For example, in Ref.~\onlinecite{Assouline2019} $\varphi_0 \approx 0.9 \pi$ was measured at $B_y \approx 100 mT$. It was compared to the theoretical predictions for the ballistic Eq.~(\ref{phi0_ballistic}) $\varphi_{0b} \approx  0.01 \pi$ and diffusive Eq.~(\ref{phi0_diffusive}) $\varphi_{0d} \approx 0.94 \pi$. The authors concluded that their JJs agree with the theory developed in Ref.~\onlinecite{Bergeret2015}. In Ref.~\onlinecite{strambini2020josephson} it was also reported that their data could be fitted by the diffusive theory \cite{Bergeret2015}. Rather high values of $\varphi_0 \sim \pi/2$ at $B_y \approx 400 mT$ (the particular value of $\varphi_0$ depends on the gate voltage) have also been observed in Ref.~\onlinecite{Mayer2020}. The authors compare their results to the theoretical predictions expressed by Eqs.~(\ref{phi0_ballistic}) and (\ref{phi0_diffusive}) and concluded that  both results return values of $\varphi_0$ which are much smaller than the observed ones. A possible explanation is that the investigated JJs were in the short junction limit with a fully developed  proximity effect inside the interlayer, which is not described by those theoretical results.

As it was mentioned earlier, in all the discussed experiments, the exchange field generating $\varphi_0$ has been created by the externally applied field. Refs.~\onlinecite{Assouline2019} and \onlinecite{Mayer2020} report  the linear dependence of the anomalous phase shift on the in-plane magnetic field, as it is predicted by Eqs.~(\ref{phi0_ballistic}) and (\ref{phi0_diffusive}). The experimental results reported in Ref.~\onlinecite{strambini2020josephson} are more complicated. The nonmonotonic dependence of the anomalous phase shift on the applied in-plane field with a maximum shift at $B_y \approx 5 mT$ and saturation for $|B_y| > 30 mT$ has been observed, see Fig.~\ref{phi0_symmetric}(b). The authors suggest that this behavior is due to magnetic Kondo impurities in the nanowire. Due to the antiferromagnetic nature of the Kondo interaction, the effective exchange field created
by these unpaired spins is opposite to the Zeeman field generated by $B_y$ so that the two contributions are competing in the anomalous
phase with a partial cancellation. The total anomalous phase shift is divided into two contributions $\varphi_0(B_y) = \varphi_{int}(B_y)+\varphi_{ext}(B_y)$, where $\varphi_{int}(B_y)$ is an intrinsic phase shift,  which is present even in the absence
of the in-plane magnetic field if a finite $B_y$ has been previously applied. Since it stems from a ferromagnetic ordering, $\varphi_{int}$ depends only on the history of $B_y$ and shows a hysteresis in the back and forth sweep direction. The extrinsic contribution to the phase shift, $\varphi_{ext}$
stems directly from the external magnetic field. The dependence $\varphi_{ext}(B_y)$ is characterized by a linear increase at low magnetic fields  up to a maximum phase shift of $\pm \pi/2$. Based on the diffusive theory\cite{Bergeret2015} the authors of Ref.~\onlinecite{strambini2020josephson} developed a model explaining the observed behavior of $\varphi_{ext}$. It leads to $\varphi_{ext} \approx  C \alpha^3 B_y + O(B^3)$. This model provides a reasonable explanation of the obtained data and predicts a nonuniversal saturation of the anomalous phase shift at large fields with the saturation value depending on the spin-orbit strength. 

The measurement of the intrinsic anomalous phase shift generated by the magnetic impurities reported in Ref.~\onlinecite{strambini2020josephson} is an important step to the development of a new generation of anomalous phase shift JJs based on magnetic materials. It will open a great perspective for experimental investigation of exciting physics related to the coupling between the condensate phase and the magnetization of the magnet and their interplay. Several examples of the corresponding theoretical predictions are discussed below in this review.

\subsection {Anomalous phase shift in topological insulator-based S/F/S junctions}

\label{af_TI}

Here  we consider the Josephson junction through topological insulators 
where the  SOC is  so strong  that only one helical band is present.
 In such systems, one can expect the strongest possible spontaneous phase shift effect. At the same time, having only one helical band allows for the significant simplification of the theoretical description using the generalized quasiclassical theory. More particular, we consider the anomalous phase shift in  superconductor/ferromagnet/superconductor (S/F/S) Josephson junctions  designed on a three dimensional topological insulator (3D TI) surface. At present, great progress has been made in the experimental implementation of F/TI hybrid structures. In particular, a structure was successfully implemented experimentally, in which a sufficiently strong exchange field was induced in the surface states of the TI due to the proximity effect with a high-$T_c$ ferromagnetic insulator \cite{Jiang2014,Wei2013,Jiang2015,Jiang2016}.

\begin{figure}[!tbh]
         \centerline{\includegraphics[clip=true,width=3.0in]{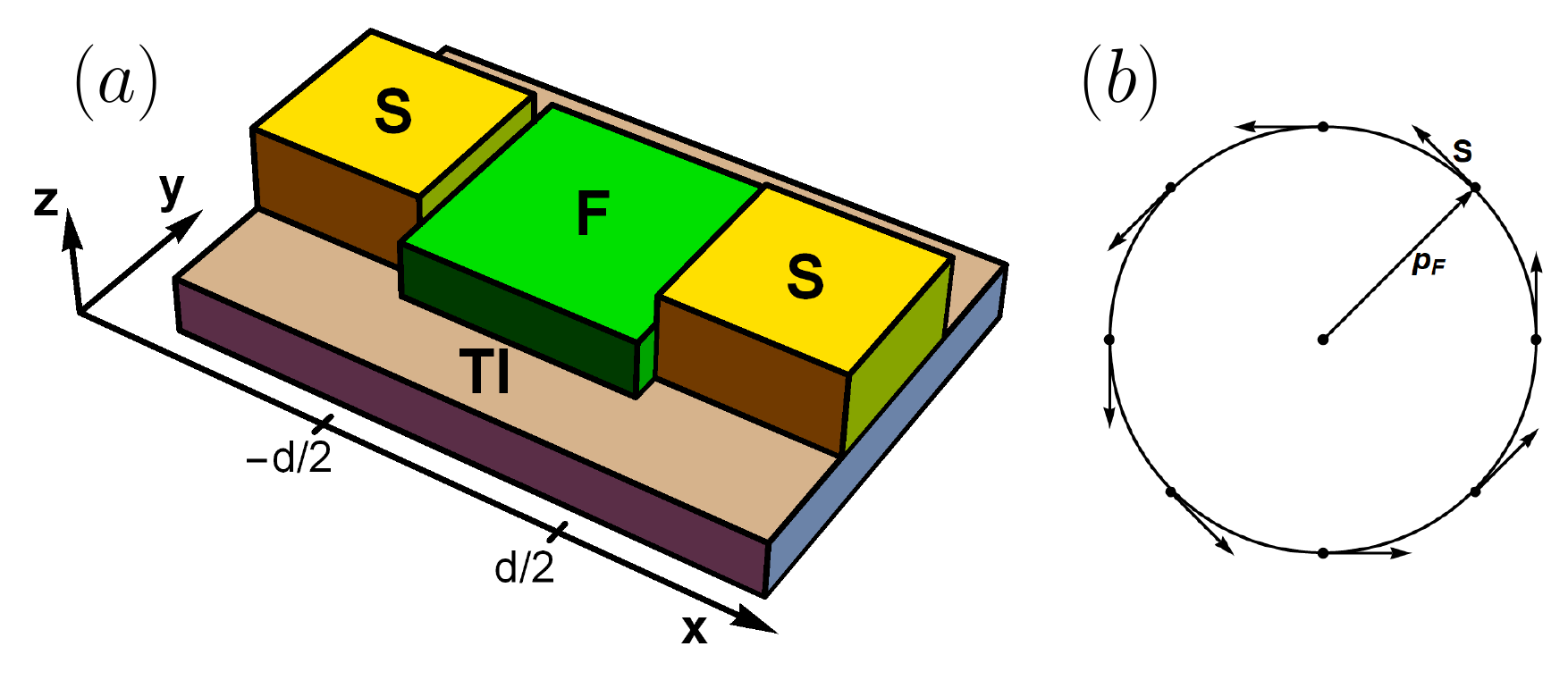}}
        \caption{(a) Sketch of the superconductor/ferromagnet/superconductor(S/F/S) Josephson junction on top of the  3-dimensional topological insulator (3D TI). The length of the interlayer $d$ is assumed to be of the order of the normal state coherence length $\xi_N$ of the 3D TI conductive surface layer. (b) Helical Fermi surface of the conductive surface states of the 3D TI. For a given momentum direction the electron spin $\bm s$ makes a right angle with the momentum direction. The opposite spin direction is not allowed.}
\label{TI_JJ}
\end{figure}

The 3D TI surface states host Dirac quasiparticles,  exhibiting full spin-momentum locking: an electron spin always makes a right angle with its momentum. The sketch of the system under consideration and the Fermi surface of the conducting electrons in the 3D TI surface states are demonstrated in Fig.~\ref{TI_JJ}. The spin-momentum locking gives rise to a very strong dependence of the CPR on the magnetization direction  \cite{Tanaka2009,Linder2010,Zyuzin2016,Nashaat2019}. In particular, the anomalous ground state phase shift proportional to the in-plane magnetization component perpendicular to the supercurrent direction was reported. 

It is assumed that the magnetization $\bm M(\bm r)$ of the ferromagnet induces an effective exchange field $\bm h(\bm r) \sim \bm M (\bm r)$ in the underlying conductive surface layer. The discussed theory is also applicable if the Zeeman term in the hamiltonian of the TI surface states is induced by the applied magnetic field. The hamiltonian that describes the TI surface states in the
presence of an in-plane exchange field $\bm h(\bm r)$ reads:
\begin{equation}
\hat H=\int d^2 \bm r' \hat \Psi^\dagger (\bm r')\hat H(\bm r')\hat \Psi(\bm r')
\label{H},
\end{equation}
\begin{equation}
\hat H(\bm r)=-iv_F (\bm \nabla \times \bm e_z)\hat {\bm \sigma}+\bm h(\bm r)\hat {\bm \sigma} -\mu
\label{h},
\end{equation}
where $\hat \Psi=(\Psi_\uparrow, \Psi_\downarrow)^T$, $v_F$ is
the Fermi velocity, $\bm e_z$ is a unit vector normal to the surface
of TI, $\mu$ is the chemical potential, and $\hat {\bm \sigma}=(\sigma_x, \sigma_y, \sigma_z)^T$ is a vector of
Pauli matrices in the spin space. It was shown \cite{Zyuzin2016,Bobkova2016} that in the quasiclassical approximation $(h,\varepsilon,\Delta) \ll \mu$ the Green's function has the following spin structure: $\check g (\bm n_F, \bm r, \varepsilon) = \hat g (\bm n_F, \bm r, \varepsilon) (1+\bm n_\perp \bm \sigma)/2$, where $\bm n_\perp = (n_{F,y},-n_{F,x},0)$ is the unit vector perpendicular to the direction of the quasiparticle trajectory $\bm n_{F} = \bm p_F/p_F$ and $\hat g$ is the {\it spinless} $4 \times 4$ matrix in the particle-hole and Keldysh spaces containing normal and anomalous quasiclassical Green's functions. The spin structure above reflects the fact that the spin and momentum of a quasiparticle at the surface of the 3D TI are strictly locked and make a right angle. It was demonstrated\cite{Zyuzin2016,Bobkova2016,Hugdal2017} that the spinless retarded Green's function $\hat g(\bm n_F, \bm r, \varepsilon)$ obeys the following transport equations in the ballistic limit:
\begin{eqnarray}
-i v_F \bm n_F \hat \nabla \hat g = \Bigl[ \varepsilon \tau_z - \hat \Delta, \hat g \Bigr]_\otimes,
\label{eilenberger}
\end{eqnarray}
where $[A,B]_\otimes = A\otimes B -B \otimes A$ and $A \otimes B = \exp[(i/2)(\partial_{\varepsilon_1} \partial_{t_2} -\partial_{\varepsilon_2} \partial_{t_1} )]A(\varepsilon_1,t_1)B(\varepsilon_2,t_2)|_{\varepsilon_1=\varepsilon_2=\varepsilon;t_1=t_2=t}$. $\tau_{x,y,z}$ are Pauli matrices in particle-hole space with $\tau_\pm = (\tau_x \pm i \tau_y)/2$. $\hat \Delta = \Delta(x)\tau_+ - \Delta^*(x)\tau_-$ is the matrix structure of the superconducting order parameter $\Delta(x)$ in the particle-hole space. We assume $\Delta(x)=\Delta e^{-i\chi/2}\Theta(-x-d/2)+\Delta e^{i\chi/2}\Theta(x-d/2)$. The spin-momentum locking allows for including $\bm h$ into the gauge-covariant gradient $\hat \nabla \hat A = \bm \nabla \hat A + (i/v_F)[(h_x \bm e_y - h_y \bm e_x)\tau_z, \hat A]_\otimes$.
Eq.~(\ref{eilenberger}) should be supplemented by the normalization condition $\hat g\otimes \hat g = 1$ and the boundary conditions at $x=\mp d/2$. It is assumed that the JJ is formed at the surface of the TI, the superconducting order parameter $\Delta$ and $\bm h$ are effective quantities induced in the surface states of TI by proximity to the superconductors and a ferromagnet. In this case, there are no reasons to assume the existence of potential barriers at the $x=\mp d/2$  interfaces and, therefore, these interfaces are considered as fully transparent.  In this case, the boundary conditions are reduced to continuity of $\hat g$ for a given quasiparticle trajectory at the interfaces.

Our derivation closely follows Ref.~\onlinecite{Nashaat2019}. To obtain the simplest sinusoidal form of the current-phase relation we linearize Eq.~(\ref{eilenberger}) with respect to the anomalous Green's function. In this case the retarded component of the Green's function $\hat g^R = \tau_z + f^R \tau_+ + \tilde f^R \tau_-$. The solution of the linearized Eilenberger equation satisfying asymptotic conditions $f^{R} \to (\Delta/\varepsilon)e^{\pm i \chi/2}$ at $x \to \pm \infty$ and continuity conditions at $x=\mp d/2$ takes the form:
\begin{eqnarray}
f^R_{\pm} = \frac{\Delta e^{\mp i \chi/2}}{\varepsilon}\exp\Bigl[{\frac{\mp 2i (\bm h\bm n_\perp - \varepsilon)(d/2 \pm x)}{v_x}}\Bigr],\nonumber \\
\tilde f^R_{\pm} = -\frac{\Delta e^{\mp i \chi/2}}{\varepsilon}\exp\Bigl[{\frac{\mp 2i (\bm h\bm n_\perp - \varepsilon)(d/2 \mp x)}{v_x}}\Bigr],
\label{f_sol}
\end{eqnarray}
where the subscript $\pm$ corresponds to the trajectories ${\rm sgn}\! ~v_x =\pm 1$.  

The density of electric current along the $x$-axis is
\begin{eqnarray}
j_x = -\frac{e N_F v_F}{4} \int \limits_{-\infty}^{\infty} d \varepsilon \int \limits_{-\pi/2}^{\pi/2} \frac{d \phi}{2 \pi} \cos \phi \times \nonumber \\
\Bigl[(g^R_+ \otimes \varphi_+ - \varphi_+ \otimes g^A_+)-(g^R_- \otimes \varphi_- - \varphi_- \otimes g^A_-)\Bigr],
\label{current}
\end{eqnarray}
where $\phi$ is the angle, which the quasiparticle trajectory makes with the $x$-axis. $\varphi_{\pm}$ is the distribution function corresponding to the trajectories ${\rm sgn}\!~v_x =\pm 1$. In equilibrium $\varphi_\pm = \tanh[\varepsilon/2T]$. 
Exploiting the normalization condition one can obtain $g^{R}_{\pm} \approx 1 - f^{R}_{\pm} \tilde f ^{R}_\pm/2$. Taking into account that $g^A_{\pm} = -g^{R*}_{\pm}$ the following final expression for the Josephson current has been obtained:
\begin{eqnarray}
j_s = j_c \sin (\chi - \chi_0), \label{Josephson_CPR}\\
j_c = ev_F N_F T \sum \limits_{\varepsilon_n >0} \int \limits_{-\pi/2}^{\pi/2} d \phi \cos \phi \frac{\Delta^2}{\varepsilon_n^2}  \times \nonumber \\
\exp\left[-\frac{2\varepsilon_n d}{v_F \cos \phi}\right] \cos \left[\frac{2h_xd \tan \phi}{v_F}\right], 
\label{critical_current} \\
\chi_0 = 2 h_y d/v_F \label{chi_0},
\label{josephson_final}
\end{eqnarray}
where $\varepsilon_n = \pi T(2n+1)$. It is seen that the CPR Eq.~(\ref{Josephson_CPR}) contains the anomalous phase shift $\chi_0$. At high temperatures $T \approx T_c \gg \Delta$ the main contribution to the current comes from the lowest Matsubara frequency and Eq.~(\ref{critical_current}) can be simplified further 
\begin{eqnarray}
j_c = j_b \int \limits_{-\pi/2}^{\pi/2} d \phi \cos \phi  \times \nonumber \\
\exp\left[-\frac{2\pi T d}{v_F \cos \phi}\right] \cos \left[\frac{2h_xd \tan \phi}{v_F}\right], 
\label{critical_current_T} 
\end{eqnarray}
where $j_b = ev_F N_F \Delta^2/(\pi^2 T)$. Similar expression has also been obtained for Dirac materials \cite{Hugdal2017}.

Eqs.~(\ref{josephson_final}) and (\ref{critical_current_T}) demonstrate that the dependence of the Josephson current on the exchange field (and, consequently, the magnetization direction in S/F-TI/S junctions) is highly nontrivial. In contrast to the well-known S/F/S junctions via ordinary ferromagnets, where the critical current is suppressed by exchange field and {\it does not depend on its direction}, here the critical Josephson current is only suppressed by the $x$-component of the exchange field. The $y$-component of the field does not lead to the suppression. Instead, it gives rise to the anomalous phase shift. This statement is also valid for the diffusive case. The Josephson current in 3D TI-based diffusive Josephson junction has been considered in Ref.~\onlinecite{Zyuzin2016}, and exactly the same expression for the anomalous phase shift $\chi_0$ has been obtained. The result for the critical current is different in the diffusive case, but it still only depends on the $x$-component of the exchange field. Below we explain the qualitative physics of this effect. 

\begin{figure}[!tbh]
         \centerline{\includegraphics[clip=true,width=3.0in]{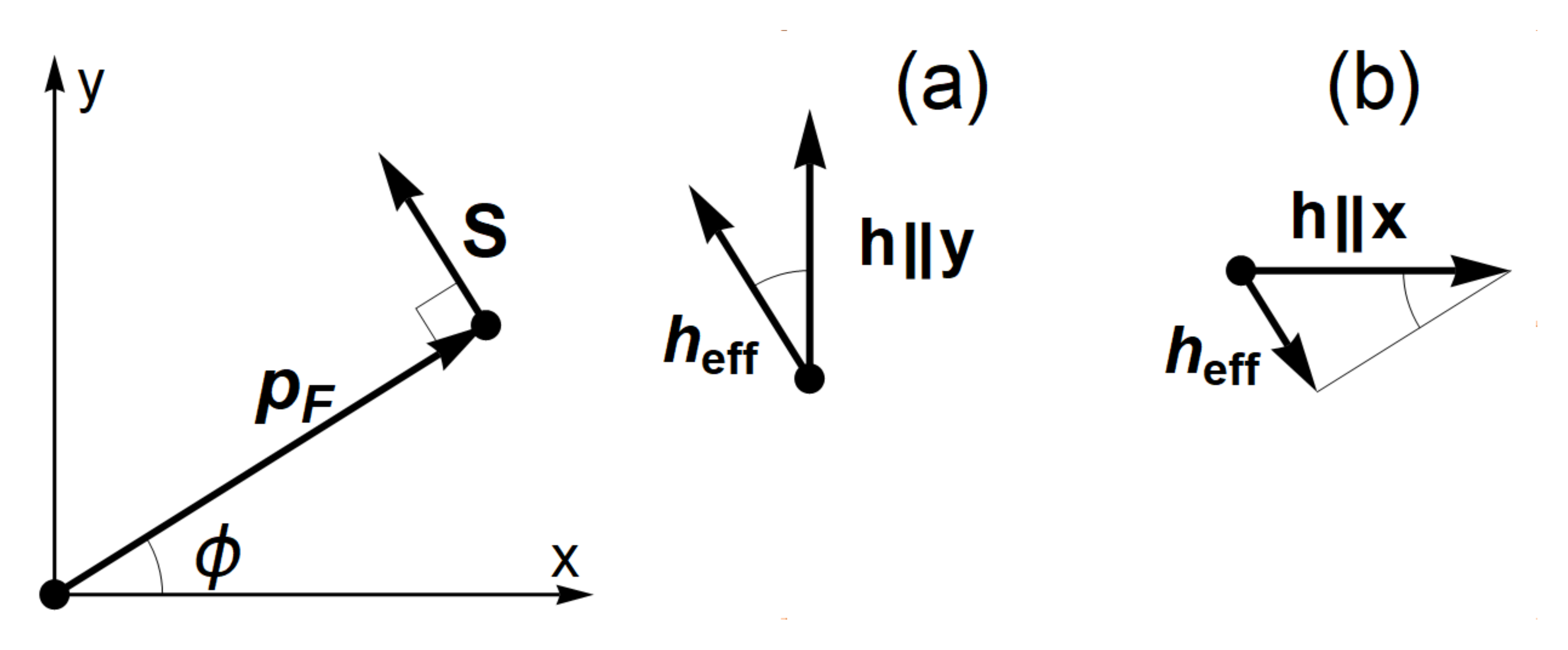}}
        \caption{Illustration of the effective exchange field seen by an electron in the conductive surface states of the 3D TI. $\bm p_F$ is the electron momentum and $\bm S$ is its spin. (a) Exchange field $\bm h$ induced by the proximity to a ferromagnet is along the $y$-axis. (b) $\bm h$ is along the $x$-axis, that is aligned with the Josephson current direction. In both cases the electron only feels effective exchange field $\bm h_{eff}$, which is along its spin direction.}
\label{TI_explain}
\end{figure}

Let us consider a pair travelling in the TI surface states  in the interlayer region of the JJ and carrying the Josephson current. The pair consists of  electrons with opposite momenta and opposite spin directions. At first, we assume that the exchange field is along the $y$-axis, that is $\bm h = h \bm e_y$. Because the spin of an electron is strictly perpendicular to its momentum, each of the electrons forming the pair "sees" its own effective exchange field, which is equal to the projection of $\bm h$ onto its spin direction. If the electron momentum makes an angle $\phi$ with the $x$-axis, the effective field seen by the electron is $h \cos \phi$, see Fig.~\ref{TI_explain}(a). After passing the distance $x$ along the $x$-axis the electron acquires a phase $\Phi \approx p_{F,x}x+(h\cos \phi/v_{F,x})x$. The second electron from the pair with the opposite values of the momentum and spin sees the opposite effective exchange field and, therefore, acquires the same phase. The total phase gained by the pair is $(2 h\cos \phi/v_{F,x})x$  in agreement with Eq.~(\ref{f_sol}). Taking into account that $v_{F,x}=v_F \cos \phi$ we see that for the given exchange field direction the phase acquired by the pair does not depend on the trajectory direction. Therefore, the contributions from the different trajectories to the Josephson current do not cancel each other due to the momentum averaging. As a result, the critical current is not suppressed by the exchange field, and the influence of the field only appears via the overall phase $\pm 2h_y d/v_F$ acquired by all the pairs travelling through the interlayer to the right (left). Because of the independence of the phase gain on the momentum direction between two scattering events, the consideration is also applicable to the diffusive case resulting in the same answer. Interestingly that the same phase factors can result in the electric current oscillations through the
Andreev interferometer \cite{Malshukov2018}. 

Now let us compare the previous consideration to the case of the exchange field along the $x$-axis, that is $\bm h = h \bm e_x$. In this case the effective exchange field which is seen by the electron is $-h \sin \phi$ and the same reasoning results in the phase acquired by the pair $\Phi = -(2h \tan \phi/v_F)x$. The phase depends on the trajectory direction and, therefore, the Josephson current carried by all the pairs is greatly reduced due the averaging of these acquired phases. It has also been reported that due to the finite width of the junction along the $y$-direction, the decay of the critical current upon the increase of the $x$-component of the exchange field is accompanied by the Fraunhofer-like oscillations \cite{Malshukov2020}.

In summary, we can see that the Josephson current via the 3D TI-based S/F/S junction depends strongly on the exchange field direction. The physical reason for this dependence is the spin-momentum locking. In particular, the Josephson current exhibits the anomalous phase shift $\chi_0 = 2h_yd/V_F$. It is determined by the exchange field component perpendicular to the current direction analogously to the case of Rashba materials, described above. However, the anomalous phase shift in TI-based Josephson junctions is much larger than in the materials with Rashba SOC because it does not contain the reducing factor $\Delta_{so}/\varepsilon_F \sim \alpha p_F/\varepsilon_F$. It is directly connected to the fact that the 3D TI has only one helical Fermi surface [see Fig.~\ref{TI_JJ}(b)] in contrast to the Fermi surface of a Rashba material, which consists of two helical bands with opposite helicity [see Fig.~\ref{Direct_SO_1}(a)].

\subsection{Anomalous phase shift in S/F/S junctions via inhomogeneous ferromagnets}

The interaction of conduction electron spin with a magnetic texture can be described using an artificial spin-orbital coupling potential. 
In general the local transformation
 $ \hat U ({\bm r})= e^{i\bm {\hat\sigma }\bm{\theta}( {\bm r})/2} $
 rotates spin axes to the local frame where ${\bm h} \parallel {\bm z}$. It is  parametrized by the  spin vector $\bm \theta = \theta \bm n$ defined by the spatial texture of the exchange field
 distribution  $\bm h(\bm r)=\hat R( \bm \theta(\bm r))\bm h$, where $\hat R$ is the spatially-dependent rotation
 matrix and we choose ${\bm h}= h {\bm z}$.
 This transformation 
 generates the  spin-dependent potential 
  $ \hat\sigma_a \{ M_{k}^a,  \hat p_k \}/2$
 with  pure gauge SU(2) field
 $
 M_{k}^a = -i {\rm Tr} \left(\hat\sigma_a \hat U^\dagger \nabla_k \hat U\right) /2 m
 $. This artificial SOC leads to the spontaneous current\cite{Bobkova2017} in the form (\ref{II_4}) 
 with  $\chi_{k}^a \propto  M_{k}^a$ so that $\bm j\propto h \chi_{k}^z$. This spontaneous current is nonzero provided that $M_{k}^z \neq 0$ which physically means that the magnetic texture is  non-coplanar. On the qualitative level, the non-coplanarity is needed to obtain  the different orbital states of conduction electrons in magnetic textures $\pm \bm h(\bm r)$ connected by the time-reversal transformation $\bm h  \to -\bm h$. In the non-coplanar case it is not possible to compensate this sign change by the global, that is coordinate-independent spin rotation. 
  The minimal non-coplanar texture consists of three magnetic moments $\bm m_{1,2,3}$ with nonzero scalar spin chirality\cite{tatara2019effective} $\chi = \bm m_1  (\bm m_2\times \bm m_3)$. As demonstrated in Ref.\onlinecite{Rabinovich2018}
 already the arrangement of three point-wise non-coplanar magnetic impurities leads to the generation of the spontaneous supercurrent and superconducting phase gradients $j, \nabla \varphi \propto \chi$. 
  
 The anomalous phase shifts has been studied in Josephson junctions through various non-coplanar magnetic textures, including the three-layer systems \cite{Braude2007, Grein2009, Liu2010,margaris2010zero,Kulagina2014, Mironov2015, Silaev2017,silaev2017theta,Kuzmanovski2016}, magnetic helices\cite{Bobkova2017,Rabinovich2018} and skyrmions\cite{Rabinovich2018}. In most cases the magnetic systems are assumed to be metallic consisting either of strong ferromagnets or the half-metals. However, in several papers it has been shown that magnetoelectric effects can be engineered  even with the help of ferromagnetic insulators\cite{Silaev2017,silaev2017theta} (FI) which are currently considered as the promising platform for coupling of the superconductivity and ferromagnetism\cite{liu2019semiconductor,vaitiekenas2021zero,strambini2017revealing,rouco2019charge,heikkila2019thermal}.  
 Below we discuss the particular example of of such system\cite{silaev2017theta} . 
 
 \begin{figure}[h!]
 \centerline{$
 \begin{array}{c}
 \;\;
 \includegraphics[width=0.25\linewidth]{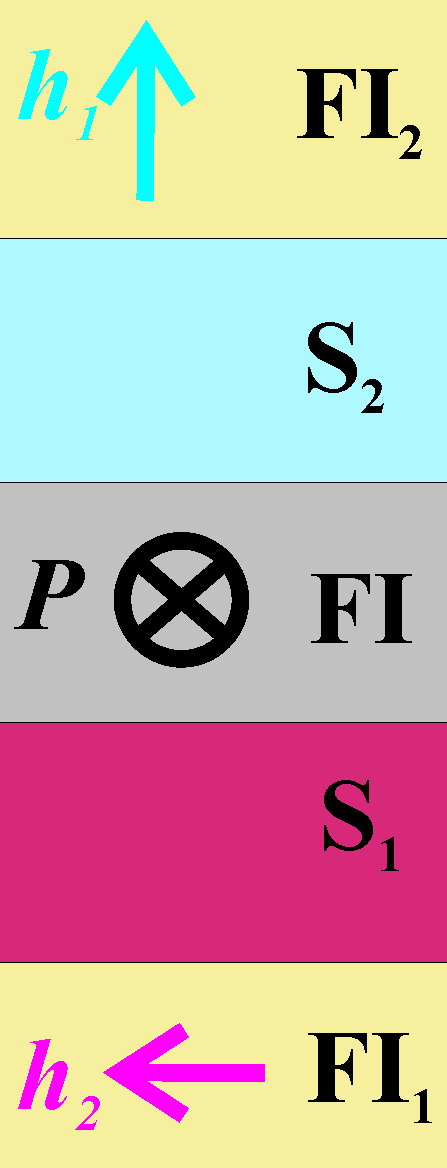}
 \;\;
  \includegraphics[width=0.6\linewidth]{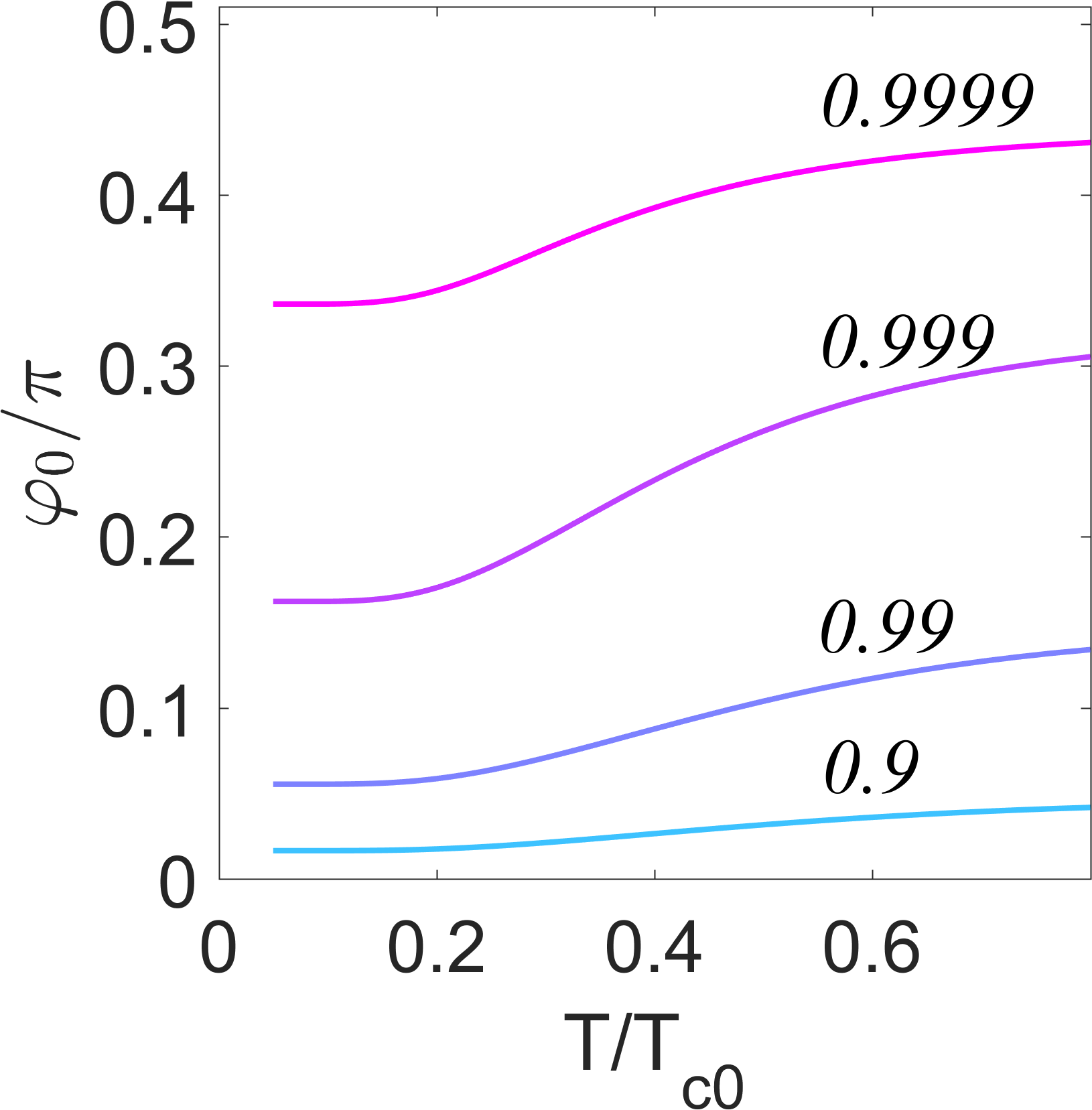}
 \;\;
 \end{array}$}
 \caption{\label{Fig:FSFSF} 
 (Left panel) Sketch of FI-S-FI-S-FI [FI stands for a ferromagnetic insulator] non-coplanar spin valve system with the superconducting electrodes S$_{1,2}$ having different phases  $\varphi_{1,2}$. The exchange fields ${\bm h_{1,2}}$ in FI$_{1,2}$
 form a non-coplanar system with the spin polarization of central barrier ${\bm P}$. Lengths of S layers are of the order of the superconducting coherence length to ensure well-pronounced proximity-induced exchange field in the S-layers.
 (Right panel) Temperature dependencies of the anomalous phase shift at $P=0.9;\; 0.99;\; 0.999;\; 0.9999$ (curves from top to bottom) for $\bm h_1\perp \bm h_2 \perp \bm P$.  }
 \end{figure}

 Let us consider the system shown in Fig.\ref{Fig:FSFSF}  consisting of two superconducting electrodes S$_{1,2}$ separated by a single FI interlayer acting as a spin-filtering barrier with  polarization $\bm P$. The outer FI layers generate Zeeman fields $\bm h_{1,2}$ in S$_{1,2}$ due to the magnetic proximity effect\cite{Tokuyasu1988,millis1988quasiclassical}.  
 To calculate the currents across spin-filtering barriers we use generalized  Kuprianov-Lukichev boundary conditions \cite{Kupriyanov1988}, 
 that include  spin-polarized 
 tunnelling at the SF interfaces \cite{bergeret2012electronic,Eschrig2015,Moor2015,Moor2015_2}. The matrix tunnelling current
 from S$_1$ to S$_2$ is given by  
 \begin{equation}
 \check I_{12}=[\check\Gamma  \check g_{1} \check\Gamma^\dagger, \check g_{2} ],
 \label{Eq:KupLuk}
 \end{equation} 
 where $\check g_{k}$ for $k=1,2$ are the matrix Green's functions in the superconducting electrodes S$_k$. 
 The spin-polarized tunnelling matrix has the form $\check\Gamma= t_+\hat\sigma_0\hat\tau_0 + t_-(\bm { m \hat\sigma})\hat\tau_3 $, 
 where ${\bm m}$ is the direction of  barrier  magnetization, $t_\pm = \sqrt{( 1\pm\sqrt{1-P^2} )/2} $ and $P$ 
 being the spin-filter efficiency of the barrier that ranges from $0$ (no polarization)  to $1$ (100\% filtering efficiency).
 The matrix Green' functions are determined by the equation \cite{Bergeret2005}
 \begin{equation} \label{Eq:Usadel}
   [\omega_n \tau_3+i({\bm{ h\cdot \sigma})}\tau_3 + \check{\Delta} - \check\Sigma_{s}, \check{g}] =0.
 \end{equation}  
  Here  $\omega_n = (2n+1)\pi T$ is the Matsubara frequency, $ \check{\Delta}=\Delta\tau_1 e^{i\tau_3\varphi}$ 
   is the order parameter with the amplitude $\Delta$ and phase $\varphi$,
   ${\bm h } $ is the exchange field. 
   We include the spin-orbital (SO) scattering process which lead to the spin relaxation described by  \cite{Bergeret2005}
   $\check\Sigma_{s} = ({\bm {S}}\cdot\check{g} {\bm {S}}) /8\tau_{so} $, where $\tau_{so}$ is the SO scattering time. 
 Due to the  normalization condition $\check g^2=1$ . 
 
 {
 The solution of Eq.~(\ref{Eq:Usadel}) can be found in the form
 }
 \begin{equation} \label{Eq:GR}
 g = \tau_{3} \left[ g_{03} + g_{33} (\bm \sigma \bm h) \right] +
       \tau_{1}\left[ g_{01} + g_{31} (\bm \sigma \bm h) \right]
  \end{equation}
  The terms diagonal in Nambu space ($\tau_3$) correspond to the normal correlations 
 which determine the  density of states.
 The  off-diagonal components ($\tau_1$) describe spin-singlet $g_{01}$ and spin-triplet   
 $g_{31}$ superconducting correlations which appear as a result of the exchange splitting  \cite{Bergeret2001}.

 The
  CPR   (\ref{Eq:CPRgen}) can be  calculated for the spin-valve shown in Fig.(\ref{Fig:FSFSF}) 
  using the general matrix current (\ref{Eq:KupLuk}).   
  The usual $j_0 = j_c \cos\varphi_0$ and anomalous $j_{an} = j_c \sin\varphi_0$ 
 Josephson currents through tunnel barrier are given by
   \begin{align} \label{Eq:I0}
   & \frac{R_N j_0}{\pi e T} =  \sum_{\omega_n} \left[ r (g_{01}^2 + 
    h_{1\parallel}h_{2\parallel} g_{31}^2) + ({\bm h_{1\perp}}{\bm h_{2\perp}}) g_{31}^2  \right]
    \\ \label{Eq:Ian}
   & \frac{R_N j_{an}}{\pi e T} = \chi P\sum_{\omega_n}  g_{31}^2,
   \end{align}
  where $\chi = {\bm P} ({\bm h_1\times \bm h_2})$ is the spin chirality, 
   ${\bm h_{i \perp}}$ are the projections of ${\bm h_{i }}$ on the plane perpendicular to $\bm P$.

 Expressions 
 (\ref{Eq:I0}, \ref{Eq:Ian}) show that
 the anomalous current $j_{an}$ is mediated by spin-triplet component $g_{31}$. 
 Physically the phase-shifting term  $j_{an}$ appears as a result of the additional phase picked up by the spin-triplet 
     Cooper pairs when tunnelling  between two superconductors with non-collinear exchange fields through the spin-polarising barrier.  Therefore $\varphi_0$ Josephson effect is the directly observable signature of the spin-triplet superconducting current
  across the junction. 
For the ideal spin filter  $P=1$  Eqs.~(\ref{Eq:I0}), (\ref{Eq:Ian}) yield a temperature-independent
 phase shift of CPR $\varphi_0 = \theta_h$, where $\theta_h$ is the geometric angle between  
 the vectors $\bm h_{\perp 1}$ and $\bm h_{\perp 2}$. In the general case  $\varphi_0$
 can be quite different, as shown in Fig.(\ref{Fig:FSFSF}).

\subsection{Direct magnetoelectric effect in Josephson junctions via spin-orbit materials}

\label{SO_direct}

The direct magnetoelectric effect was predicted for
superconducting systems in the presence of spin-orbit coupling, where it represents the generation of an equilibrium spin polarization in response to
supercurrent. The spin-orbit coupling can be both of intrinsic type, that is arising due to the inversion symmetry breaking \cite{Edelstein1995,Edelstein2005,Malshukov2008,Bobkova2017_2} and of extrinsic type, that is impurity-induced \cite{Bergeret2016,Virtanen2021}. Physically, the effect is the same for superconductors in the presence of the spin-orbit coupling and for the Josephson junctions and can be expressed by the general equation:
\begin{equation}
S^a = \frac{1}{ev_F}\kappa_k^a j_{s,k}     
\end{equation}
where $j_{s,k}$ is the $k$-component of the supercurrent density. However, in this review we describe the effect and the used theoretical approaches focusing on the Josephson junctions. The direct magnetoelectric effect was considered both for ballistic \cite{Bobkova2017_2} and diffusive Josephson junctions \cite{Malshukov2008} via spin-orbit coupled materials.

\begin{figure}[!tbh]
     \begin{minipage}[b]{\linewidth}
     \centerline{\includegraphics[clip=true,width=2.4in]{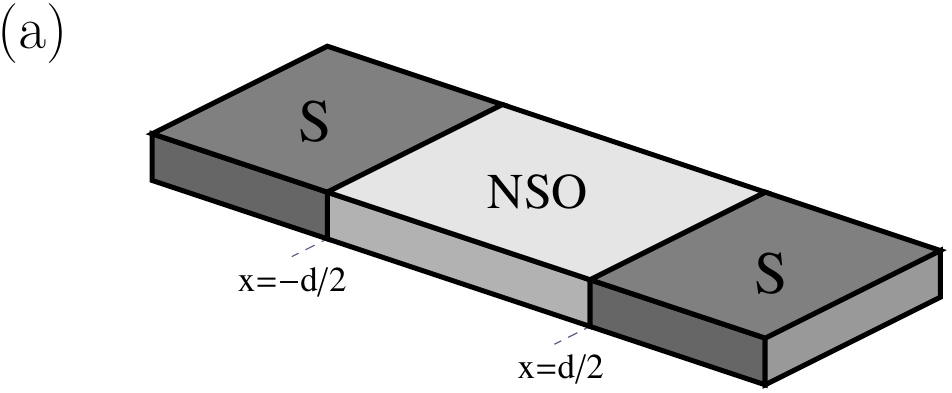}}
     \end{minipage}\hfill
    \begin{minipage}[b]{\linewidth}
   \centerline{\includegraphics[clip=true,width=2.4in]{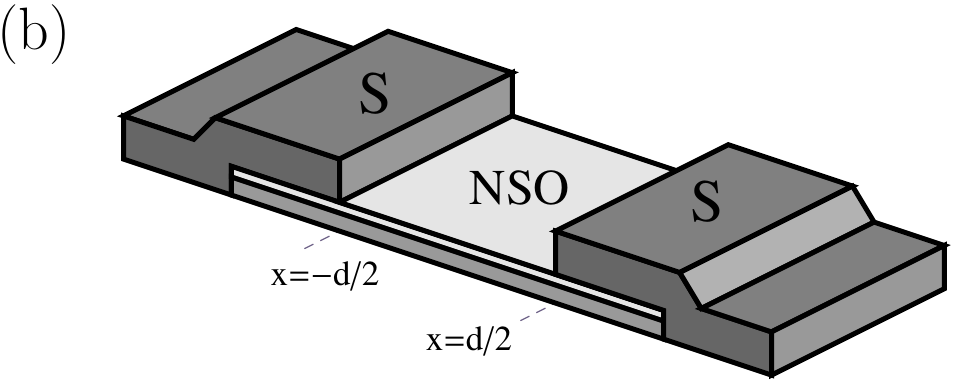}}
  \end{minipage}
   \caption{Sketches of possible realizations of the  S/NSO/S junction [NSO means normal metal with spin-orbit coupling]. Adopted from Ref.~\onlinecite{Bobkova2017_2}.}
\label{junction}
\end{figure}

We will focus on the case of ballistic S/NSO/S junction with Rashba-type spin-orbit coupling in the interlayer and discuss the value of the electron spin polarization induced by the applied supercurrent and the theoretical approach used for solving the problem in Ref.~\onlinecite{Bobkova2017_2}.
The sketch of the system under consideration is shown
in Fig.~\ref{junction}. The S/NSO interfaces are at $x = \mp d/2$.
The Hamiltonian of a singlet superconductor in the presence of an arbitrary linear in momentum spin-orbit (SO) coupling \cite{Bergeret2013,Bergeret2014}:

\begin{eqnarray}
\hat H=\int d^2 \bm r' \hat \Psi^\dagger (\bm r')\hat H_0(\bm r')\hat \Psi(\bm r')+  \nonumber \\
\Delta(\bm r)\Psi_\uparrow^\dagger (\bm r)\Psi_\downarrow^\dagger (\bm r)+\Delta^*(\bm r)\Psi_\downarrow (\bm r)\Psi_\uparrow (\bm r)
\label{H_ballisticSO},
\end{eqnarray}
\begin{equation}
\hat H_0(\bm r)=\frac{\hat {\bm p}^2}{2m}-\frac{1}{2}\hat {\bm A} \hat {\bm p} -\mu
\label{H0},
\end{equation}
where $\Delta(\bm r)$ is the superconducting parameter, which is nonzero only in the superconducting leads. $\hat H_0$ is the Hamiltonian of the normal metal in the presence of the spin-orbit coupling (NSO). The general linear in momentum SO is expressed by the term $\frac{1}{2}\hat {\bm A} \hat {\bm p}=\frac{1}{2} A_{j}^\alpha p_j\hat \sigma^\alpha$, where  $\hat \sigma^\alpha$ are Pauli matrices in spin space. $\hat \Psi=(\Psi_\uparrow, \Psi_\downarrow)^T$, $\mu$ is the chemical potential. For particular case of the Rashba SOC $A_x^y = -A_y^x = \alpha$ and for the Dresselhaus SOC $A_x^x = -A_y^y = \beta $. 

In the superconducting systems under the applied supercurrent the Cooper pairs acquire nonzero total momentum, which leads to the analogous shift of the Fermi surfaces in the momentum space, as it was discussed in the introduction fot nonsuperconducting materials. But in the superconducting case the polarization is provided by the averaged polarization of triplet pairs, as it supported below by the corresponding expression of the polarization in terms of the anomalous Green's function. Therefore, the supercurrent-induced polarization is intimately connected to another manifestation of the magnetoelectric effect in superconducting systems: generation of triplet pairs under the applied electric current. Now our goal is to describe this effect in Josephson junctions.

\begin{figure}[!tbh]
         \centerline{\includegraphics[clip=true,width=1.7in]{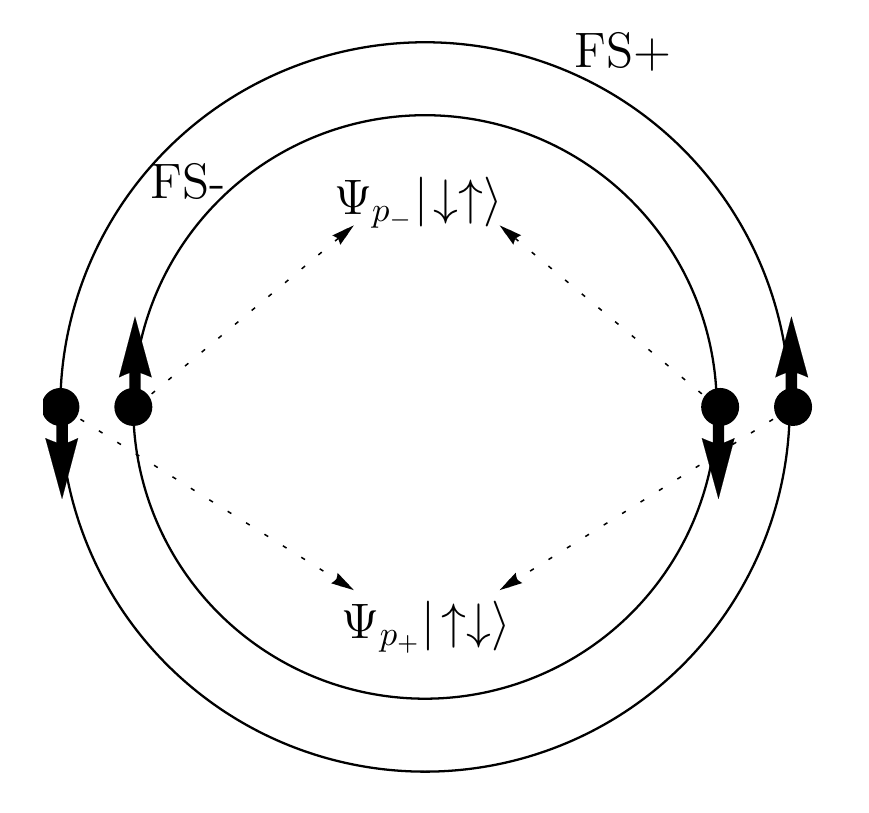}}
        \caption{Opposite-spin and opposite-momenta pairing for the Rashba material. Two spin-split Fermi-surfaces are denoted by ${\rm FS}_\pm$.}
\label{Direct_SO_2}
\end{figure}

Superconducting hybrid mesoscopic systems are often considered within the framework of the quasiclassical theory, which makes it possible to effectively solve spatially inhomogeneous problems. The
SOC can be treated in the quasiclassical approximation when its characteristic energy $\Delta_{so}$ is much less than the Fermi energy $\varepsilon_F$. This situation is typical. However the quasiclassical consideration does not capture the magnetoelectric effects, which are of the first order with respect to the parameter $\Delta_{so}/\varepsilon_F$. In particular, it is known that the standard quasiclassical equations in the presence of SOC and in the absence of an exchange field do not provide a 
transformation of singlet correlations to the triplet ones \cite{Bergeret2013,Bergeret2014}. The physical reason for this can be immediately seen from Fig.~\ref{Direct_SO_2}. The opposite-momenta pairing occurs between the electrons of the same helical band. It is obvious that for a given momentum direction one can compose a pair $\Psi_{p_+}|\!\uparrow \downarrow \rangle$ at one of the subbands and a pair $\Psi_{p_-}|\!\downarrow \uparrow \rangle$ for the other subband, where $p_\pm$ are Fermi momenta of the both helical subbands. Each of the pairs is a singlet-triplet mixture. Therefore, taking into account the fact that wave functions of the pairs differ at different Fermi surfaces $\Psi_{p_+} \neq \Psi_{p_-}$, we obtain a triplet admixture of the pair wave function \cite{Gor'kov2001}. However, the quasiclassical approximation disregards the difference between the values of the Fermi momenta for the both helical subbands, that is $p_+ = p_-$ with the quasiclassical accuracy. In this case the singlet pairs do not accompanied by such a triplet admixture. Therefore, in order to describe the triplet correlations in SOC materials and the resulting magnetoelectric effects, a generalization of the quasiclassical theory to include the first order corrections with respect to the parameter $\Delta_{so}/\varepsilon_F$ is required. Several approaches have been reported in the literature \cite{Malshukov2008,Konschelle2015,Bobkova2017_2}. We focus on the approach developed in Ref.~\onlinecite{Bobkova2017_2}.

The generalized quasiclassical equation for the quasiclassical Green's function can be written as follows:
\begin{eqnarray}
i {\bm v}_F \bm \nabla \check g +
\left[ \varepsilon \hat \tau_z + \check \Delta(\bm r) + \frac{1}{2}\hat {\bm A} {\bm p}_F, \check g\right]+ \nonumber \\
\frac{p_y}{4v_{F,x}} \bigl[ \hat A_x, \hat A_y \bigr]\bigl(\check g-{\rm sgn}v_{F,x} \bigr)+\frac{i\hat A_y p_y}{2p_{F,x}}\partial_x \check g=0.
\label{eq_g_final}
\end{eqnarray}
The second line of Eq.~(\ref{eq_g_final}) represents the corrections of the order of $\Delta_{so}/\varepsilon_F$ to the standard Eilenberger equation \cite{Eilenberger1968}. Eq.~(\ref{eq_g_final}) should be supplied by the normalization condition. In usual quasiclassical theory, the normalization condition is $\check g^2=1$. However, in the framework of the discussed approach, it should be modified if one would like to take into account the terms of the order of $\Delta_{so}/\varepsilon_F$. It takes the form:
\begin{eqnarray}
\check g^2 - \frac{\hat A_y p_y {\rm sgn}v_{F,x}}{p_{F,x}v_{F,x}}\biggl[ \check g - {\rm sgn}v_{F,x}\biggr]=1
\label{norm_final}
\end{eqnarray}
Quasiclassical equations are not valid in the vicinity
of interfaces, where the normal state Hamiltonian of the
system changes over the atomic length scales. Therefore they
should be supplied by the boundary conditions. Corrections to the boundary conditions appear even for the simplest   “absolutely
transparent interfaces” case when the boundary conditions take the form:
\begin{eqnarray}
\check g^l - \check g^r =\pm \Biggl\{ {\rm sgn}v_{F,x}-\check g, \frac{\hat A_y p_y}{4 v_{F,x}p_{F,x}} \Biggr\},
\label{boundary_final}.
\end{eqnarray}
The signs $\pm$ correspond to the NSO/S and S/NSO interfaces, respectively. It is seen that neglecting the right hand side of the above equation, which is of the first order in $\Delta_{so}/\varepsilon_F$, we obtain the well-known quasiclassical boundary condition at a fully transparent interface: $\check g^{l} = \check g^{r}=\check g$, that is just the continuity of the Green's function. In the framework of this modified quasiclassical theory, the boundary condition is reduced to the  standard continuity condition $\check g^l=\check g^r$ only for the case of  equal SO coupling in  both materials.

For the considered S/NSO/S junction the $y$-component of the triplet anomalous Green's function in the NSO interlayer, which determines the electron polarization $S_y$, takes the form:
\begin{eqnarray}
f_\pm^y=-(x \pm \frac{d}{2})M_\pm \frac{p_{F,x}}{p_F}e^{2i\varepsilon x/v_{F,x}}
\label{f_inter_y},
\end{eqnarray}
\begin{eqnarray}
M_\pm = \frac{i\alpha p_y^2 \Delta}{p_{F,x} v_{F,x}^2 p_F}e^{\mp i \chi/2 + i \varepsilon d/|v_{F,x}|}
\label{M_pm}.
\end{eqnarray}
It is seen that the triplet correlations are of the first order with respect to the parameter $\Delta_{so}/\varepsilon_F=\alpha p_F/\varepsilon_F$. Result (\ref{f_inter_y}) coincides with the result of the exact calculation in terms of  Gor'kov Green's functions beyond the quasiclassical approximation \cite{Reeg2015} up to the first order with respect to the parameter $\Delta_{so}/\varepsilon_F$. Please note that the statement of Ref.~(\onlinecite{Reeg2015}) that the triplet pairing vanishes to the first order in  $\Delta_{so}/\varepsilon_F$ is not correct. The correct expansion contains terms of the first order by this parameter.

Making use of the above equations it is possible to evaluate the average spin polarization:
\begin{eqnarray}
\bm S =\frac{1}{2} \big\langle \hat \Psi^\dagger(\bm r,t)\hat {\bm \sigma} \hat \Psi (\bm r,t)  \big\rangle 
\label{spin_general}.
\end{eqnarray}

The supercurrent-induced spin polarization is directly related to the triplet pairing. It can be written as \cite{Bobkova2017_2}
\begin{eqnarray}
\bm S \propto \int d \varepsilon dp_y \tanh \frac{\varepsilon}{2T}{\rm Re}\Bigl[\frac{1}{g_{+,s}^R}\Bigl( f_{+,s}^R \tilde {\bm f}_{+,t}^R + \tilde f_{+,s}^R \bm f_{+,t}^R  \Bigr) + \nonumber \\
\frac{1}{g_{-,s}^R}\Bigl( f_{-,s}^R \tilde {\bm f}_{-,t}^R + \tilde f_{-,s}^R \bm f_{-,t}^R  \Bigr)
\Bigr].~~~~~~~~~
\label{g_f_connection}
\end{eqnarray}
$g_s(f_{s})$ and $\bm g_t(\bm f_{t})$ are the singlet and triplet components of the normal (anomalous) Green's function, which can be found from the matrix Green's function structure:
\begin{eqnarray}
\check g =
\left(
\begin{array}{cc}
g_s + \bm g_t \bm \sigma & f_s + \bm f_t \bm \sigma \\
{\tilde f}_s + \tilde {\bm f}_t \bm \sigma & {\tilde g}_s + \tilde {\bm g}_t \bm \sigma
\end{array}
\right),
\label{norm_ph}
\end{eqnarray}
where the structure in the particle-hole space is shown explicitly and the matrix structure in the spin space is encoded in the Pauli matrices $\bm \sigma$. The anomalous Green's function $\tilde f$ can be expressed as ${\tilde f}_\pm^y=-f_\mp^y(-\Delta,-\chi,-v_{F,x})$. The only component of the triplet anomalous Green's function $\bm f$, which is not an odd function of $p_y$ and, therefore, survives after integration over $p_y$, is $f_y$, expressed by Eq.~(\ref{f_inter_y}). As a result, only the $y$-component of the spin polarization is nonzero in the Josephson junction via the Rashba SOC material, according to the qualitative consideration above. In a vector form the induced spin polarization can be represented as
\begin{eqnarray}
\bm S = \kappa \bigl[ \bm c \times \frac{\bm j_s}{e v_F} \bigr]
\label{spin_phys}.
\end{eqnarray}
For the considered case $\alpha p_F/2 \pi T = \Delta_{so}/2\pi T \gg 1$
\begin{eqnarray}
\kappa = \frac{\alpha p_F}{8\varepsilon_F}
\label{d1}.
\end{eqnarray}
It is  the same value as for homogeneous superconductors \cite{Edelstein1995}. Whether this universal behavior holds for S/NSO/S tunnel junctions has not yet been investigated. It is worth to note here that the direct magneto-electric effect should also take place in S/NSO/S junctions with very strong SO coupling in the interlayer ($\Delta_{so} \sim \varepsilon_F$), but the described theory is not able to consider this case quantitatively. This problem can be solved on the basis of the different quasiclassical formalism, where the SO interaction is so strong that the coupling between the two helical subbands is disregarded \cite{Agterberg2007,Houzet2015}.

Another technique accounting for the magnetoelectric effects is a gauge-covariant approach to establish the
transport equations \cite{Konschelle2015}. In this approach, both the electromagnetic and spin interactions are described in terms
of $U(1)$ Maxwell and $SU(2)$ Yang-Mills equations, respectively \cite{}. The additional terms to the Eilenberger equation are expressed via the gauge field $\bm F_{\mu \nu} = \partial_\mu \bm A_\nu - \partial_\nu \bm A_\mu + i [\bm A_\mu, \bm A_\nu]$. In the framework of this approach the well-known Edelstein's results for supercurrent induced spin polarization in homogeneous superconductors, which were obtained in the diffusive \cite{Edelstein2005} and ballistic limits\cite{Edelstein1995},  have been generalized for the case of arbitrary disorder strength \cite{Ilic2020}.

The supercurrent-induced spin polarization in the S/NSO/S Josephson junctions has also been calculated in the diffusive limit \cite{Malshukov2008}. The first-order corrections in $\Delta_{so}/\varepsilon_F$ were added to the Usadel equation for the Green's functions. The triplet anomalous Green's function has been obtained in the form
\begin{eqnarray}
f_t = -i \frac{\alpha \tau}{\sqrt 2}\frac{\partial f_s}{\partial x}.
\end{eqnarray}
The supercurrent-induced electron spin polarization is
\begin{eqnarray}
\bm S = \frac{eN_F \alpha \tau}{\sigma}[\bm c \times \bm j_s],
\label{pol_diffusive}
\end{eqnarray}
where $\sigma$ - is the conductivity of the NSO interlayer, $N_F$ is the density of states at the Fermi level and $\tau$ is the elastic-scattering time. The ratio $Sev_f/(j_s) \sim \alpha p_f /\varepsilon_F$ in Eq.~(\ref{pol_diffusive}) is also of the first order in $\Delta_{so}/\varepsilon_F$. In the subsequent paper \cite{Malshukov2011} of the same group, the S/NSO/S junction under the applied voltage has also been considered. It was predicted that the spin-Hall current does not turn to zero in contrast to the stationary Josephson effect. The physical reason is that besides a direct proximity effect caused by a Cooper pair’s transition into a triplet state, the spin current and polarization are also driven by a periodic electric field associated with the charge imbalance.

\subsection{Direct magnetoelectric effect in S/TI/S junctions}

\label{TI_direct}
As it is discussed in the previous section, the value of the supercurrent-induced electron spin polarization in Josephson junctions via SO-materials is rather small and $\propto \Delta_{so}/\varepsilon_F$. The physical reason for this  is the presence of $\it two$ helical Fermi surfaces in the material with intrinsic SOC, which contribute to the direct magnetoelectric effect in opposite directions.
In contrast, the 2D Fermi surface of the conductive surface states of the 3D topological insulators (TIs) consists of the only helical Fermi surface, see Fig.~\ref{TI_JJ}(b). It leads to the absence of the partial compensation of the current-induced electron spin polarization produced by the helical Fermi surfaces with the opposite helicities. As a result, the current-induced spin polarization does not contain the reducing factor $\Delta_{so}/\varepsilon_F$ and takes the form:
\begin{eqnarray}
\langle \bm s \rangle = -\frac{1}{2ev_F}[\bm e_z \times \bm j_s].
\label{spin_current}
\end{eqnarray}
Eq.~(\ref{spin_current}) has been obtained, making use of the quasiclassical theory for TI-based superconducting hybrid structures (assuming no Zeeman field in the system)\cite{Bobkova2016}, which has already been described in Sec.~\ref{af_TI}. Eq.~(\ref{spin_current}) coincides with the result for the current-induced spin polarization in normal state TI surface states \cite{Shiomi2014}. 

\begin{figure}[!tbh]
         \centerline{\includegraphics[clip=true,width=3.0in]{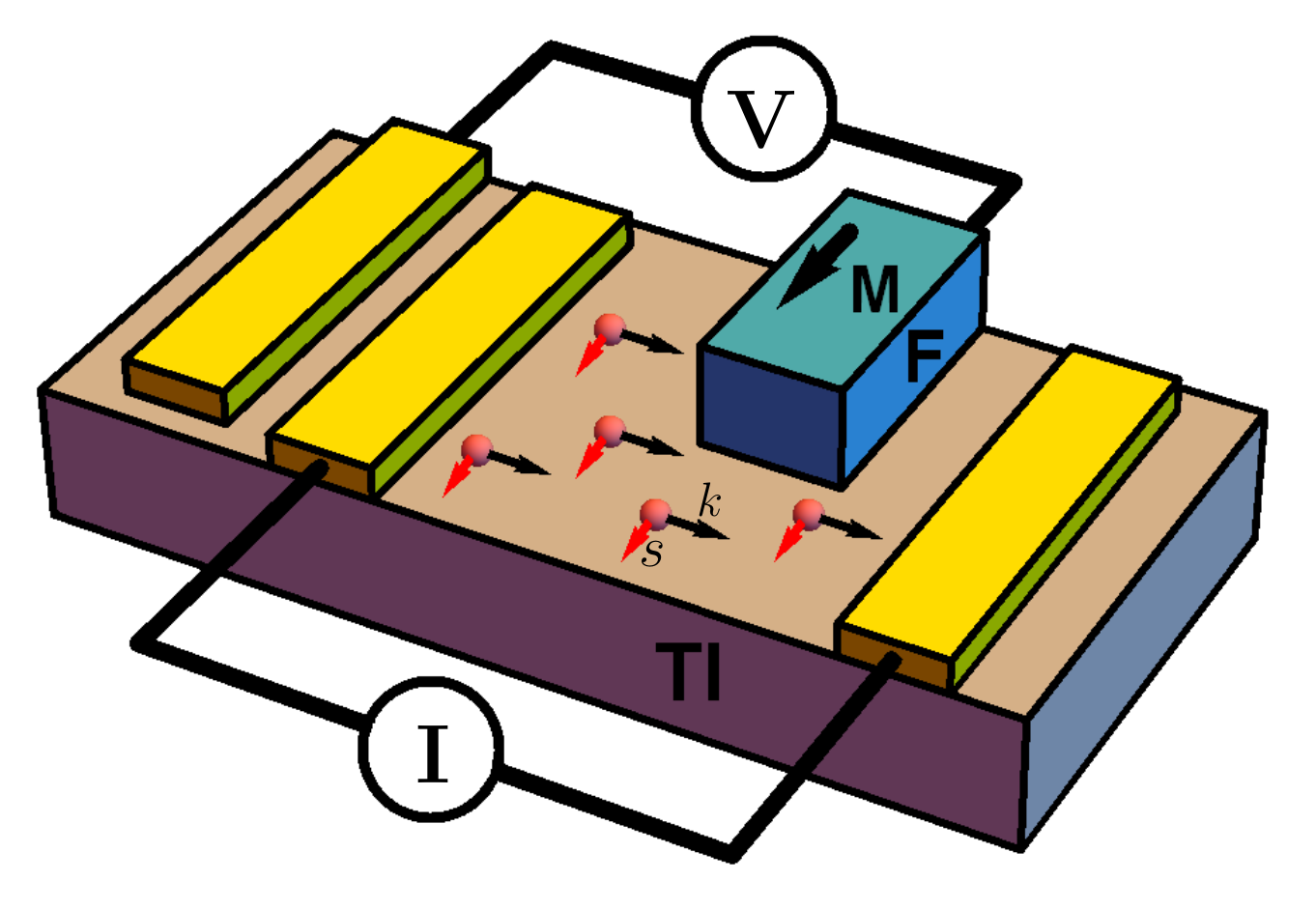}}
        \caption{Sketch of the experimental setup for the electrical measurement of the direct magnetoelectric effect ai the surface of the 3D TI. If the magnetization ($\bm M$) of the ferromagnetic detector (F, blue) has a component along the conductivity electron polarization (red arrows), the voltage $V$ is nonzero. If the magnetization of the detector is along the current axis, $V=0$.}
\label{measurement_TI}
\end{figure}

While the supercurrent-induced spin polarization in TIs has not been measured yet, the current-induced spin polarization in normal state TIs has been directly measured  as a voltage  on a ferromagnetic
metal tunnel barrier surface contact \cite{Li2014}. The voltage measured at the contact is proportional to the projection of the spin polarization of the TI onto the direction of the ferromagnet magnetization.  An unpolarized bias current is
applied between two nonmagnetic contacts, see Fig.~\ref{measurement_TI}.  When the charge
current is orthogonal to the magnetization of the ferromagnetic
detector contact, the TI spin is parallel (or antiparallel) to the magnetization, and a spin-related signal is detected at
the ferromagnetic contact proportional to the magnitude of the
charge current. When the direction of the charge current is reversed,
the measured voltage changes sign. When the
contact magnetization is rotated in-plane $90^\circ$ so that the charge
current is parallel to the magnetization, no spin voltage is detected,
because the TI spin polarization is now perpendicular to the contact
magnetization.

\label{st_direct}

\section{Magnetoelectric effects and magnetization dynamics}

\label{dynamics}

\subsection{Direct coupling between magnetization and superconducting condensate: basic physics and equations}

The   discussed above direct and inverse magnetoelectric effects are of great interest for Josephson junctions with ferromagnetic interlayers because they result in the coupling between the ferromagnet magnetization and the Josephson current. This leads to several interesting ways of electrical control of magnetization and electrical detection in S/F/S junctions. The present section aims to a discussion of those effects.

First of all we formulate the basic equations and describe qualitatively the physics underlying the direct coupling between the supercurrent and the magnetization. In the considered S/F/S junction the  coupled dynamics of magnetization 
$\bm M$
and Josephson phase difference
 $\chi$ is determined by the following 
 closed set of equations 
\begin{align} \label{josephson_modified} 
  & j=j_{c}(\bm M)\sin \left(\chi-\chi_0\{\bm M\} \right) + \frac{\dot \chi-\dot \chi_0 \{\bm M\}  }{2eRS}.
 \\ \label{LLG}
 & \frac{\partial\bm M}{\partial t} = -\gamma \bm M \times \bm H_{eff} + \frac{\alpha}{M} \bm M \times \frac{\partial\bm M}{\partial t} + \bm T,
\end{align}
 Eq.(\ref{josephson_modified}) represents the 
 non-equilibrium current-phase relation (CPR)
 generalizing 
 resistively shunted Josephson junction (RSJ) model. The capacitive term is neglected here.
 This relation is written \cite{Rabinovich2019_2} in a gauge-invariant 
 form amended to include the anomalous phase shift $\chi_0\{\bm M\}$ defined by 
 SOC and magnetic texture. In contrast to the previously used gauge non-invariant formulations\cite
{Konschelle2009,Shukrinov2017} Eq.(\ref{josephson_modified}) describes the normal spin-galvanic effects when $j_c=0$ such as the
 electromotive force and charge current generated in the ferromagnet due to the  time
derivative of the Berry phase
\cite{Volovik1987, Barnes2007, Saslow2007, Duine2008,
Tserkovnyak2008,Schulz2012,Nagaosa2013}.  The analogous equation is also valid for a more general nonsinusoidal CPR. 

\begin{figure}[!tbh]
         \centerline{\includegraphics[clip=true,width=3.5in]{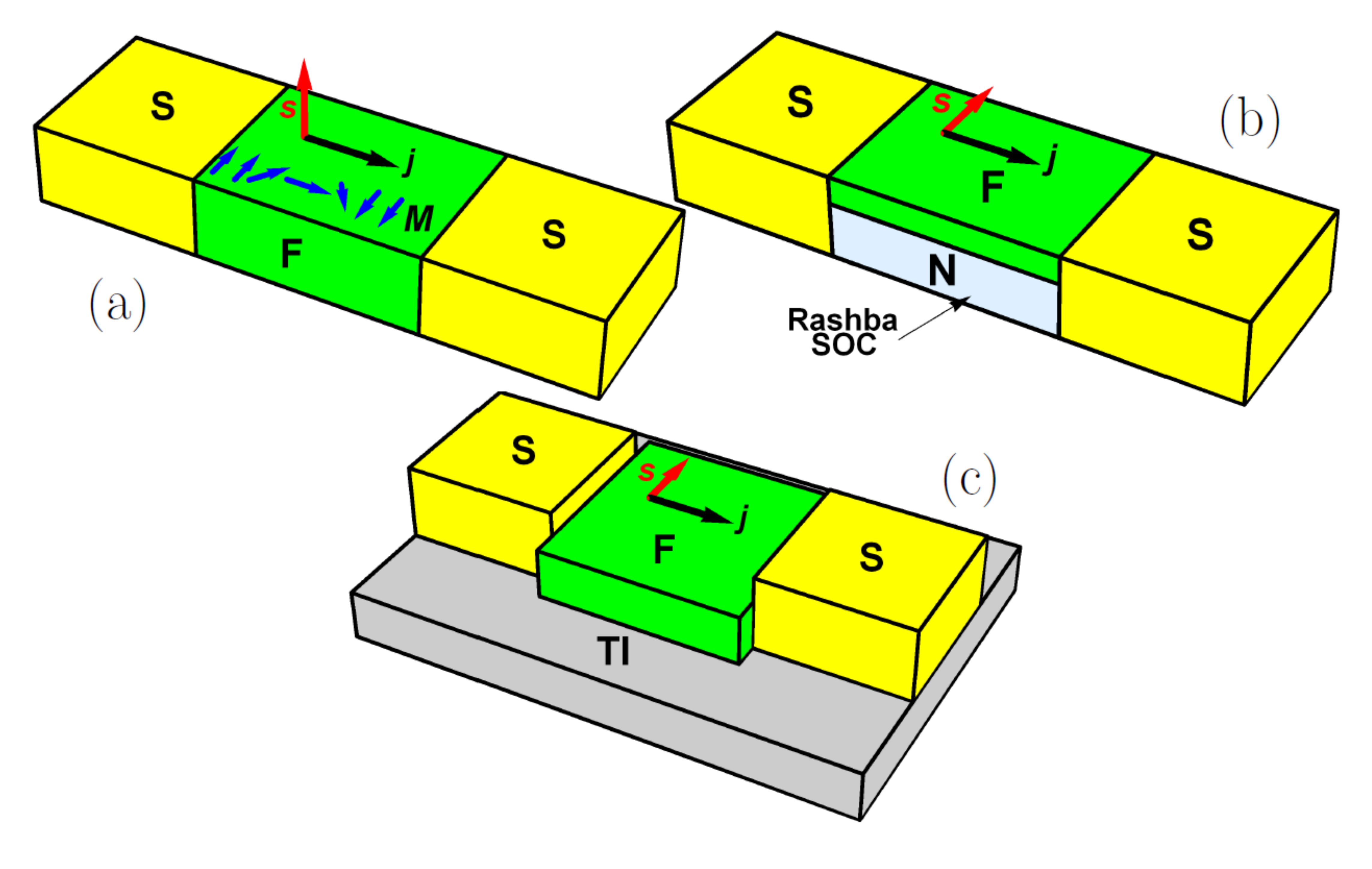}}
        \caption{Examples of Josephson junctions, where the current-induced electron polarization is possible. The direction of the spin polarization relative to the current is shown by red arrows. (a) a JJ via a spin-textured ferromagnet (F); (b) a JJ via a combined ferromagnet (F) + SOC material interlayer (N); (c) a JJ via a ferromagnet on top of a 3D topological insulator (TI). For all the cases it is assumed that the interlayer length should not exceed a few normal state coherence lengths of the interlayer material to ensure a sizeable Josephson current through the JJ.}
\label{examples_hybrids}
\end{figure}
 
The dynamic is described by the Landau-Lifshitz-Gilbert (LLG) equation (\ref{LLG}), where $\alpha$ is the Gilbert damping constant. The polarization of the conductivity electrons, which occurs in the interlayer of the JJ due to the direct magnetoelectric effect discussed above, interacts with the ferromagnet magnetization by the exchange mechanism. Suppose  this interaction is internal for the ferromagnet. In that case, it can be accounted for in the framework of the s-d model \cite{Tsymbal2012}, which describes the exchange interaction between the conductivity electrons and localized electrons responsible for the magnetization according to $\hat H_{s-d} = -J\sum \limits_i \hat S_i \hat s$, where $\hat S_i$ is the spin operator of the localized spin at site $i$ and $\hat s$ is the spin operator of the conductivity electron. The examples are interlayers made of spin-textured ferromagnets or ferromagnets with intrinsic SOC. In hybrid interlayers consisting of a ferromagnet and a nonmagnetic material with SOC or a topological insulator, see examples in Fig.~\ref{examples_hybrids}, the spin polarization induced by the current in the nonmagnetic material (NM) due to the direct magnetoelectric effect, see Secs.~\ref{SO_direct} and \ref{TI_direct}, interacts with the ferromagnet magnetization via the exchange interaction at the F/NM interface:
\begin{eqnarray}
H_{int} = - \int d^2 \bm r \hat \Psi^\dagger (\bm r) J_{ex} \bm S \bm \sigma \hat \Psi(\bm r),~~~~~~~~
\label{interface_ham}
\end{eqnarray}
where $\hat \Psi=(\Psi_\uparrow, \Psi_\downarrow)^T$, $\bm S$ is the localized spin operator in the ferromagnet, $J_{ex}$ is the interface exchange constant and the integration is performed over the 2D interface.

In both cases the torque can be represented in the form:
\begin{eqnarray}
\bm T =  J \bm M \times \langle \bm s \rangle.
\label{torque}
\end{eqnarray}
where $\langle \bm s \rangle$ is the averaged polarization of conductivity electrons. $J$ is the exchange constant of the s-d model or $J=J_{ex}/d_F$ for the hybrid F/HM interlayers with $d_F$ standing for the ferromagnet thickness.

Below, the particular examples of the electrically induced magnetization dynamics in JJ via ferromagnets and the backward influence of the dynamics on the JJ are discussed.

\subsection{Supercurrent-induced magnetization dynamics in S/F/S junctions via spin-textured ferromagnets}

In the context of inhomogeneous ferromagnetic interlayers, the supercurrent-induced
spin torques have been calculated theoretically in
Josephson junctions through single spins\cite{Holmqvist2018,Nussinov2005,Zhu2004} two \cite{Waintal2002,Linder2011,Halterman2016}, three \cite{Kulagina2014} FM layers and bulk inhomogeneous ferromagnets \cite{Bobkova2018}. The supercurrent-induced dynamics has also been studied. In particular, the supercurrent-induced magnetization switching in S/F/N/F/S JJs between parallel and antiparallel configurations has been predicted in  Ref.~\onlinecite{Linder2011}. The supercurrent-induced DW motion has been studied in Ref.~\onlinecite{Bobkova2018}. The spin torque in these systems is commonly viewed as a spin transfer from the spin-polarized triplet supercurrent to the magnetization. However, it can also be understood in terms of the supercurrent-induced conductivity electron spin polarization. From this point of view, it can be considered in a unified manner with the supercurrent-induced torque in JJs with SOC in the interlayer. Below we illustrate this concept by the example of the supercurrent-induced DW motion in S/F/S JJs.

Let us consider an S/F/S JJ via a strong ferromagnet with the exchange energy $h$ of the order of the Fermi energy. This model is relevant for  most  classical ferromagnets, including Fe, Ni, Co and permalloy. It is well-known that in this case, the opposite spin pairs decay very rapidly into the depth of the ferromagnet (on the length scale of a few nanometers), and the Josephson current is presumably carried by equal spin pairs\cite{Buzdin2005,Bergeret2005}. These pairs can be generated via different mechanisms, including rotation of the pair spin at magnetic inhomogeneities  and spin-flip scattering at S/F interfaces\cite{Eschrig2015}. The theory of equal-spin Josephson current via homogeneous strong ferromagnets has been developed in Refs.~\onlinecite{Grein2009,Eschrig2003}. In order to describe the Josephson current in strong inhomogeneous ferromagnets and its influence on the magnetic texture, this theory has been generalized for the case of inhomogeneous magnetization in Refs.~\onlinecite{Bobkova2017_3,Bobkova2018}.

The hamiltonian of the ferromagnetic interlayer is
\begin{align} \label{Eq:Gorkov1}
 \hat H (t,\bm r)= -\frac{ \hat{\bm\Pi}_{\bm r}^2}{2m_F} +
  \left( \bm{ \hat\sigma} {\bm h} (\bm r,t) \right) - i (\bm{\hat \sigma} \hat B \hat{\bm\Pi}_{\bm r} ) ,
   \end{align}
  where $\hat{\bm\Pi}_{\bm r} = \nabla - i(e/c) \bm A (\bm r) $ and $\bm A$ is the vector potential of the electromagnetic field. The last term in Eq.(\ref{Eq:Gorkov1}) is the general form of a linear in momentum spin-orbit coupling (SOC) determined by
 the constant tensor coefficient $\hat B$.

 The quasiclassical theory is formulated in terms of the spinless quasiclassical Keldysh-Green's function $\hat g_\sigma(t_1,t_2)$ ($\sigma = \pm 1$), which is defined separately at each of the Zeeman split Fermi-surfaces for spin-up and spin-down electrons and in general depends on the spatial coordinates $\bm r$ and the two time variables $t_{1,2}$. This theory only accounts for the equal-spin Cooper pairs residing at the same Fermi surface. In contrast, the opposite spin correlations residing at different Fermi surfaces are neglected due to their strong suppression resulting from the large Zeeman splitting of the Fermi surfaces. The Usadel equation takes the form:
\begin{equation} \label{Eq:UsadelEquation}
 \{\hat \tau_3\partial_t, \hat{g}_\sigma\}_t  -
 D_{\sigma} \hat\partial_{\bm r}( \hat g_{\sigma}\circ \hat\partial_{\bm r} \hat g_{\sigma})  =0 ,
 \end{equation}
 where $D_{\sigma}$ are the spin-dependent diffusion coefficients, in the
 isotropic case given by  $D_{\sigma} = \tau_{\sigma}  v^2_{\sigma}/3$. The spin-dependent Fermi velocities $v_\pm = \sqrt{2(\mu \pm h)/m_F}$ are
 determined on each of the spin-split Fermi surfaces with $m_F$ being the electron mass. The time dependence appears due to the dynamical character of the problem under consideration and $\circ$-product is defined as
 $(\hat A \circ \hat B) (t_1,t_2)= \int_{-\infty}^{\infty} dt \hat A(t_1,t)\hat B(t,t_2)$. The commutator of the Green's function with an arbitrary operator $\hat C$ is defined as $[\hat C, \check{g}]_t = \hat C(t_1,\bm r_1)\check{g} -\check{g} \hat C(t_2,\bm r_2)$. The anticommutator $\{\hat C, \check{g}\}_t$ is defined analogously with the plus sign. 
   $\hat \sigma_i$ and  $\hat \tau_i$ are Pauli matrices in spin and Nambu spaces, respectively. and 
 \begin{align}
    \label{Eq:GradientCovariant}
  & \hat \partial_{\bm r} = \nabla - ie[\bm A\hat \tau_3, .]_t + i\sigma [\bm Z\hat \tau_3, .]_t .
  \end{align}
  
 The spin-dependent gauge field is given by the superposition of two terms
 $\bm Z= \bm Z^{m} + \bm Z^{so}$, where $Z^{m}_i = -i {\rm Tr} \Bigl( \hat \sigma_z \hat U^\dagger \partial_i \hat U \Bigr)/2 $ is the texture-induced part. $\hat U= \hat U ({\bm r,t}) $ is in general the time- and space-dependent unitary $2\times 2$ matrix that
   rotates the spin quantization axis $\bm z$ to the local frame determined by the exchange field, so that
    ${\bm h} \parallel {\bm z}$. The
 term $Z^{so}_i = m_F (\bm{m }\bm B_i)$ (where $\bm m = \bm M/M$) appears due to the SOC.
  
 Eq.~(\ref{Eq:UsadelEquation}) is a spin-scalar equation but do not describe conventional spin-singlet superconducting correlations, unlike the standard spin-scalar form of the non-stationary Usadel equation. It is only applicable for strong ferromagnets and describes equal-spin triplet correlations.
 
 In the framework of this theory the torque Eq.~(\ref{torque}) consists of two terms
 \begin{align} \label{torque_adiabatic}
   & \bm T = \bm T_{st} + \bm T_{so},
   \\ \label{torque_st}
   & \bm T_{st} =  2 \mu_B (\bm{\tilde J}^z \nabla) \bm{ m},
   \\  \label{torque_so}
   & \bm T_{so} = 4 \mu_B m_ F( \bm{ m}\times \bm B_j) {\tilde J}^z_{j},
  \end{align}
  where $\bm m = \bm M/M$,  $\bm B_j = (B_{xj},B_{yj},B_{zj})$ is a vector, which is determined by $j$-th
  { coordinate} component of the
  tensor $\hat B$ and $\tilde {\bm J}^z$ is the spin current in the local frame determined by the exchange field, which is represented by difference between the Josephson current carried by spin-up and spin-down pairs. In Eqs.~(\ref{torque_adiabatic})-(\ref{torque_so})
  $\bm T_{st}$
  is the supercurrent spin transfer torque\cite{Berger1996,Slonczewski1996,Ralph2008}. In the considered approximation it takes the form of the adiabatic torque and does not contain a non-adiabatic torque term. $\bm T_{so}$ is the spin-orbit torque \cite{Miron2010,Gambardella2011,Manchon2008}. Its particular structure strongly depends on the type of
  the spin-orbit coupling, realized in the system. 
  
 Eqs.~(\ref{torque_adiabatic})-(\ref{torque_so}) can be viewed in terms of the supercurrent induced spin polarization. If the Josephson current flows along the $x$ direction, the spin polarization takes the form
 \begin{equation}
 \langle \bm s_\perp \rangle =
 -\frac{2\mu_B \tilde J_x^z}{JM}
 \Bigl[ \bm m \times \partial_x \bm m 
  + 
 2 m_F \bm m \times(\bm m \times \bm B_x)\Bigr]
 \label{torque_polarization}
 \end{equation}
 where only perpendicular to $\bm m$ component of the polarization is considered. The first term is the current-induced spin polarization originated from the spin texture and is analogous to the first term in Eq.~(\ref{II_2}). The second term of Eq.~(\ref{II_2}) associated with the nonadiabatic torque does not appear here because locally spin-up and spin-down pairs are not coupled in the framework of this theory and, therefore, the pair analogue of spin-flip processes, which account for  this term, is not allowed. The second term represents the current-induced polarization due to the SOC. 
 
 It was shown \cite{Bobkova2018} that the current-induced torques, Eqs.~(\ref{torque_st}) and (\ref{torque_so}) allow for the DW motion in the S/F/S junction in full analogy with the case of a nonsuperconducting ferromagnets. In the absence of the spin-orbit torque, the DW motion driven by the adiabatic torque only occurs if the Josephson current exceeds a threshold value $\tilde J_x^z>\tilde J_x^{z,crit}$, where $\tilde J_x^{z,crit}$ corresponds to the electric current density $eK d_{DW}/\hbar  \sim eK_\perp d_{DW}/\hbar$, where $K$ and $K_\perp$ are the easy- and hard-axis anisotropy constants of the considered ferromagnet, respectively, and $d_{DW}$ is the DW width. This estimate is in full agreement with the result obtained in Ref.~\onlinecite{Tatara2004} for nonsuperconducting ferromagnetic strips under the action of the adiabatic torque, where it has been concluded that $j^{crit} \sim K_\perp d_{DW}$, while in the numerical analysis of Ref.~\onlinecite{Bobkova2018} only the case $K_\perp \sim K$ has been investigated. Taking for estimations the material parameters of $CrO_2$ nanowires \cite{Zou2007} one can obtain that $j^{crit} \sim 10^{10}-10^{11} A/m^2$, what is one-two orders of magnitude larger than the Josephson critical current density $j_c \sim 10^9 A/m^2$, measured in such nanowires \cite{Singh2016}. 
 Other S/F/S devices with large critical current density use usual Co and Ni ferromagnets \cite{Khaire2010,Robinson2010_2,
bhatia2021nanoscale}, rare earth ferromagnet Ho \cite{Robinson2010} and half-metal manganite \cite{Srivastava2017}. Recently the Josephson S/F/S junction thorough the pinned domain wall in Ni was realized \cite{bhatia2021nanoscale}. In principle, it should be possible to realize the supercurrent-controlled domain wall motion in such systems.  
 
\begin{figure}[!tbh]
         \centerline{\includegraphics[clip=true,width=3.0in]{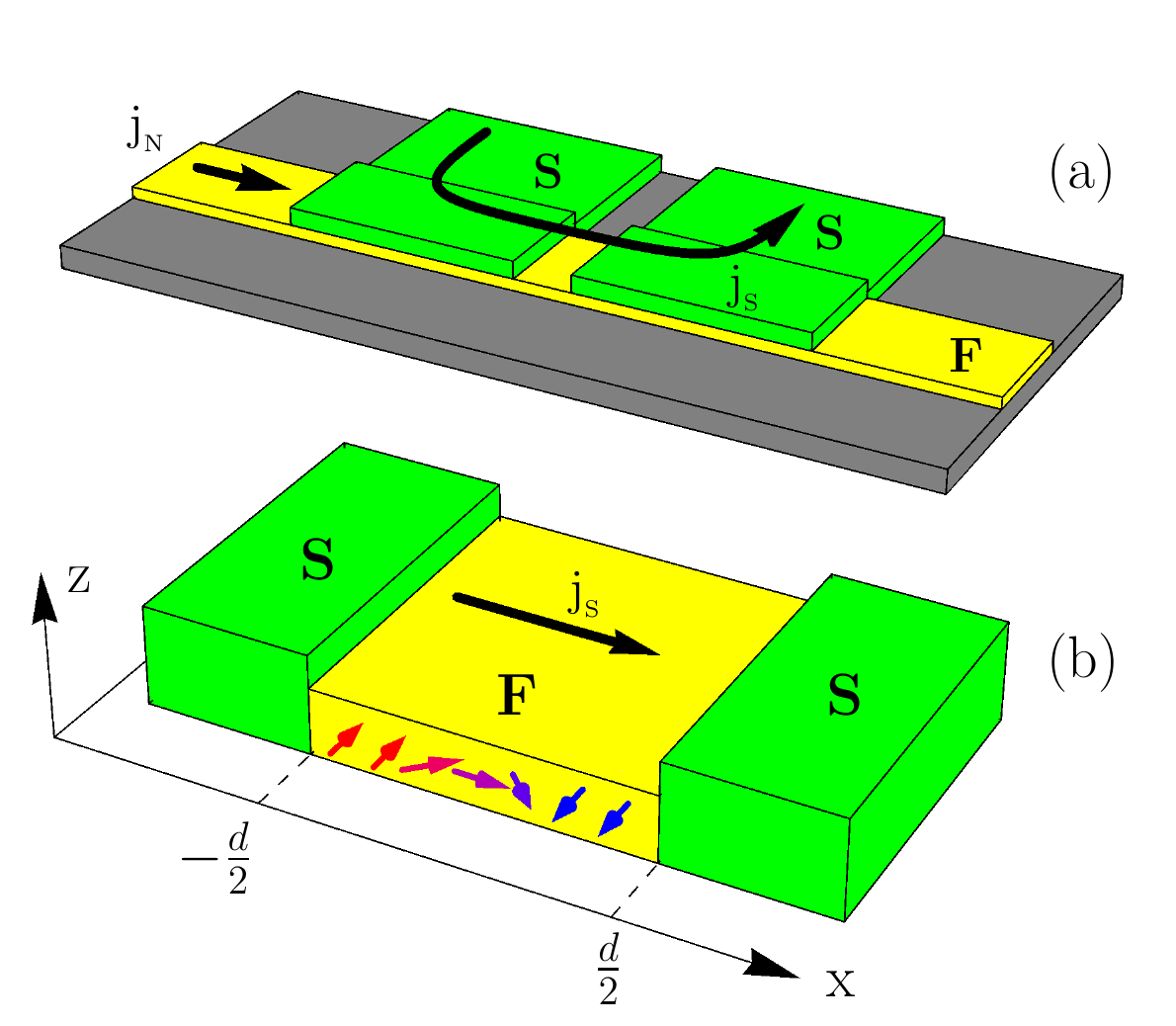}}
        \caption{(a) Sketch of the system, where the Josephson current-induced DW motion can be realized.  Superconducting electrodes forming a Josephson junction are fabricated over a ferromagnetic strip. The normal current $j_N$ flowing through the strip controls the position of the DW and can be used to move it inside the Josephson junction.
       (b) A simplified model of the Josephson junction region. There is a Neel-type DW in the interlayer.  The Josephson supercurrent flows in the F region in $x$ direction. Adopted from Ref.~\onlinecite{Rabinovich2019_2}.}
\label{direct_spin}
\end{figure} 
 
 The DW motion of an unpinned DW under the action of the spin-orbit torque  is possible for arbitrary small values of the applied supercurrent. For example, if we consider a Neel DW in the $(x,y)$-plane, see Fig.~\ref{direct_spin}, and a Rashba-type spin-orbit coupling with $\bm B_x = (0, B_R, 0)$ the spin-orbit torque Eq.~(\ref{torque_so}) at small applied supercurrents leads to the stationary motion of the DW with the velocity
 \begin{equation}
 v_{st} = -u \frac{\beta}{\alpha},
    \label{velocity_DW_so} 
 \end{equation}
where $u = 2 \mu_B J_{s} /M$ is the characteristic velocity associated with the value of spin current $J_{s}=(1/2e)(J_\uparrow - J_\downarrow)$ flowing through the ferromagnetic interlayer and $\beta = 2 m_F B_R d_{DW}$ is the dimensionless SO coupling parameter.
 
 \subsection{Resistive state of the Josephson junctions in the presence of magnetization dynamics}

The reciprocal effect to the torque is the appearance of a gauge spin-dependent vector potential $\bm Z$ in the local spin basis \cite{Frohlich1993,Rebei2006,Jin2006,Bernevig2006,Hatano2007,Leurs2008,Tokatly2008,Bergeret2013}. The gauge spin-dependent vector potential generates an anomalous phase shift. In case the magnetization depends on time, it also produces an electromotive force\cite{Kim2012,Tatara2013,Yamane2013}. It has been shown \cite{Rabinovich2019_2} that due to the presence of this electromotive force, the magnetization dynamics deprives the Josephson junctions of the nondissipative regime, i.e. they cannot support a supercurrent because of the appearance of a nonzero voltage between the leads. The voltage appears to compensate the  electromotive force. This voltage that maintains the DW motion in spin-textured interlayers or  precession of a homogeneous magnetization in JJs, compensating the dissipation power occurring due to Gilbert damping by the work done by a power source.

\begin{figure}[!tbh]
 \begin{minipage}[b]{\linewidth}
   \centerline{\includegraphics[clip=true,width=3.0in]{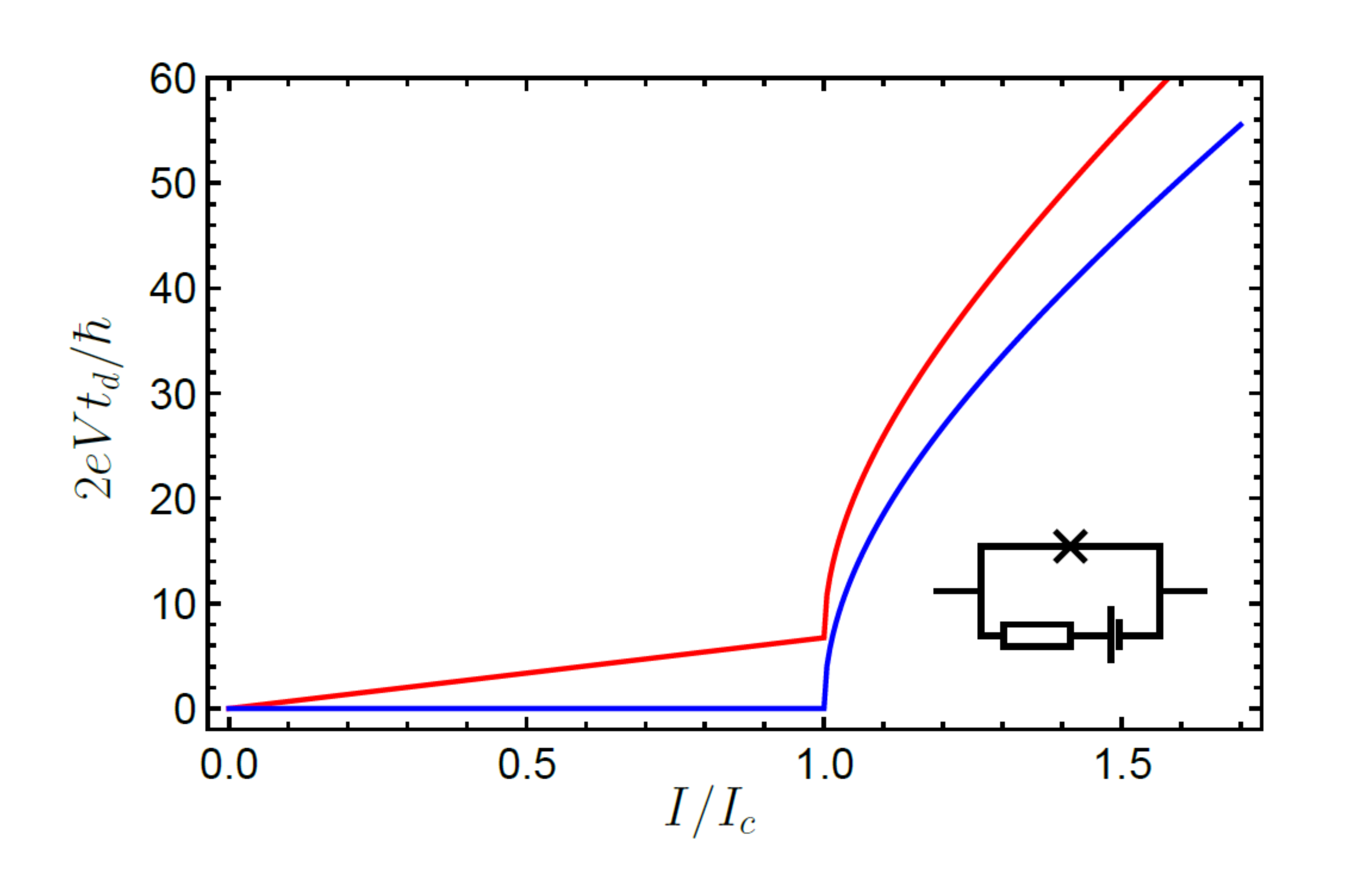}}
   \end{minipage}
      \caption{IV-characteristics of the SFS junction with a DW at rest (blue) and a moving DW (red). $\beta = 1$, $\alpha = 0.1$, $eKd_W/(\pi j_c)=5$. Data are taken from Ref.~\onlinecite{Rabinovich2019_2}. Insert: the equivalent circuit scheme of the junction.}
 \label{dc_j}
 \end{figure}

 Suppose that we apply a constant electric current $I=jS$  (here $S$ is the junction area) to the Josephson junction and consider a steady motion of the DW across the junction with a constant velocity defined by Eq.~(\ref{velocity_DW_so}) under the action of the spin-orbit torque. In this case the  time-averaged voltage induced at the junction can be calculated from Eq.~(\ref{josephson_modified}) and takes the form
{ \begin{eqnarray}
\overline{V(t)} = RS\sqrt{j^2-j_c^2}+\dfrac{\pi \beta^2 u}{e\alpha d_W},
\label{V_averaged}
\end{eqnarray}}
where the first term is well-known and represents the conventional Josephson voltage appearing at $j>j_c$. The second term $V_M $ is nonzero both at $j>j_c$ and at $j<j_c$ and leads to the fact that the Josephson junction is in the resistive state if the DW is driven by current. The corresponding IV- characteristics of the junction are shown in Fig.~\ref{dc_j}. { For numerical estimates of $V_M$ one can take $\alpha=0.01$, $d_W = 60 nm$, $u \approx 1 m/s$, what corresponds to the maximal Josephson current density\cite{Singh2016}  through the $\rm {CrO}_2 $ nanowire $j_c \sim 10^9$ ${A/m}^2$. The dimensionless SOC constant $\beta$ can vary in wide limits. Having in mind that experimentally the predictions can be realized, for example, for hybrid interlayers consisting of a ferromagnet/heavy metal bilayers, $\beta = 1-10$ considering that the SOC $\alpha_R$ ranges from $10^{-11}$ to $ 10^{-10} eV m$ at interfaces of heavy-metal systems \cite{Ast2007}. Then one can obtain $V_M|_{j=j_c}$ up to $10^{-5} - 10^{-3}V.$ }

The resistance of the junction at $j<j_c$ caused by the DW motion is given by
{ \begin{eqnarray}
R_{DW} = \left(\frac{\partial V}{\partial I}\right)_{I<I_c} = \frac{\pi \gamma \beta^2\hbar}{2e^2 S\alpha d_W M},
\label{R_averaged}
\end{eqnarray}}
It is important that according to Eq.~(\ref{R_averaged}) $R_{DW}$ per unit area does not depend on the Josephson junction parameters, such as $j_c$ and $R$, and is determined only by the characteristics of the magnetic subsystem. In  this case, the work done by a power source is exactly equal to the energy losses in the magnetic subsystem due to the Gilbert damping and is not spent on compensating the Joule losses in the interlayer. Indeed, at $j<j_c$ the normal current through the Josephson junction is zero despite nonzero voltage generated at the junction. It is seen directly from Eq.~(\ref{josephson_modified}) because 
 for $j<j_c$ it has the solution $\dot \chi(t)=\dot \chi_0(t)$. The equivalent circuit scheme of the junction is presented in the insert to Fig.~\ref{dc_j}. The voltage is compensated by the electromotive force induced in the junction by the emergent electric field $(\hbar/e) \dot{\bm Z}^{so}$. 

In real setups 
the time of the DW motion through the junction is limited by the finite junction length: $t_{DW} \approx d/v_{st} = (\alpha/\beta) (d/u)$. Therefore, the voltage should be averaged over  $t<t_{DW}$. 
Although experiments on the DW motion in Josephson junctions have not yet been carried out, 
 the estimates $t_{DW} \geq 0.5 (\alpha/\beta) \times 10^{-6}$s has been obtained \cite{Rabinovich2019_2} for $j_c \sim 10^9 A/m^2$. For other setups, which report the Josephson current carried by equal-spin triplet correlations \cite{Khaire2010,Robinson2010}, this time can be several orders of magnitude higher due to much less values of the critical current density.  

\begin{figure}[!tbh]
   \centerline{\includegraphics[clip=true,width=3.0in]{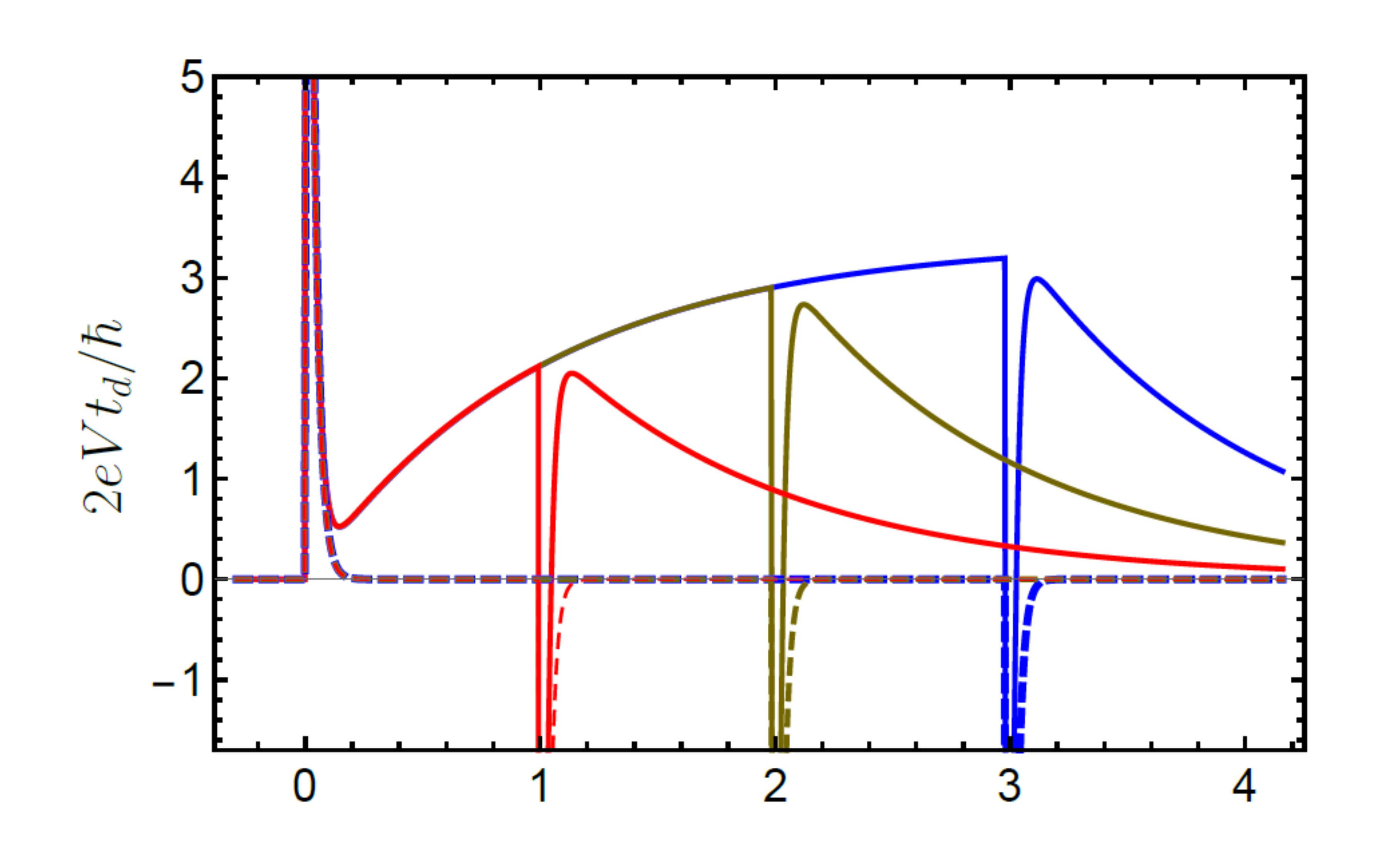}}
        \caption{$V(t)$ for rectangular current impulses. Different curves correspond to different impulse periods $T$; The solid lines correspond to $\beta = 1$ (the anomalous phase due to the DW motion is nonzero) and the dashed lines are for $\beta = 0$ (the anomalous phase shift is zero). Different colors correspond to different values of a characteristic time scale $t_d $, where the DW velocity  reaches its stationary value: (a) $j=0.5j_c$, $T=3t_d$ (blue), $T=2t_d$ (yellow), $T=t_d$ (red).   $\alpha = 0.1$, $eKd_W/(\pi j_c)=5$, $t_d= 40 t_J$. Data are taken from Ref.~\onlinecite{Rabinovich2019_2}.}
 \label{pulse_j}
 \end{figure}
 
 From the practical point of view, it is convenient to induce DW motion by large current pulses.  For short pulses
 $j(t)=j\theta(t)\theta(T-t)$  with $T<t_{DW}$, the DW does not leave the junction during the impulse time. The resulting voltage signal consists of two parts of different physical origins. The first part is the conventional Josephson response with the characteristic time $t_J=1/2eRI_c$. The other part is of purely magnetic origin and vanishes if there is no motion of the domain wall in the junction. The resulting voltage signals  for $j<j_c$ 
are shown in Fig.~\ref{pulse_j}.
In this regime, the typical $V(t)$ curve consists of an initial sharp Josephson voltage impulse, a final sharp impulse of the same nature and a gradual voltage increase and decrease of purely magnetic origin, which takes the form $ V(t) = - \pi \beta v(t)/e d_W$. This gradual voltage increase does not occur if the DW does not move.

\subsection{Electrical control of magnetization in S/F/S junctions}

\subsubsection{Magnetization dynamics under the applied voltage}

In general, a Josephson current can induce  magnetization dynamics. The coupling is described by Eq.~(\ref{torque}) and realized via the generation of the electron spin polarization in the interlayer region, carried by triplet pairs. 
The pair spin should be misaligned with the ferromagnet magnetization to exert a torque on the magnetization. It can be achieved in JJs with misaligned or spin-textured ferromagnets or in the presence of the SOC.

The torque can be obtained  from a direct calculation of the average electron polarization $\bm s$ \cite{Halterman2016}. However, there is another widely used approach for calculating the torque. Its strategy is to determine the Josephson energy of the system $E_J$ and then find the additional contribution to the effective field $\delta \bm H_{eff}$ according to the relation $\delta \bm H_{eff} = -(1/V_F)\delta E_J/\delta \bm M$, where $V_F$ is the ferromagnet volume \cite{Waintal2002,Linder2011,Kulagina2014,Konschelle2009,Nashaat2019,Shukrinov2017,Guarcello2020}. One should understand that this approach only takes into account the torque caused by the {\it supercurrent}. At the same time, under the applied voltage, the supercurrent via the JJ is accompanied by the normal current described by the second term in Eq.~(\ref{josephson_modified}). The normal current  also contributes to the current-induced electron polarization and, therefore, to the torque. The same thing should also be considered for problems under the applied current if  the situation cannot be considered stationary, for example in the case of finite current pulses. However, for complex interlayers composed of ferromagnetic and non-ferromagnetic materials, the supercurrent and the normal current can presumably flow via the different layers.  Therefore the approach based on the Josephson energy  is applicable. The example is an S/F/S JJ via a metallic ferromagnet on top of a 3D TI \cite{Nashaat2019}, where the supercurrent flows via the TI conductive surface states and the normal current flows via the ferromagnet and does not contribute to the torque.

The magnetization dynamics has been analyzed in voltage-biased JJs via single spins \cite{Zhu2004,Nussinov2005}, where Josephson nutations were predicted. In addition, a possibility of obtaining supercurrent-induced magnetization switching in voltage-biased JJ, where the  interlayer is composed of two misaligned ferromagnets separated by a normal spacer, has been reported \cite{Linder2011}. The effect is based on the generation of triplet superconducting correlations in the misaligned configuration of the ferromagnet magnetizations. The pair spin polarization of these triplet correlations is not aligned with the magnetizations of the layers exerting a torque on the free layer. 
The magnetization dynamics has also been considered for the interlayer with three misaligned ferromagnets, two of which are fixed in misaligned configurations \cite{Braude2008}. It has been demonstrated that the magnetic system exhibits a range of
different behaviors, from simple harmonic oscillations to
fractional-frequency periodic behavior and chaotic motion depending on the ratio between the Josephson frequency $\omega_J = 2eV$ and the characteristic frequency of the magnetic system.

\begin{figure}[!tbh]
   \centerline{\includegraphics[clip=true,width=3.0in]{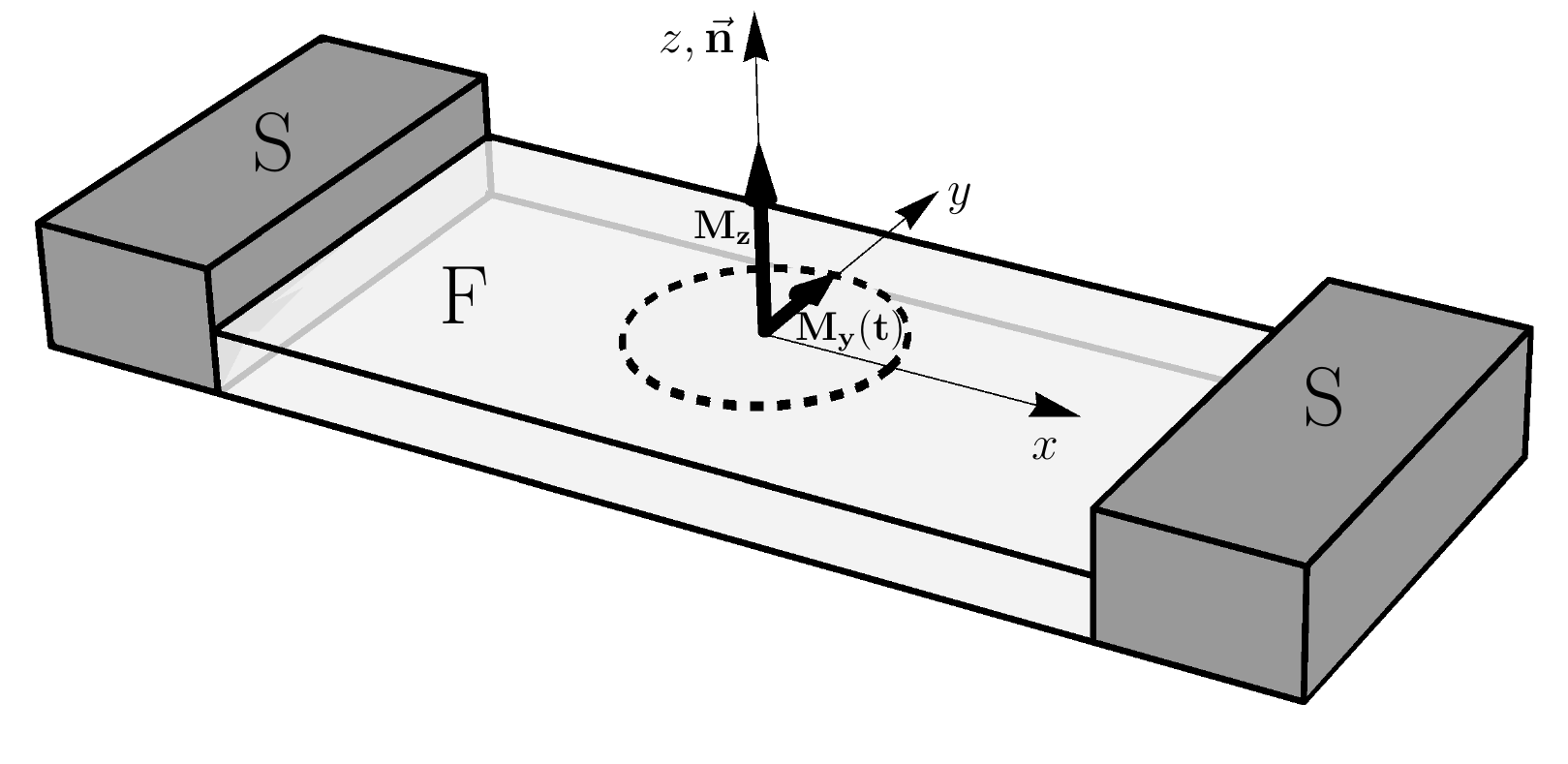}}
        \caption{Sketch of the Josephson junction via an $z$-easy axis ferromagnet with SOC. Redrawn after Ref.~\onlinecite{Konschelle2009}. The voltage-induced magnetization dynamics at $\Gamma \ll 1$ is shown.}
 \label{setup_buzdin}
 \end{figure}

Magnetization dynamics has also been studied in S/F/S Josephson junctions in the presence of the Rashba  SOC inside the F interlayer \cite{Konschelle2009}. The sketch of the considered setup is presented in Fig.~\ref{setup_buzdin}. The ferromagnet has been assumed to be  an easy-axis type with the easy axis  along the $z$-direction. It was obtained that in the low-frequency regime $\hbar \omega_J \ll T_c$ the Josephson current can cause rich dynamics of the magnetization. The influence of the Josephson subsystem on the magnetic subsystem is controlled by the parameter $\Gamma = r E_J/E_M $, where $E_J$ is the Josephson energy, $E_M $ is the energy of the easy-axis magnetic anisotropy and $r \propto (\Delta_{so}/\varepsilon_F)d$ - is the parameter characterising the SOC strength and the magnitude of the anomalous phase shift, see Sec.~\ref{af_FSO}. In the "weak coupling regime" $\Gamma \ll 1$, the magnetic moment precesses around the $z$-axis with the Josephson frequency $\omega_J$. In the limit of the strong coupling $\Gamma \gg 1$, the solution of the LLG equation yields
\begin{eqnarray}
m_y \approx 0,~~~~~~~~~~~ \nonumber \\
m_x (t)= \sin \Bigl[ \frac{\Gamma}{\omega}(1-\cos \omega_J t)\Bigr], \nonumber \\ 
m_z (t)= \cos \Bigl[ \frac{\Gamma}{\omega}(1-\cos \omega_J t)\Bigr],
\end{eqnarray}
where $\omega = \omega_J/\omega_F$ is the ratio of the Josephson frequency to the  ferromagnetic resonance one. It is seen that in this regime, the magnetic moment precesses around the $y$-axis, which is the direction of the pseudo magnetic Rashba field induced by the current. If $\Gamma/\omega>\pi/2$, the full magnetization reversal occurs in the system.  For general coupling regimes, the magnetic dynamics was found to be complicated and strongly nonharmonic.

\subsubsection{Electrical control of the magnetization easy axis}

\begin{figure}[!tbh]
   \centerline{\includegraphics[clip=true,width=2.5in]{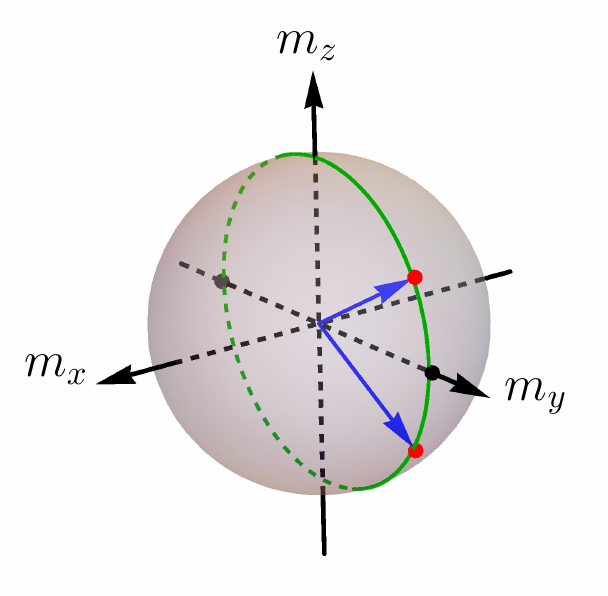}}
        \caption{Dynamical stability points (red circles) of the ferromagnet easy axis for the S/F/S JJ with the Rashba SOC in the interlayer. Redrawn after Ref.~\onlinecite{Shukrinov2018}.}
 \label{fig_reorientation}
 \end{figure}

\begin{figure}[!tbh]
   \centerline{\includegraphics[clip=true,width=3.0in]{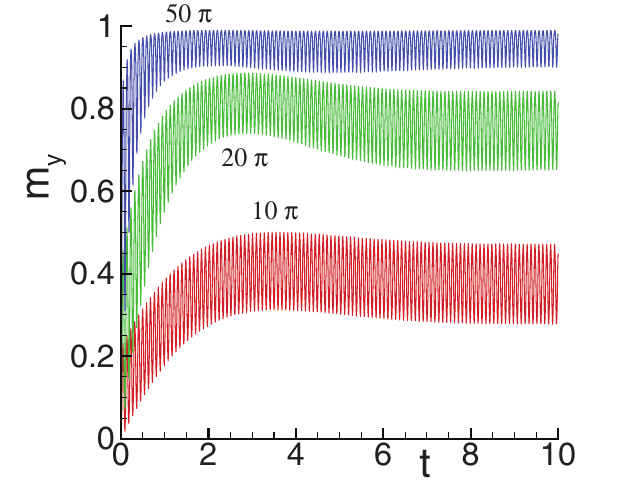}}
        \caption{Dynamical evolution of $m_y$ for the voltage-biased S/F/S JJ with the Rashba SOC in the interlayer. Different colors represent the data corresponding to different values of the parameter $\Gamma/r$. The numerical data and picture are provided by I.Rahmonov.}
 \label{fig_m_y}
 \end{figure}

It has been demonstrated that in S/F/S JJs with SOC in the interlayer,  the stable position of the ferromagnet easy axis can be dynamically reoriented under the applied voltage. The system in which such a feature was first studied is called the Kapitza pendulum. Particularly, in a pendulum with a vibrating point of suspension, the external sinusoidal force can invert the stability position of the pendulum \cite{Kapitza1951}. In Ref.~\onlinecite{Shukrinov2018} it was predicted that the S/F/S JJ with SOC exhibits the analogous behavior: the unstable fixed point, which does not coincide with the equilibrium ferromagnet easy-axis, can 
become dynamically stable under the applied voltage. It was found that if the equilibrium easy axis of the magnet is the $z$-axis (the sketch of the considered system is shown in Fig.~\ref{setup_buzdin}), then under the applied voltage, the new stability point is in the $(z,y)$-plane. The angle $\Theta$ between the stability direction and the $z$-axis is determined by
\begin{equation}
\sin \Theta = \frac{-1+\sqrt{1+4\beta^2}}{2\beta},
\label{reorientation}    
\end{equation}
where $\beta = \Gamma^2 r \alpha/2\omega (1+\alpha^2)$. Eq.~(\ref{reorientation}) determines two dynamical stability points, as it is shown in Fig.~\ref{fig_reorientation} instead of the equilibrium stability points $m_z = \pm 1$. The physical origin of the phenomenon is clear: in the presence of the Josephson current, the magnetic moment is influenced not only by the magnetic anisotropy field, but also by the pseudo magnetic Rashba field aligned with the $y$-axis. In the limit $\Gamma \gg 1$, the dynamical easy axis lies along the $y$-direction. The full time evolution of the $y$-component of the magnetization is shown in Fig.~\ref{fig_m_y} for different values of the coupling between the magnetic and Josephson subsystems.

\begin{figure}[!tbh]
   \centerline{\includegraphics[clip=true,width=3.0in]{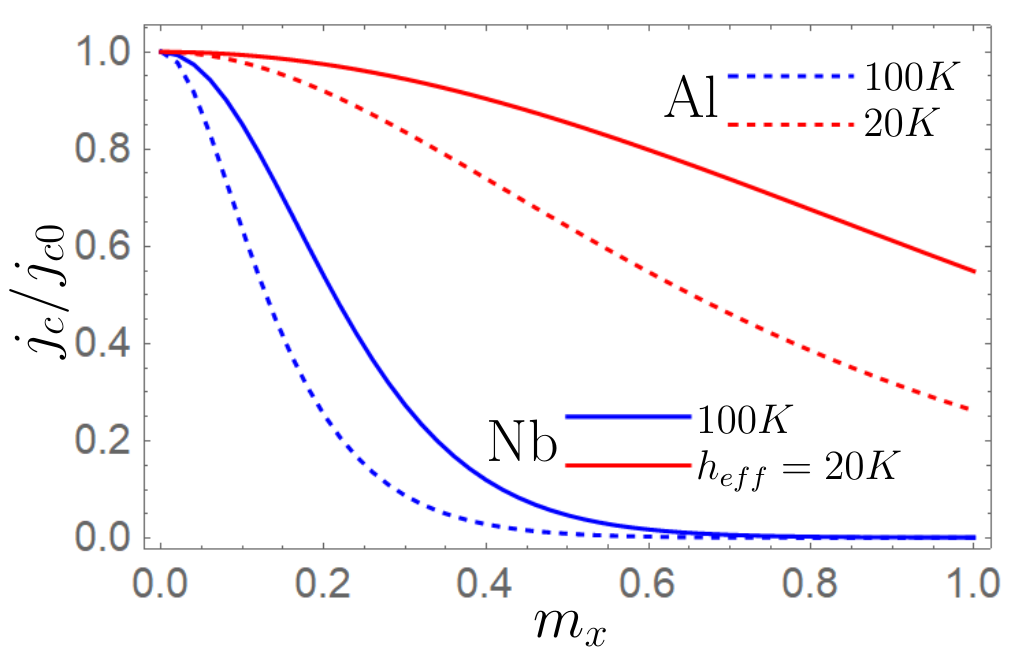}}
        \caption{Results of the numerical calculation of the critical current of the S/F/S JJ on top of the 3D TI  as a function of $m_x$. The parameters are chosen to be relevant for a JJ with Nb superconducting leads (solid lines) and Al leads (dashed lines). The value of the effective exchange field is $h_{eff} = 20K$ for red lines. In this case, the suppression of the critical by $m_x$ is not very strong. For comparison the blue lines represent the critical current at $h_{eff}=100K$. It is seen that in this case it is strongly suppressed. $j_c$ is normalized to $j_{c0} \equiv j_c(m_x=0)$. Data are taken from Ref.~\onlinecite{Rabinovich2020}.}
 \label{current_orientation}
 \end{figure}

Another interesting effect related to the electrical control of the ferromagnet easy axis is the easy axis splitting in S/F/S JJs on top of the 3D TI \cite{Nashaat2019}. The qualitative difference of this system  from the case of the S/F/S JJ with the SOC in the interlayer is that in 3D TI-based JJs the critical current demonstrates strong dependence on the $x$-component of magnetization. At the same time in S/F/S JJ with the SOC it has been considered as independent on the magnetization direction. The dependence and its physical origin were discussed in Sec.~\ref{af_TI}.  The suppression of the critical current as a function of $m_x \equiv M_x/M_s$  has been discussed in Ref.~\onlinecite{Rabinovich2020} and is presented in Fig.~\ref{current_orientation}. For estimates we take  $d=50nm$, $v_F = 10^5m/s$  and $T_c = 10K$, what corresponds to the parameters of $Nb/Bi_2Te_3/Nb$ Josephson junctions\cite{Veldhorst2012}. In this case $\xi_N = v_F/2\pi T_c \approx 12nm$.  $j_c(m_x)$ for $T_c = 1.8K$ has also been plotted, what corresponds to the Josephson junctions with $Al$ leads.

\begin{figure}[!tbh]
   \centerline{\includegraphics[clip=true,width=4.0in]{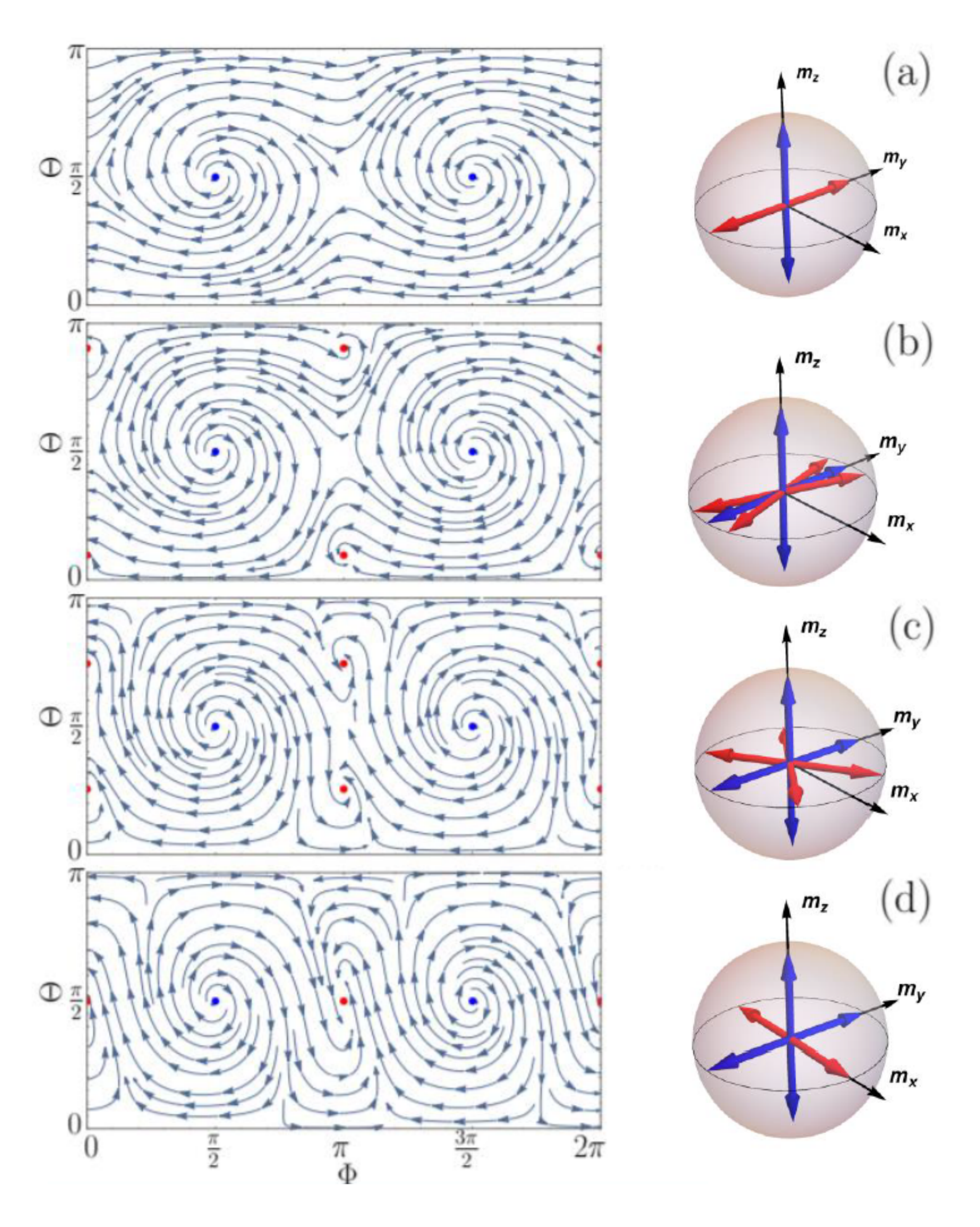}}
        \caption{Dynamical easy axis positions for the ferromagnet in the interlayer of the S/F/S JJ on top of the 3D TI are shown in the right column by the red arrows. The corresponding vector fields demonstrating the time evolution of the magnetization are represented in the left column. The stable points are shown by red circled, see text. Different panels corresponds to different values of $A$. Data are taken from Ref.~\onlinecite{Nashaat2019}.}
 \label{fig_splitting}
 \end{figure}

Now let us assume that the ferromagnetic interlayer of the JJ has an easy axis along the $y$-direction. The dependence of the anomalous phase shift on $m_y$ Eq.~(\ref{josephson_final}) results in the additional contribution to the $y$-component of the effective magnetic field $\delta H_{eff,y} = -(1/V_F)\delta E_J/\delta M_y$, where $E_J = \Phi_0 (j_c/2 \pi)[1-\cos (\chi-\chi_0)]$ with $j_c$ and $\chi_0$ determined by Eqs.~(\ref{critical_current}) and (\ref{josephson_final}). The effective magnetic field is added to the magnetic anisotropy field and does not cause the magnetization dynamics if the magnetization is along the easy axis. At the same time the dependence of the critical current on $m_x$ leads to nonzero $H_{eff,x} = A m_x$ at small $m_x$. This means that the easy $y$-axis can become unstable in a voltage-driven or current-driven junction, while this axis is always stable if the critical current does not depend on magnetization direction. Moreover, there is no difference for the system between $\pm m_x$-components of the magnetization. This leads to the fact that in the driven system an easy the axis does not reorient, keeping two stable magnetizations directions, as it was already obtained earlier, but splits. As a result, {\it four stable} directions of magnetization appear.  This splitting effect can be realized in a range of the parameter $A$ values, which can be achieved experimentally according to the estimates of Ref.~\onlinecite{Nashaat2019}.

Fig.~\ref{fig_splitting} demonstrate the corresponding four stability points (red arrows). The dynamically stable easy axes are in the $(x,y)$-plane, therefore if the magnetization direction is parametrized as $\bm m = (\sin \Theta \cos \Phi, \cos \Theta, \sin \Theta \sin \Phi)$, the stable points correspond to $\Phi = 0,\pi$. Red circles shown them at the vector fields, which demonstrate the time evolution of the magnetization starting from an arbitrary initial magnetization position. Panel (a) corresponds to the small value of the parameter $A$ when the easy-axis is not split yet. Panels (b)-(c) are in the parameter range where the splitting occurs, and for panel (d), the $A$ value already exceeds the upper boundary of the range. In this case, the easy axis is reoriented to the $x$-direction.   

\subsubsection{Magnetization reversal by electric current pulses. Cryogenic memory elements.}

Another intriguing effect related to the electrical control of the magnetization in S/F/S JJs is the ferromagnet magnetization reversal by the Josephson current pulses. One of the key challenges towards  developing ultra-low-power computers
is the fabrication of a reliable and scalable cryogenic memory architecture. Superconductor-ferromagnet-superconductor
(S-F-S) junctions are promising structures suggested for such memories \cite{Feofanov2010,Ryazanov2012,Baek2014,Golod2015,Gingrich2016,Niedzielski2018,Dayton2018,DeSimoni2018,deAndresPrada2019,Bolginov2012,Larkin2012,Bakurskiy2013,Bakurskiy2018}. The magnetization reversal by the electric current pulses discussed here has also been suggested as one of the possible realizations of the JJ-based cryogenic memory elements.

\begin{figure}[!tbh]
   \centerline{\includegraphics[clip=true,width=3.0in]{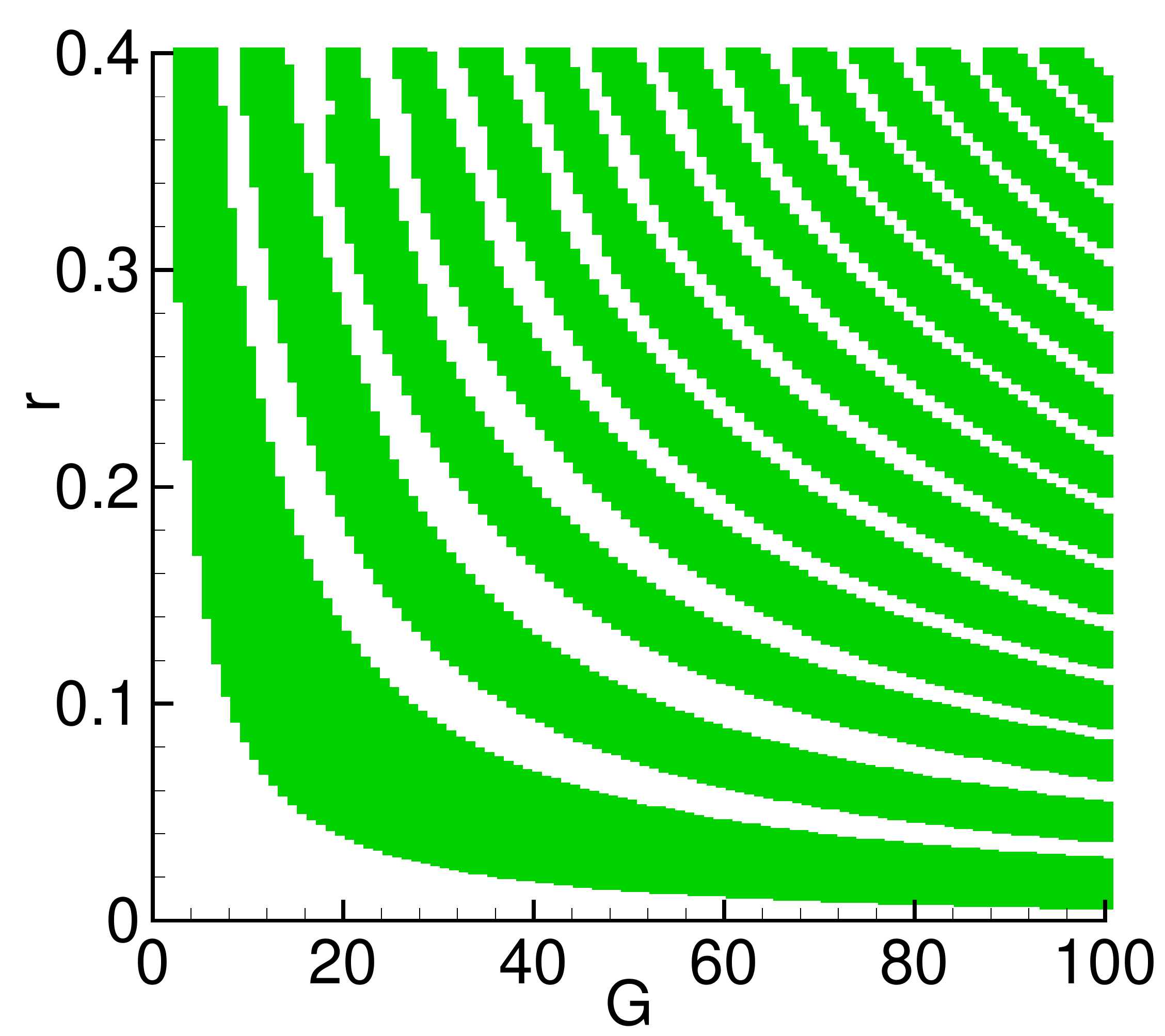}}
        \caption{Periodicity of
reversal intervals in the plane $(G,r)$, where $G=\Gamma/r$. The numerical data and picture are provided by I.Rahmonov.}
 \label{fig_reversal_1}
 \end{figure}

The magnetization reversal by current pulses has been originally discussed in Ref.~\onlinecite{Shukrinov2017} for the S/F/S Josephson junction with the Rashba SOC in the interlayer. The magnetization dynamics in the system shown in Fig.~\ref{setup_buzdin} has been considered in the regime of applied electric current pulses.  The torque acting on the magnetization has been calculated via contribution to the effective magnetic field, produced by the supercurrent $\delta \bm H_{eff} = -(1/V_F)\delta E_J/\delta \bm M$. The possibility of the moment reversal has been demonstrated. The possibility depends strongly on the pulse amplitude and duration, as well as on the value of the SOC parameter $r$ and the coupling strength between the magnetic and Josephson subsystems $\Gamma$. The ideas suggested in Ref.~\onlinecite{Shukrinov2017} have been developed in several subsequent papers. In particular, a periodicity in the appearance of intervals of the reversal of the magnetic moment under the variation of the spin–orbit coupling $r$, Gilbert damping parameter, and the coupling parameter $\Gamma$ \cite{Atanasova2019} has been predicted. An example of the corresponding periodic patterns is represented in Fig.~\ref{fig_reversal_1}. Furthermore, an analytical criterion of the most efficient reversal for the easy-axis magnet has been formulated \cite{Mazanik2020}. The criterion allows for optimization of the pulse parameters. 

In Ref.~\onlinecite{Guarcello2020} the idea to exploit the S/F/S JJ with the Rashba SOC as a cryogenic memory element has been investigated. The two memory states are encoded in the direction of the out-of-plane
magnetization and the current pulses switch between them. The robustness of the current-induced magnetization reversal against thermal fluctuations has been explored \cite{Guarcello2020,Guarcello2021}. It has also been suggested \cite{Guarcello2020} that the readout of the memory state can be nondestructively
performed by direct measurement of the magnetization
state through a dc SQUID inductively coupled to the junction.

The magnetization reversal by the electric current pulses has also been investigated for the 3D TI-based S/F/S JJs \cite{Bobkova2020}. The advantage of this system is the very strong value of the SOC parameter $r$ provided by the spin-momentum locking in the TI surface states, what allows to use low-current pulses for the reversal of the magnetization. The full spin-orbit torque containing the contributions both from the supercurrent and the normal current can be found in terms of the current-induced polarization of the surface states conductivity electrons Eq.~(\ref{spin_current}) and takes the form
\begin{eqnarray}
\bm N = \frac{J_{ex}}{d_F} \bm M \times \langle \bm s \rangle = -\frac{\gamma h_{TI}j}{eMv_F d_F}[\bm m \times \bm e_y]. 
\label{torque_TI}
\end{eqnarray}

\begin{figure}[!tbh]
   \centerline{\includegraphics[clip=true,width=3.5in]{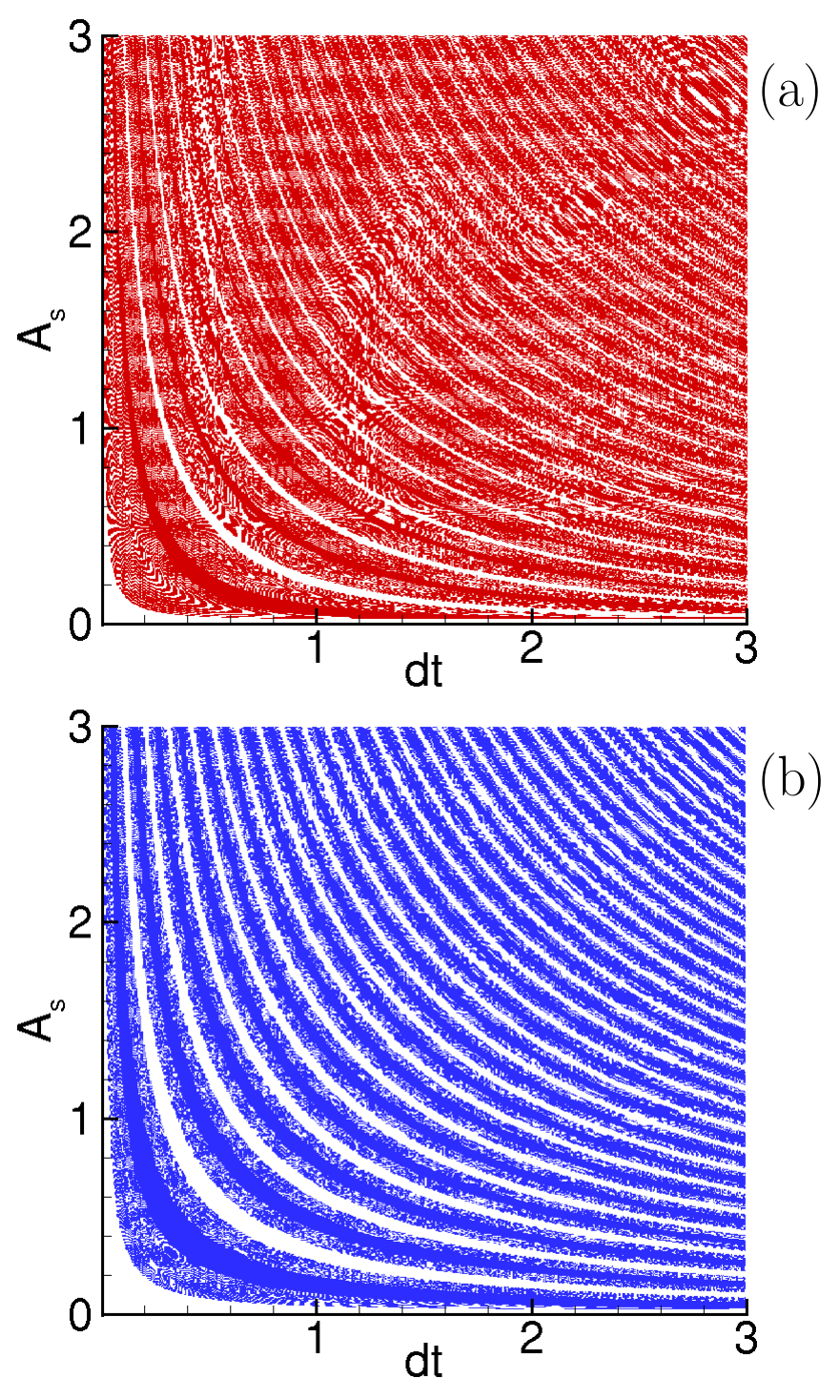}}
        \caption{Numerically calculated yes/no reversal diagram in the $(dt, A_s)$-plane, where $dt$ is the current pulse duration and $A_s$ is its amplitude, normalized to the value of the critical current at $T=0$ and $m_x=0$. $dt$ is measured in units of $M/\gamma K_u$. The regions, where the reversal occurs are colored and where it does not occur are white. They are separated by the white/colored striped regions, which represent an "uncertainty" regime and and discussed in the text. (a) $k=10$, (b) $k=1$. Data are taken from Ref.~\onlinecite{Bobkova2020}.}
 \label{fig_reversal_2}
 \end{figure}

Whether there will be a reversal of the magnetic moment under the action of a given current pulse - depends strongly on the amplitude and duration of the pulse. It is illustrated by numerical data represented in Fig.~\ref{fig_reversal_2}. This diagram shows regions where the reversal occurs/does not occur in the $(dt, A_s)$-plane, where $dt$ is the current pulse duration, and $A_s$ is its amplitude. It is seen that the regions where the reversal occurs (colored) and does not occur (white) are separated by striped regions, where the behavior of the system is very difficult to predict. Therefore, the result of the operation (yes/no reversal) is very sensitive to the pulse parameters.  It was reported that the widths of the uncertainty regions depend on the magnetic anisotropy of the ferromagnet. In Ref.~\onlinecite{Bobkova2020} the magnetic anisotropy field in the ferromagnet was chosen as follows:
\begin{eqnarray}
\bm H_{eff} = -\frac{K}{M}m_z \bm e_z + \frac{K_u}{M}m_x \bm e_x, \label{effective_field}
\end{eqnarray}
where $K$ and $K_u$ are the hard axis and the easy axis anisotropy constants, respectively. Therefore, an easy-plane anisotropy was considered in addition to the easy axis anisotropy, investigated earlier. This situation corresponds to the experimental data reported for YIG thin films\cite{Mendil2019}. 
The parameter
 $k=K/K_u$ can describe the ratio of the hard amd easi-axis anisotropy parameters. The nonzero value of this parameter results in the appearance of the uncertainty regions in the reversal diagrams, Fig.~\ref{fig_reversal_2}.  These regions grow with an increase of $k$ and disappear at $k \to 0$.

\begin{figure}[!tbh]
   \centerline{\includegraphics[clip=true,width=2.8in]{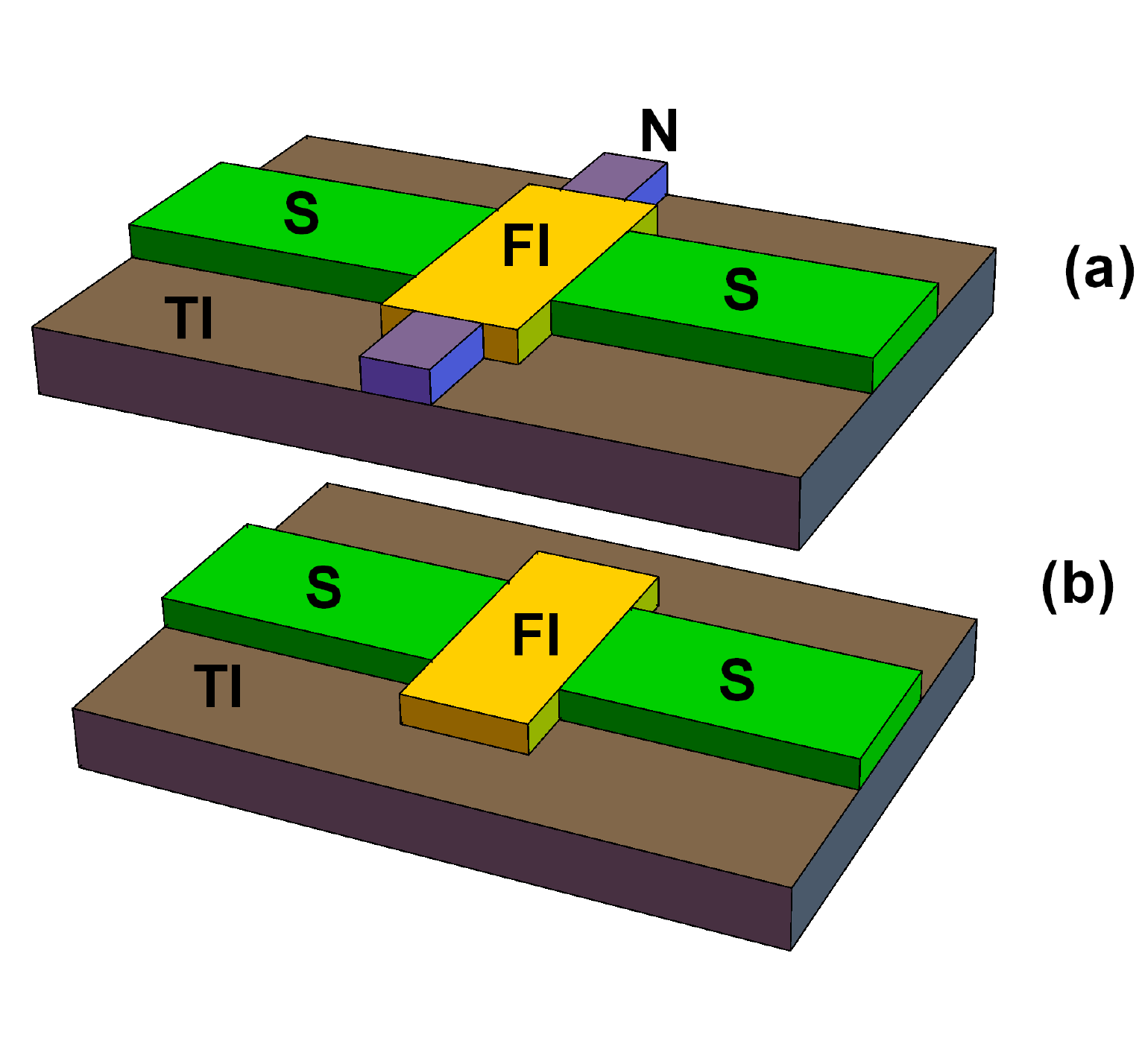}}
        \caption{(a) Sketch of the superconductor/ferromagnetic insulator/superconductor (S/FI/S) JJ on top of the 3D topological insulator (TI) with additional normal (N) electrodes, which are used for the electrical detection of $\bm m = \bm e_x \to -\bm e_x$ reversal in comparison to (b) the basic S/FI/S JJ. Taken from Ref.~\onlinecite{Bobkova2020}.}
 \label{sketch_2}
 \end{figure}

In addition, it has been suggested in Ref.~\onlinecite{Bobkova2020}  to exploit the voltage induced at the junction due to the magnetization dynamics for electrical detection of the magnetization reversal. To detect the reversal $\bm m = \bm e_x \to -\bm e_x$ it is efficient to measure the transverse voltage generated between the additional leads, as it is shown in Fig.~\ref{sketch_2}. This voltage is measured in the open circuit geometry when the electric current between the additional transverse electrodes is zero. In this case the solution of Eq.~(\ref{josephson_modified}) takes the form $\dot \chi = \dot \chi_0$. Then the voltage generated between the additional electrodes due to magnetization dynamics is determined by the dynamics of $m_x$ and can be written as follows\cite{Rabinovich2020}:
\begin{eqnarray}
V_t = \dot h_{TI,x} d/e v_F.
\label{voltage}
\end{eqnarray}
This voltage is the same both for superconducting additional electrodes and for nonsuperconducting electrodes and is only determined by the electromotive force. If the magnetization dynamics is caused by the pulse of electric current applied in the $x$-direction, then 
\begin{eqnarray}
\int V_t(t)dt = r  \frac{\hbar}{e} \frac{\Delta m_x}{2},  
\label{voltage_integral}
\end{eqnarray}
where $\Delta m_x$ is the full change of $m_x$ caused by the pulse. If the magnetization reversal $\bm m = \bm e_x \to -\bm e_x$ occurred, then $\Delta m_x = -2$, otherwise it is zero. Therefore, the integrated over time value of the voltage between the additional electrodes can be used as a criterion of the magnetization reversal. 

\section{Triplet correlations generated by the moving condensate}

\label{sec:triplets}

Another type of magnetoelectric effect, specific for superconducting systems, is the generation of triplet $S=1$ Cooper pairs by the condensate motion. It is well-established that the nonzero momentum of moving superconducting condensate is a pair-breaking factor. The reason is that it makes the momenta of two paired electrons to be not exactly opposite, and such a finite-momentum pairing has less binding energy. This fundamental mechanism called the orbital depairing effect  exists in any superconducting system and  leads to the suppression of superconductivity by the magnetic field or by the supercurrent \cite{TinkhamBook}.
 
\subsection{Triplets induced by the static Meissner currents}

In Ref.~\onlinecite{Silaev2020} it has been demonstrated that in superconductor/ferromagnet hybrids with interfacial SOC controllable condensate motion can {\it induce} superconducting correlations. More specifically, it was predicted that the condensate motion provides effective  manipulation of the odd-frequency spin-triplet pairing states \cite{Berezinskii1974} which have attracted continual interest for several decades 
\cite{Belitz1992,Belitz1999,Balatsky1992,Abrahams1995,Coleman1994,volkov2003,Bergeret2005,Fominov2015,Linder2019,Black-Schaffer2012,Black-Schaffer2013,DiBernardo2015,Komendova2017,Cayao2017,Triola2018,Cayao2018,Dutta2019,Sukhachov2019,Alidoust2017,Tanaka2007,Asano2007,Asano2013,Yokoyama2008,Banerjee2018}. It has been suggested that this mechanism should generate equal-spin triplet  Cooper pairs in currently available experimental setups with SOC \cite{Satchell2018,Satchell2019,Jeon2020,Jeon2019,Jeon2019_2,Jeon2019_3,Jeon2018,Petrzhik2019}. 
In the context of superconductor/ferromagnet hybrid structures
these correlations are known as 
long-range triplets (LRT) because they can penetrate at large distances into the ferromagnetic material \cite{Braude2007,Mironov2015,Singh2016,Robinson2010,Khaire2010,Bergeret2005,Bergeret2001,Bergeret2001_2,Houzet2007,Keizer2006,Fominov2007,Eschrig2008,Eschrig2003,Robinson2010_2,Halterman2016_2,Alidoust2018,Srivastava2017}. Consequently, in S/F/S JJs the magnetic field can in fact stimulate Josephson current by generating long-range equal-spin odd-frequency triplet correlations.

\begin{figure}[!tbh]
   \centerline{\includegraphics[clip=true,width=3.0in]{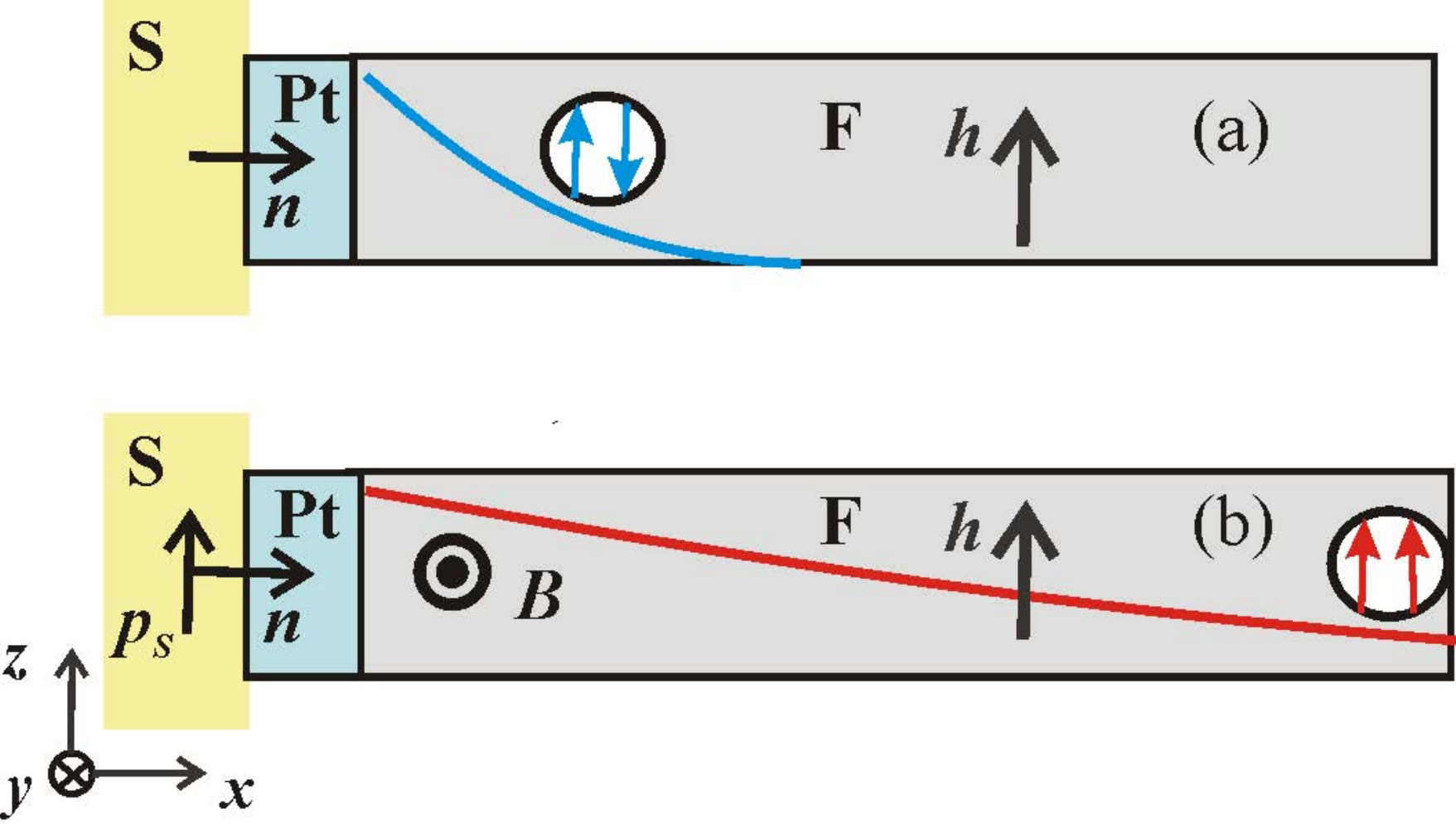}}
        \caption{Schematic picture of the simplest system, where the LRTs are generated by the moving condensate: a diffusive superconductor/ferromagnet junction with Rashba SOC
at the S/F interface induced by the thin heavy metal Pt
layer. (a) In the absence of the moving condensate, only short-range superconducting correlations are
present. (b) The condensate motion along the exchange field direction is induced, for example, by the magnetic field through the Meissner effect. The interplay of condensate momentum $p_s$, SOC and exchange field $h$ leads to the generation of
long-range s-wave spin-triplet component. Taken from Ref.~\onlinecite{Silaev2020}.}
 \label{triplets_1}
 \end{figure}

The sketch of the basic structure is shown in Fig.~\ref{triplets_1}. Without the supercurrent, LRT are absent in the generic  S/F structures such as
shown in Fig.\ref{triplets_1}(a). 
Here the exchange field $\bm h\parallel \bm z$ produces only short-range triplets (SRT) with $S_z=0$,  
shown schematically by the blue arrows, which decay at a short  length of the order $\xi_F \sim 1$ nm in usual ferromagnets.  It has been demonstrated \cite{Bergeret2013,Bergeret2014} that, in principle, the SOC in combination with the exchange field can induce LRTs in diffusive systems. However, pure Rashba or Dresselhaus SO coupling does not induce the LRTs in a transversal geometry with an in-plane magnetization \cite{Bergeret2013}. It is this situation that is the most common experimental setup and is depicted in Fig.~\ref{triplets_1}.  The generation of LRT with $S_z=\pm 1$
shown schematically by red arrows in Fig.\ref{triplets_1}(b)
can be achieved by inducing
 the superconducting condensate momentum $\bm p_s$
satisfying the condition $\bm h\times (\bm n\times \bm p_s)\neq 0$ . 
The required condensate motion can be achieved, for example, by applying the external magnetic field along the $y$-direction, which causes the Meissner currents along the $z$-direction.

 The qualitative physics of the effect can be described as follows. The general structure of  anomalous function describing the pairing in a spin-triplet channel can be parameterized as
$ (\hat{\bm\sigma}\cdot \bm d)$, where $\hat{\bm\sigma}$ is the vector of Pauli matrices. 
 The  role of $\bm h\parallel \bm z$ is to  generate $S_z=0$ spin triplet correlations with $\bm  d_{SRT} \propto \bm h$. 
The role of SOC $\hat H_{soc} = \alpha [\bm n \times \bm p]$ is to convert them to 
$S_z=\pm 1$ correlations  due to the momentum-dependent spin rotation by the SOC. 
  The resulting $p$-wave spin vector is $\bm d_{pw} = F_{pw}(\omega) \bm h\times (\bm n\times {\bm p}) $ 
with the amplitude $F_{pw}(\omega)$ which is an even function of the Matsubara frequency $\omega$. Such momentum-odd correlations are greatly suppressed in diffusive systems due the efficient impurity-induced momentum averaging. 

The  externally induced superflow $\bm p_s \neq 0$  induces Doppler shift of the quasiparticle energy levels\cite{Kohen2006,Budzinski1973} $\bm v_F\cdot \bm p_s $. 
 It results in the  suppression of pairing on one part of Fermi surface, namely for electrons with momentum $\bm p\parallel \bm p_s$. 
In the simplest case of homogeneous system 
this leads to the shift of imaginary frequencies so that the amplitude of triplet correlations is given by  $F_{pw}(\omega - i\bm v_F\cdot\bm p_s) \approx F_{pw}(\omega) - i(\bm v_F\cdot\bm p_s) \partial_\omega F_{pw}$. This modification of the pairing amplitude 
results in the additional component of the 
spin vector $\delta \bm d = -i \partial_\omega F_{pw} (\bm v_F\cdot\bm p_s)\bm h\times (\bm n\times {\bm p})$.
The s-wave component $\bm d_{sw} = \langle\delta \bm d \rangle_p$ is 
given by $\bm d_{sw} = (2/3i)E_F (\partial_\omega F_{pw})  \bm h\times (\bm n\times {\bm p}_s)$. 

By the order of magnitude 
$ |\bm d_{pw}| \sim  h v_F\alpha /\Delta^2 \gg v_F\alpha /E_F$. Then the typical amplitude of the s-wave correlations is 
$|\bm d_{sw}|\sim  (p_s \xi) h v_F\alpha /\Delta^2 $ where $\xi$ is the coherence length.  It means that the magnitude of this magnetoelectric effect can be considerably larger than that of the effects discussed before because the magnitude of the electron spin polarization and current-induced triplet correlations due to the Rashba SOC is governed by the parameter $ v_F\alpha /E_F$, see Sec.~\ref{SO_direct}.

Technically the triplet correlations can be calculated on the basis of the quasiclassical Usadel equation \cite{Silaev2020}. If the Rashba SOC is only present at the S/F interfaces, the (SOC+supercurrent)-induced SRT-LRT conversion can be described by the effective boundary condition at the S/F interface, which in the framework of the linearized with respect to the anomalous Green's function approach takes the form\cite{Silaev2020}:
\begin{align} 
    n_x \nabla_x \hat{\bm f}_{LRT} =  4 i  \tilde \alpha \hat\tau_3\hat{\bm f}_{SRT}\times (\bm p_s\times \bm n)\label{bc_triplet}
  \end{align}
where the surface SOC strength $\tilde{\alpha} = \int dx \alpha (x)$. 
The solution of the linearized Usadel equations for the LRT anomalous Green's function
\begin{align}
          & \frac{D}{2} \nabla^2_x \hat{\bm f}_{LRT} =
      |\omega|  \hat {\bm f}_{LRT}  \label{usadel_triplet}
        \end{align}
supplemented by the boundary condition (\ref{bc_triplet}) gives the LRT anomalous Green's function, which in the $\xi_F\ll d_F\ll \xi_N$ takes the form:
\begin{align}
      \hat{\bm f}_{LRT}  \propto \frac{\gamma\xi_F^2\tilde \alpha }{d_F \omega }
      \frac{\Delta}{\sqrt{\Delta^2+\omega^2}}
      \bm h\times (\bm n\times \bm p_s),
      \label{ft_2}
  \end{align}
where $\xi_F = \sqrt{D/h}$ is the coherence length of the SRTs in the ferromagnet and $\gamma$ is the S/F interface transparency\cite{Kupriyanov1988, Bergeret2005}.  The amplitude of long-range spin-triplets is proportional to the condensate momentum 
   $\bm p_s$. If it is generated by the external magnetic field through the  Meissner effect, then $p_s \propto B$.
   
\begin{figure}[!tbh]
   \centerline{\includegraphics[clip=true,width=3.5in]{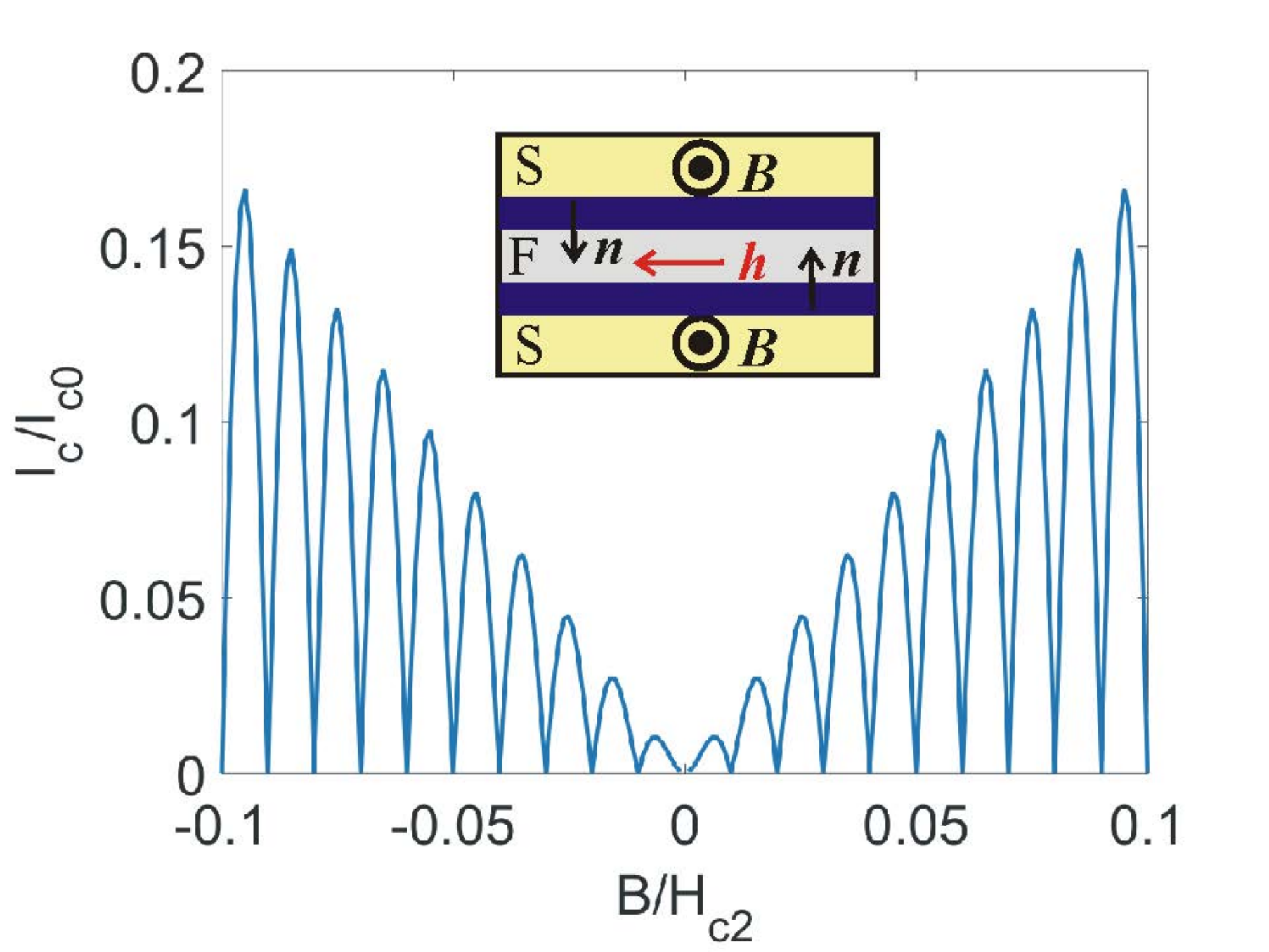}}
        \caption{Interference patterns of the critical current $I_c(\Phi) $ for the magnetic-field induced Josephson effect through the magnetic and SOC interlayers as shown in inset. The results are calculated in Ref.~\onlinecite{Silaev2020} and the figure is adopted  from the same paper.}
 \label{triplets_2}
 \end{figure}
   
If now two S/F interfaces with Rashba SOC are combined into the S/F/S JJ, the amplitude of critical current grows as $I_c\propto B^2$ for a small external magnetic field, when the total flux through the junction area $\Phi= 2\lambda_L L B$ is small $\Phi\ll \Phi_0$.  Here $L$ is the length of the junction.
 For larger fields,  one needs to take into account
 phase variation along the junction
 which leads to the usual factor $(L \sin\phi)/\phi$ in the critical current, where $\phi= 2\pi \Phi/\Phi_0$. It results in $I_c\propto B$ envelope dependence of the critical current shown in Fig.~\ref{triplets_2}. This growth is bounded from above by the depairing effects.  The $I_c(B)$ pattern in Fig.~\ref{triplets_2} drastically differs from the ones observed previously in non-ferromagnetic Josephson junctions  with SOC \cite{Suominen2017,Assouline2019} and ferromagnetic ones without SOC \cite{Kemmler2010} . 
This behaviour can be considered as the fingerprint of the LRT produced by the moving condensate in the presence of SOC.

\subsection{Dynamic triplets induced by alternating electric fields}

\begin{figure}
 \centerline{$
 \begin{array}{c}
 \includegraphics[width=3.5in]
 {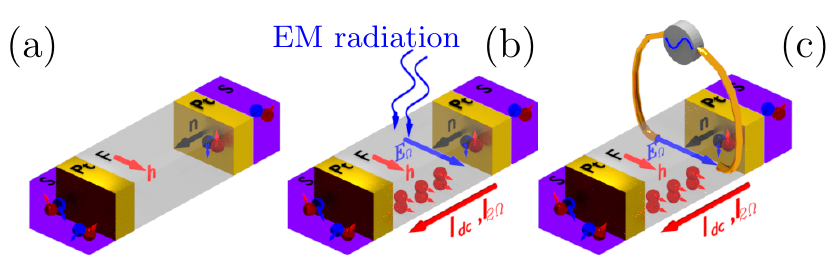} 
 \end{array}$}
 \caption{\label{Fig:1} 
   S/F/S junction with Rashba SOC at the interfaces induced by the thin layer of heavy metal Pt. (a) Only short-range superconducting correlations are present shown by the blue and red spheres with opposite arrows. Therefore, if the interlayer length is of the order of several normal state coherence lengths, the Josephson current through the junction is strongly suppressed.  (b,c) Generation of long-range triplet (LRT) correlations due to the irradiation of the setup with electromagnetic wave (b) and 
   by applying the  ac current source (c) both producing the electric field $\bm E (t)= \bm E_\Omega e^{i\Omega t}$ in the ferromagnetic interlayer. The LRT are shown schematically by the red spheres with co-directed arrows corresponding to the spin states aligned with the exchange field $\bm h$.  They sustain the Josephson current. Adopted from Ref.~\onlinecite{Bobkova2021}.
   }
 \end{figure}

The moving condensate can be also induced by the alternating electric field $\bm E(t)$, which is described by the time-dependent vector potential $\bm E = -(1/c)\partial_t \bm A$. It produces an oscillating condensate motion with the momentum $\bm p_s = -(2e/c)\bm A(t)$. It has been demonstrated in Ref.~\onlinecite{Bobkova2021} that in the S/F/S Josephson junction sketched in Fig.~\ref{Fig:1} with Rashba SOC at the S/F interfaces this oscillating condensate motion produces triplet correlations with the energy and time-dependent spin vector constructed as follows 
\begin{align}\label{Eq:SpinVector}
     \bm d (\varepsilon,t) = \int dt^\prime K_d (\varepsilon,t-t^\prime) (\bm E (t^\prime) \times \bm n) \times \bm h.
 \end{align}
 The scalar kernel $K_d (\varepsilon, t-t')$ is determined  in the framework of a particular microscopic model. 
 
 The triplets are long-range and result in the controllable appearance of the Josephson effect in the setups shown in Fig.~\ref{Fig:1} under irradiation or by applying an ac current source. This mechanism can help to achieve switching rates in the terahertz and even the visible light frequency domains. If the applied electric field has zero time-average value, the time-average value of the dynamic triplets also vanishes.  In spite of this fact, they result in nonzero dc component of the Josephson current via the JJ. For the case of a harmonic electromagnetic wave, the following current-phase relation has been obtained:
     \begin{align} \label{Eq:CurrentJJ}
     I (\chi,t)= [I_{dc}^c + I_{2\Omega}^c \cos(2\Omega t) ]\sin\chi .  
   \end{align}
      Both the dc and double-frequency critical current amplitudes are determined by the alternating  electric field $I_{dc}^c \propto E_\Omega E_{-\Omega}$ and $I_{2\Omega}^c \propto  E_{\Omega}^2$.  By the order of magnitude $I_{dc}^c,\; I_{2\Omega}^c \sim I_0 $ , where 
     \begin{align}\label{Eq:Ic}
    I_0 = - \sigma_F S (\Delta/ed_F) (2\tilde\alpha\gamma\xi_F/\pi)^2   (\Delta /T)^2  P/P_c
  \end{align}
  where $S$ is the junction area, 
 $P = c |E_\Omega|^2$ 
is the radiation power,  
 $P_c = (c\hbar /e^2) \hbar \Omega^2/\xi_S^2 $
is the radiation power needed to speed up the Cooper pairs to the depairing velocity. In Ref.~\onlinecite{Bobkova2021} it has been estimated that $I_0/(P/P_c)^2 \sim 10^{-1} - 10^{-3}$ A for typical parameters of JJs with ferromagnetic interlayers and taking $\tilde \alpha \sim 0.1-1$ 
 \cite{lo2014spin,Banerjee2018, ast2007giant,Triola2016,Cayao2018}.
Assuming $\xi_S \approx 30$ one can estimate $P_c \approx 10 (\Omega/GHz)^2$  W/m$^2$. 
Therefore such a JJ is quite sensitive to the radio-frequency and microwave irradiation. For example, a cell phone at one 
meter distance generates microwave radiation with $\Omega \approx 3-4$ GHz and $P\sim P_c$, which 
induces rather large  currents  $I_0 \sim 10^{-1} - 10^{-3}$ A. At the same time the frequency rise strongly suppresses the power sensitivity.
For the frequency of the cosmic background radiation 
$P_c \approx 10^6 $ W /m$^2$  so that the power density $P= 10^{-5}$
W /m$^2$ induces rather small critical current $I_0 \sim 10^{-12} - 10^{-15}$ A.  
However, even THz and visible light radiation sources can induce large critical current.
For example, a THz radiation with power $1$ mW /mm$^2$ yields $I_0 \sim 10^{-5} -10^{-7}$ A. 
Laser beam of the frequency about $\Omega \sim 10^{6}$ GHz 
carrying the power 
$1$ mW focused into the spot of $1$ $\mu$m$^2$ size induces the critical current $I_0 \sim 10^{-6} - 10^{-8}$ A which is well within the measurable limits.

 \begin{figure}[h!]
 \centerline{$
 \begin{array}{c}
 \includegraphics[width=2.0in]{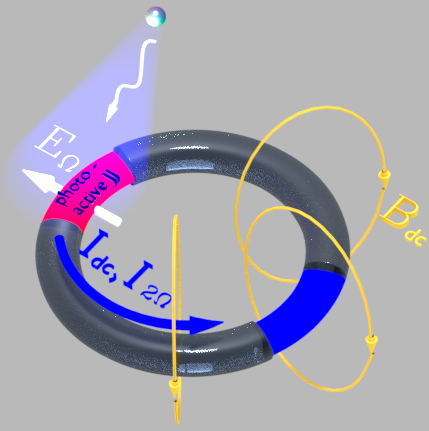} 
 \end{array}$}
 \caption{\label{fig:photo_magnetic} 
    { Sketch of the photo-magnetic SQUID}. The device consists of the photo-active Josephson junction  (red weak link) and a $\pi$-JJ (blue weak link). 
   Electric field  $\bm E_\Omega e^{i\Omega t}$ coming from the radiation source switches on both dc $I_{dc}$ and $I_{2\Omega}e^{2i\Omega t}$ components of the circulating current. The dc component produces spontaneous magnetic field $\bm B_{dc}$. Adopted from Ref.~\onlinecite{Bobkova2021}. 
}
 \end{figure}

The LRT Josephson effect, induced by  electromagnetic radiation, also provides an interesting possibility to create photo-magnetic devices based on the superconducting loops with the weak links formed by the radiation-controlled JJ, as it is sketched in Fig.~\ref{fig:photo_magnetic}. 
 In the absence of external irradiation 
 there are no currents in the loop.   Radiation switches on the photo-active JJ.  Then gradually increasing  the radiation power, it has been found that the zero-current state becomes unstable  under the following condition 
 \begin{align} \label{Eq:Instability} 
  I_{dc}^c > \frac{\Phi_0}{2\pi} \frac{\omega_0\omega_p}{\sqrt{\omega_0^2 + \omega_p^2}}
 \end{align}
  where $\omega_p = \sqrt { 2\pi I_\pi^c /C\Phi_0 }$ is the plasma frequency corresponding to the  $\pi$-JJ. Eq.~(\ref{Eq:Instability}) has been obtained under the condition $\Omega \ll \omega_0$, where $\omega_0 = 1/\sqrt{LC}$ is the eigen frequency of the superconducting loop, in order to avoid parametric effects due to the time-dependent current amplitude of the photo-assisted JJ.
  In case of the typical values $\omega_p=\omega_0 \sim 10$ GHz the threshold value in the r.h.s. of Eq.~(\ref{Eq:Instability}) about   
 $10^{-6}$ A.  Once the condition (\ref{Eq:Instability}) is satisfied the SQUID switches to the state with spontaneous dc current $I_{dc}$ and constant magnetic field $\bm B_{dc}$.
  The photo-induced magnetic flux
 magnitude was estimated as $\sim 10^{-2} \Phi_0 $. 
 
 One can also obtain the photo-magnetic response  without any threshold for the incoming power, provided the second branch of the SQUID contains the  Josephson anomalous phase junction.  Such photo-magnetic element generates dc current 
 $I_{dc} \approx I_{dc}^c \cos\varphi_0$ and the corresponding magnetic field 
 $\bm B_{dc}$ being exposed to any arbitrary small radiation power.

\section{Conclusions}

In this review, we have discussed the fundamental aspects and characteristic features of the  magnetoelectric effects, which were reported in the literature on JJs. The main focus of the review is on the manifestations of the direct and inverse magnetoelectric effects in various types of Josephson systems. The coupling of the magnetization in JJs with ferromagnetic interlayers to the Josephson current via the magnetoelectric effects and perspectives of this coupling are also discussed. 

To summarize, the direct magnetoelectric effect, that is, the current-induced spin polarization of the conductivity electrons, can arise in JJs via SOC materials, via topological insulators and also via spin-textures ferromagnets, which mathematically in the local spin basis can also  be considered as materials with SOC. The effect is a driving force of the spin torques acting on the ferromagnet inside the JJ and, therefore, is of key importance for the electrical control of the magnetization. The inverse magnetoelectric effect in JJs takes the form of the anomalous ground state phase shift and has been reported for JJs via spin-textured ferromagnets, multilayered ferromagnetic systems, homogeneous ferromagnets with SOC and combined interlayers consisting of topological insulators or materials with SOC and ferromagnets. The effect accounts for the back action of the magnetization dynamics on the Josephson subsystem, making the JJ  be in the resistive state in the presence of the magnetization dynamics of any origin. Another manifestation of the magnetoelectric effects in JJs is the generation of long-range triplet pairs in S/F/S JJs by the moving condensate, which allows for controllable and low-dissipative manipulation by the critical current of the JJ.

Although by now progress has been most pronounced on the theoretical understanding of the magnetoelectric effects in JJs, the experimental activity has in the past few years started
to catch up. In particular, a number of experiments confirmed the anomalous ground state phase shift in JJs via SOC materials and TIs under the applied Zeeman field. There is also a growing activity in the field of spin pumping experiments in superconducting hybrids and, in particular, JJs \cite{Li2018,Jeon2019_3,Golovchanskiy2020}, where some interesting results concerning the influence of the superconducting subsystem on the ferromagnetic resonance are obtained.  Nevertheless, there remains a plethora of interesting physics to investigate, and we hope that the most valuable
experiments in the near future will directly verify the role of magnetoelectric effects in S/F/S JJs thus opening a way to applications in low-dissipative spintronics. 

\label{conclusions}

\begin{acknowledgments}
The authors thank I. Rahmonov for sharing  numerical data and figures. The work was supported by RSF project No. 18-72-10135. I.V.B. also acknowledges the financial support by the Foundation for the Advancement of Theoretical Physics and Mathematics “BASIS”.
\end{acknowledgments}

\bibliography{ME_review.bib}

\end{document}